\documentclass[aps,prx,reprint,superscriptaddress]{revtex4-1}
\pdfoutput=1
\usepackage{easy-todo}
\usepackage{color}
\usepackage{graphicx}
\usepackage{amsmath}
\usepackage{amsfonts}
\usepackage{amssymb}
\usepackage{microtype}
\usepackage{verbatim}
\usepackage{bbold}
\usepackage{hyperref}
\usepackage{cleveref}
\allowdisplaybreaks
\usepackage{multirow}
\usepackage{bbm}
\usepackage{braket}
\hypersetup{
  colorlinks   = true, 
  urlcolor     = blue, 
  linkcolor    = blue, 
  citecolor   = blue 
}
\graphicspath{{./Figures/}}

\DeclareMathOperator{\tr}{tr}

\begin{document}

\author{Pieter W. Claeys}
\email{pc652@cam.ac.uk}
\affiliation{TCM Group, Cavendish Laboratory, University of Cambridge, Cambridge CB3 0HE, UK}

\author{Austen Lamacraft}
\affiliation{TCM Group, Cavendish Laboratory, University of Cambridge, Cambridge CB3 0HE, UK}

\title{Maximum velocity quantum circuits}

\begin{abstract}
We consider the long-time limit of out-of-time-order correlators (OTOCs) in two classes of quantum lattice models with time evolution governed by local unitary quantum circuits and maximal butterfly velocity $v_{B} = 1$ . Using a transfer matrix approach, we present analytic results for the long-time value of the OTOC on and inside the light cone. First, we consider `dual-unitary' circuits with various levels of ergodicity, including the integrable and non-integrable kicked Ising model, where we show exponential decay away from the light cone and relate both the decay rate and the long-time value to those of the correlation functions. Second, we consider a class of kicked XY models similar to the integrable kicked Ising model, again satisfying $v_{B}=1$, highlighting that maximal butterfly velocity is not exclusive to dual-unitary circuits.
\end{abstract}

\maketitle

\section{Introduction}

All physical systems are characterized by a maximum velocity for the propagation of influences. While the speed of light sets the ultimate bound, for many systems other velocities act as an effective speed limit. In the motion of a fluid, for example, it is the speed of sound that plays a decisive role. 

In any given nonrelativistic many-body system, however, it is not obvious that such an effective velocity exists. For quantum spin systems with finite range interactions, Lieb and Robinson \cite{Lieb:1972aa} showed that the response function giving the effect on a local observable $A(t)$ at time $t>0$ of another observable $B(0)$ is characterized by finite velocity $v_{\text{LR}}$ that is particular to the system in question. More precisely, they showed that the expectation of commutator $\langle\left[A(t),B(0)\right]\rangle$ vanishes exponentially when $A(t)$ lies outside a `light cone' originating at $B(0)$. This result has since been generalized to systems where the local Hilbert space dimension is infinite \cite{Nachtergaele:2007aa},  and interactions are long-ranged \cite{Hauke:2013aa,Foss-Feig:2015aa}. 

In recent years it has been realized that the Lieb--Robinson result does not capture every aspect of our notion of `propagation of influence'. If we perturb a fluid by displacing a single molecule, the subsequent trajectories of the surrounding molecules will begin to diverge (exponentially, since the motion is chaotic) from those of the unperturbed system. On the other hand, this effect is local, and will take time to propagate throughout an extended system. Since the reponse to the displacement will vary depending on the initial conditions, one way to quantify this effect is to look at the expectation of the square norm of the commutator $\langle\lvert\left[A(t),B(0)\right]\rvert^2\rangle$, or equivalently the \emph{out-of-time-order correlator} (OTOC) \cite{Larkin:1969aa}
\begin{equation}
  C_{AB}(t) = \langle A(t) B(0) A(t) B(0) \rangle.
  \end{equation}
The difference between these two notions of propagation can be starkly illustrated by the following example. Particles in a static disorder potential will undergo diffusion (ignoring Anderson localization in the case of quantum dynamics). Thus the density-density response is purely diffusive at times exceeding the mean free time with no notion of a finite velocity of propagation. The OTOC, however, displays ballistic propagation with a constant velocity \cite{Aleiner:2016aa,Patel:2017aa}. The velocity characterizing the growth of the OTOC is known as the `butterfly velocity' $v_{\text{B}}$ \cite{roberts_diagnosing_2015,Roberts:2015aa}.

Though it may be harder to detect, the butterfly velocity  is arguably the more fundamental measure of the spread of influence in a many-body system. A basic question is therefore: how large can $v_{\text{B}}$ be, and for what kind of system is it maximized? The purpose of this paper is to answer this question for a particular class of many body dynamical systems: those described by \emph{unitary circuits}. 

\subsection{Unitary circuits}

\begin{figure}[h]
  \centering
   \includegraphics[width=0.8\columnwidth]{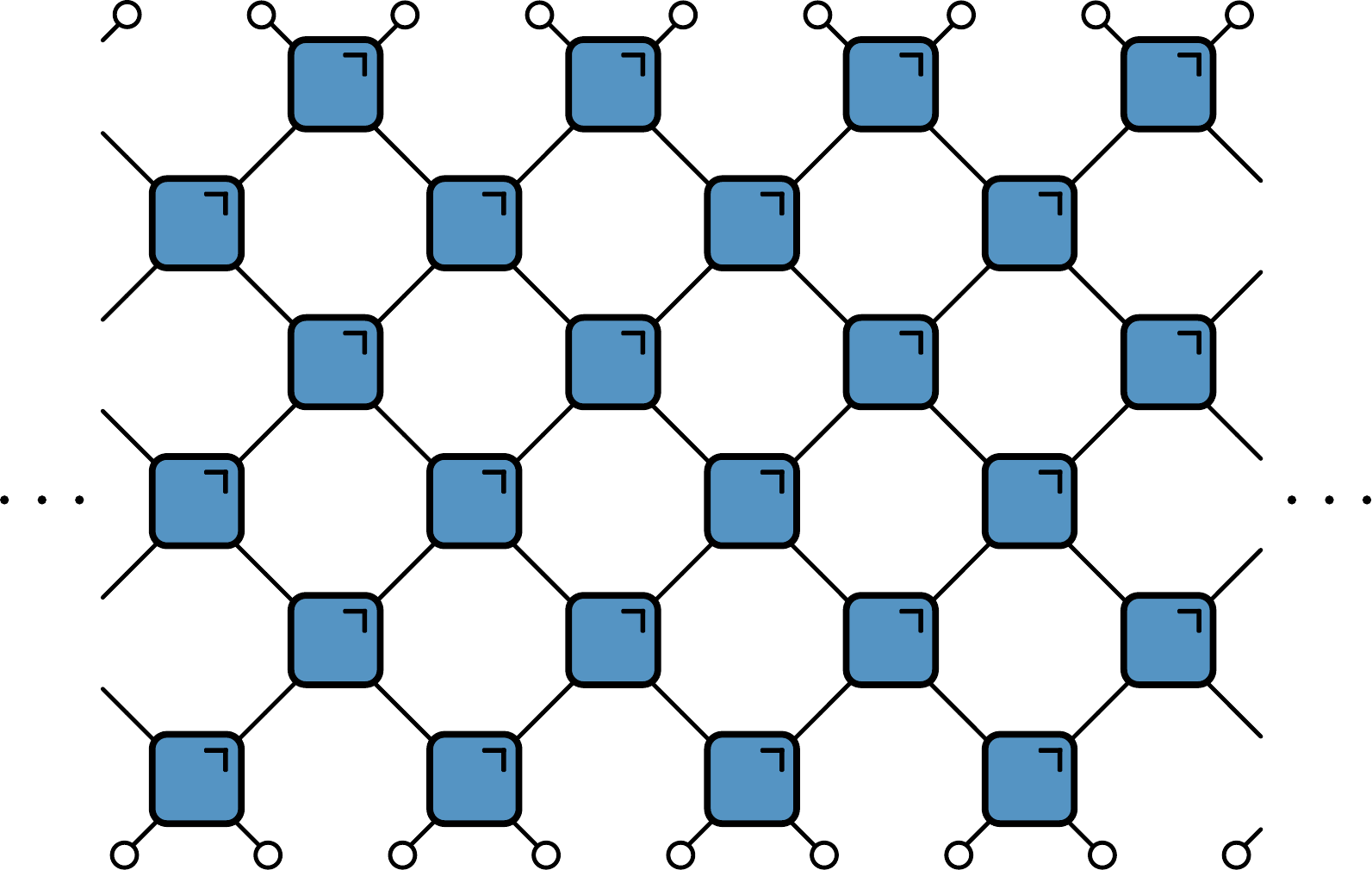}
  \caption{A `brick wall' quantum circuit. The qubits are arranged in a horiztonal row with two-qubit gates acting alternately on the even and odd links between them. Discrete time runs vertically.}
  \label{fig:brick}
\end{figure}

Consider a quantum system comprised of a large number of spin-1/2 subsystems, or qubits. The Hilbert space of the system is $\mathcal{H}=\mathbb{C}^2\otimes \mathbb{C}^2\cdots \otimes\mathbb{C}^2$,  with one factor for each qubit. A unitary circuit describes a sequence of unitary transformations -- or gates -- each acting on a subset of the qubits. Originally introduced as a model of quantum computation \cite{Nielsen:2000aa}, such circuits have been widely studied in recent years as a model for many-body dynamics \cite{Nahum:2017aa,khemani_operator_2018,von_keyserlingk_operator_2018,nahum_operator_2018,chan_solution_2018,rakovszky_sub-ballistic_2019}. The `dynamics' of a unitary circuit takes place in discrete time, but can be regarded as arising from continuous time Hamiltonian dynamics, with a Hamiltonian that may be fixed, periodically varying, or random. To impose a notion of locality on the circuit, we can insist that it consists of gates that act only on neighbouring sites in a lattice. In this work we will be concerned with `brick wall' circuits of the form shown in \Cref{fig:brick}. Note that that similar circuits, but with the qubits arranged in a square array, are the basis of Google's Sycamore processor \cite{Arute:2019aa}. 
A basic feature of these circuits is that a maximum velocity of propagation equal to 1 is intrinsic to their structure. To see this, consider the (infinite temperature) correlation function of Pauli spin operators $\sigma_\alpha$ at site $x$ and time $t\geq 0$ (both integer) and $\sigma_\beta$ at site $0$ and time 0
\begin{equation}
  c_{\alpha \beta}(x,t) = \langle \sigma_{\alpha}(x,t) \sigma_{\beta}(0,0) \rangle.
  \end{equation}
When $|x|>t$, $\sigma_{\alpha}(x,t)$ and $\sigma_{\beta}(0,0)$ commute, as none of the unitary transformations performed on $\sigma_{\alpha}(x,t)=\mathcal{U}^{\dagger}(t) \sigma_{\alpha}(x) \mathcal{U}(t)$ act on the tensor factor corresponding to site $0$. Thus $c_{\alpha \beta}(|x|>t,t)=0$ by the tracelessness of the Pauli operators. For similar reasons the OTOC
\begin{equation}
  C_{\alpha \beta}(x,t) = \langle \sigma_{\alpha}(0,t) \sigma_{\beta}(x,0) \sigma_{\alpha}(0,t) \sigma_{\beta}(x,0) \rangle.
  \end{equation}
satisfies $C_{\alpha \beta}(|x|>t,t)=1$ since $\sigma_{\alpha}(0,t)$ commutes with $\sigma_{\beta}(|x|>t,0)$. For smaller $|x|$ the OTOC will begin to deviate from 1. As $|x|$, $t\to \infty$, the value of $|x|/t$ where this deviation occurs defines the butterfly velocity $v_\text{B}$. Generally, $v_\text{B}<1$. To illustrate the range of possible behaviour we consider some examples:

\begin{enumerate}
  \item If the gates are taken to be random unitary matrices the butterfly velocity was found to be (on average)
  \begin{equation}
    v_\text{B} = \frac{q^2-1}{q^2+1},
  \end{equation}
  where $q$ is the local Hilbert space dimension ($q=2$ for qubits) \cite{Nahum:2017aa}. Thus $v_\text{B}\to 1$ only as $q\to\infty$.

  \item If the circuit is designed to simulate a Hamiltonian $H=\sum_j h_{j,j+1}$ consisting of terms acting on neighbouring sites then the time evolution operator may be approximated for short times as
  \begin{equation}
    e^{-iH\Delta t}\approx e^{-iH_\text{e}\Delta t}e^{-iH_\text{o}\Delta t},
  \end{equation}
  where $H_\text{e}=\sum_j h_{2j,2j+1}$ and $H_\text{o}=\sum_j h_{2j+1,2j}$ act on the odd and even layers of the circuit. As $\Delta t\to 0$ the circuit approximates the continuous time evolution more accurately, but any finite velocity implied by the Hamiltonian $H$ corresponds to a vanishing velocity in `gate time'. 

  \item A simple example of a circuit with $v_\text{B}=1$ is one that consists only of SWAP gates 
  \begin{equation}
    U_{\text{SWAP}}\ket{s_1}_1\ket{s_2}_2 = \ket{s_2}_1\ket{s_1}_2.
  \end{equation}
  Of course, such a circuit is not particularly interesting: it generates no entanglement between the qubits. 

\end{enumerate}

In this paper we are concerned with the question of when $v_\text{B}$ achieves the maximum velocity 1. We will refer to such circuits as \emph{maximum velocity circuits} (MVCs).

In light of above examples, one may ask whether the family of MVCs has any members with nontrivial (entangling) dynamics. In the next section we will describe a class of circuits which answers this question in the affirmative.

How does the OTOC behave for circuits with $v_\text{B}<1$? In one dimension (qubits in a row) it was established, first for random circuits \cite{nahum_operator_2018,rakovszky_diffusive_2018}, and then for continuous time models \cite{Rowlands:2018aa}, that for $|x|/t\sim v_\text{B}$ the OTOC displays diffusive broadening 
\begin{equation}
  C_{\alpha\beta}(x\sim v_\text{B}t,t)\underset{t\to\infty}{\longrightarrow} \mathcal{C}\left(\frac{x-v_\text{B}t}{2D\sqrt{t}}\right),
\end{equation}
where $\mathcal{C}(x)=\frac{1}{2}\left(1+\text{erf}(x)\right)$ is written in terms of the error fuction $\text{erf}(x)$ and $D$ is a (nonuniversal) diffusion constant.

Since MVCs have the maximal $v_\text{B}=1$, there is no room for broadening of the front, as this would lead to $C\neq 1$ outside the light cone. Our results demonstrate this by explicit calculation (see \Cref{fig:OTOC}).

\subsection{Dual-Unitary Circuits}

Analytically tractable models of many-body quantum dynamics are scarce. In Refs.~\cite{bertini_exact_2018,bertini_entanglement_2019} it was shown that the kicked Ising model (KIM) at particular values of the coupling constants was amenable to exact calculation of the spectral form factor and entanglement entropies (starting from certain initial conditions in the latter case). In particular, the entanglement calculation showed that all eigenvalues of the reduced density matrix are equal. By casting the KIM as a unitary circuit, Ref.~\cite{gopalakrishnan_unitary_2019} showed that this degeneracy of the entanglement spectrum was a consequence of a property of the model now called \emph{dual unitarity}. As well as being unitary, dual unitary gates are also unitary when interpreted as generating evolution in the spatial direction (see \Cref{sec:dual}). Ref.~\cite{bertini_exact_2019} explored the full family of dual unitary gates for qubits (which turns out to be 14-dimensional, compared to the 16 dimensions of the group $U(4)$). Subsequent work on dual unitary circuits has studied the behaviour of operator entanglement \cite{bertini_operator_2019}, the properties of matrix product state initial conditions that preserve the solubility of the dynamics \cite{piroli_exact_2019}, and new realizations of dual unitarity for local dimension $q>2$ \cite{rather_creating_2019,gutkin_local_2020}. The property of dual-unitarity is equivalent to maximal operator entanglement, meaning that the gate can create maximum entanglement when acting on product states \cite{rather_creating_2019}.

As we have already explained,  unitarity guarantees that correlations are non-zero only within the light cone with velocity 1 \cite{chan_solution_2018}. The fundamental observation of Ref.~\cite{bertini_exact_2019} was that a circuit that is both unitary and dual unitary has correlations vanishing everywhere but exactly on the light cone, where correlations may be constant, oscillating, or decaying. This raises the natural questions:

\begin{enumerate}
  \item Are dual-unitary circuits generally MVCs?
  \item If so, do dual-unitary circuits exhaust the class of MVCs, or are there MVCs that do not share the other features of dual-unitary circuits?
\end{enumerate}

\subsection{Summary of results}

\subsubsection{Dual-unitary circuits are MVCs}

We show that the answer to the first of our questions is yes: dual-unitary circuits generically have $v_\text{B}=1$. We demonstrate this by computation of the OTOC $C_{\alpha\beta}(x,t)$ inside the light cone for $x+t\to\infty$ (recall that $C_{\alpha\beta}(x,t)=1$ outside the light cone).

\begin{figure}[h]
  \centering
  \includegraphics[width=0.9\columnwidth]{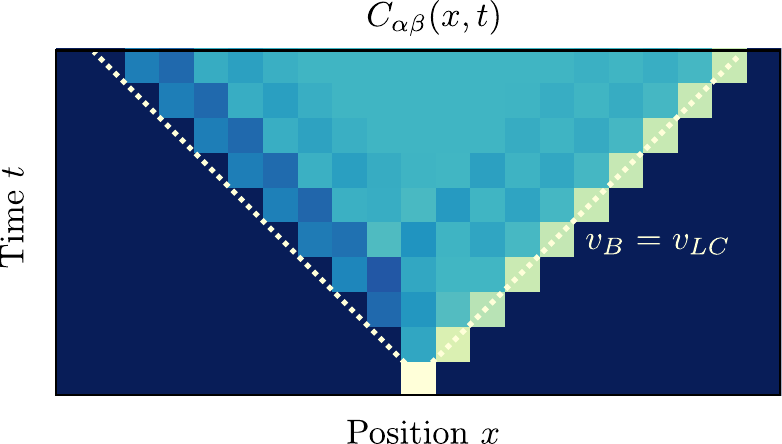}
  \caption{Typical behaviour of the OTOC in a maximal velocity circuit (in this case the kicked Ising model) with $v_B=1$, where a maximal value is reached on the light cone and the OTOC decays exponentially away from the light cone.}
  \label{fig:OTOC}
\end{figure}

In all cases the OTOC has a strong parity effect, being independent of $x-t$ for $x-t$ even and $x \neq t$. For $x-t$ odd the OTOC may be expressed in terms of the same quantum channel that determines the correlation functions on the light cone in dual-unitary circuits \cite{bertini_exact_2019}, and the possible behaviour of the OTOC moving inside the light cone is inherited from this quantum channel. For example, it may tend to zero (see \Cref{fig:OTOC}), to a non-zero constant, or oscillate. Within the class of dual-unitary circuits the kicked Ising model (KIM) is distinguished: in this case we find the action of the quantum channel explicitly and evaluate the OTOC exactly inside the light cone at $x+t\to\infty$, where it decays with increasing $t-x$. A further specialization is to the KIM at the integrable point where we evaluate the OTOC exactly for all values of $x+t$.

It is natural to conjecture that $v_\text{B}=1$ is a defining feature of the dual-unitary family that goes hand in hand with their other properties. 

\subsubsection{MVCs that are not dual unitary}

In fact, this is not the case. Our second contribution is to identify a family of models which is not dual-unitary but for which the calculation of the OTOC at $x+t\to\infty$ is tractable and yields $v_\text{B}=1$. This model can be regarded as a kind of kicked XY model. Unlike the dual-unitary case, the behaviour of the OTOC is decoupled from that of the correlation functions. While the correlation functions are generally exponentially decaying, the OTOC oscillates with period 4 without decay inside the light cone. 

\subsection{Outline}

The outline of the remainder of this paper is as follows. In \Cref{sec:CorrelationFunctions} we introduce the formalism that we use for calculations and, as a a warm-up, demonstrate how it may be used to calculate correlation functions on the light cone for arbitrary unitary circuits. \Cref{sec:OTOC} generalizes the formalism to the OTOCs and demonstrates that generically $v_\text{B}<1$, identifying the conditions required for $v_\text{B}=1$. We next calculate the OTOC for dual unitary circuits (\Cref{sec:dual}) and the new family of circuits with $v_\text{B}=1$, the kicked XY models (\Cref{sec:XY}). \Cref{sec:conclude} presents our conclusions. 

A Python implementation of all presented calculations is available online \footnote{\href{https://github.com/PieterWClaeys/UnitaryCircuits}{https://github.com/PieterWClaeys/UnitaryCircuits}}.

\section{Correlation functions}
\label{sec:CorrelationFunctions}
As a warm-up, and to introduce the graphical calculus, we consider correlation functions of the form
\begin{equation}\label{eq:def_corr}
c_{\alpha \beta}(x,t) = \langle \sigma_{\alpha}(x,t) \sigma_{\beta}(0,0) \rangle,
\end{equation}
at infinite temperature, $\langle \mathcal{O} \rangle = \tr( \mathcal{O})/\tr(\mathbbm{1})$, and where the set $\{\sigma_{\alpha}, \alpha = 0 \dots q^2-1\}$ presents an orthonormal local operator basis for a local $q$-dimensional Hilbert space, satisfying $\tr\left(\sigma_{\alpha}\sigma_{\beta}\right)/q = \delta_{\alpha \beta}$. It is particularly convenient to choose $\sigma_{0} = \mathbbm{1}$ such that all other operators within this basis are necessarily traceless, similar to the Pauli matrices for local qubits with $q=2$. 

The time evolution is governed by a unitary circuit consisting of two-site operators, where each gate $U$ can be graphically represented as
\begin{align}
\vcenter{\hbox{\includegraphics[width=0.8\linewidth]{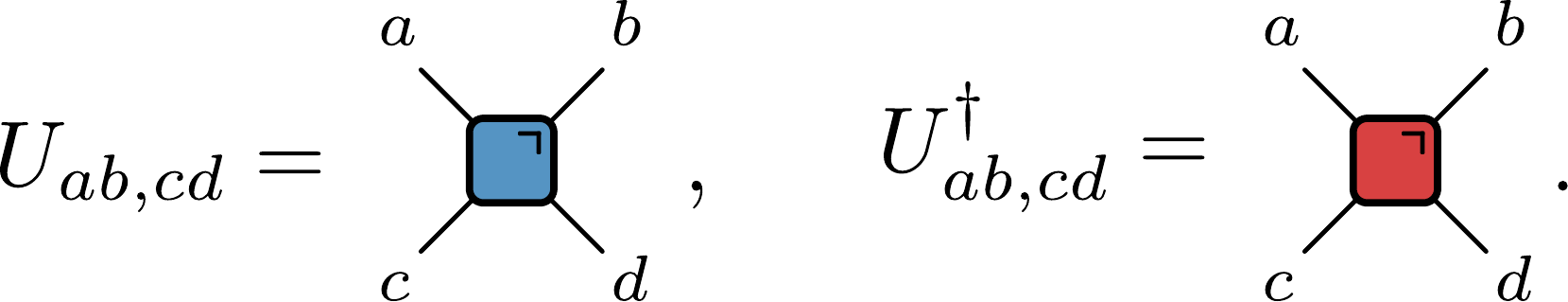}}} \label{eq:def_U_Udag}
\end{align}
In this notation each leg carries a local $q$-dimensional Hilbert space, and the indices of legs connecting two operators are implicitly summed over (see e.g. Ref.~\cite{orus_practical_2014}). With this convention, the full evolution $\mathcal{U}(t)$ at time $t$ consists of the $t$-times repeated application of staggered two-site gates
\begin{align}
\vcenter{\hbox{\includegraphics[width=0.8\linewidth]{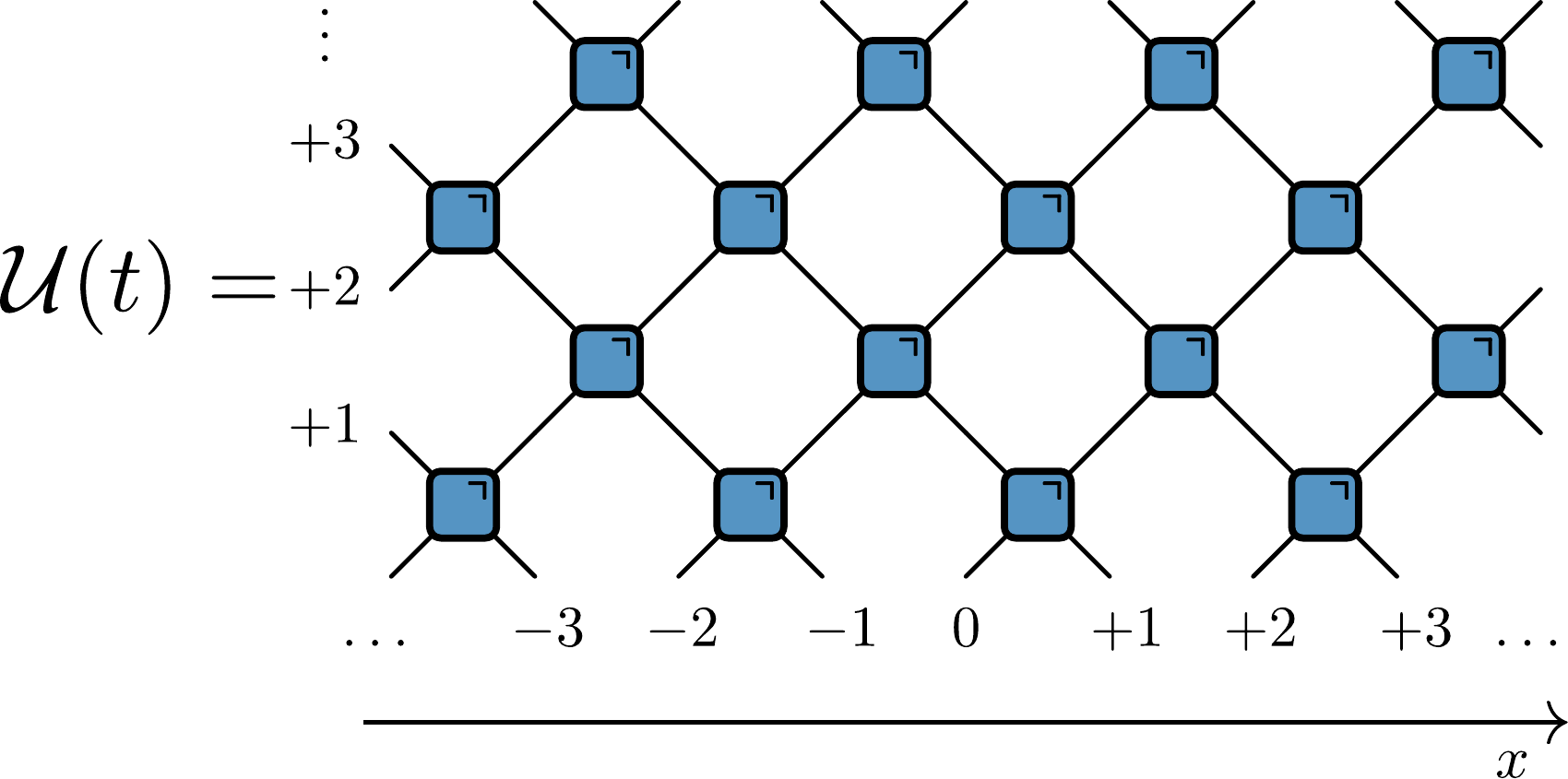}}} \label{eq:def_U_tot}\, .
\end{align}
These can extend arbitrarily far in the $x$-direction, such that all presented results will hold in the thermodynamic limit of infinite system size. A simple example that will turn out to be relevant in the context of OTOCs is given by correlators on the light cone ($x=t$). Forgetting about constant prefactors for the time being,
\begin{align}
 \langle \sigma_{\alpha}(t,t) \sigma_{\beta}(0,0) \rangle \propto \tr\left(\mathcal{U}^{\dagger}(t) \sigma_{\alpha}(x=t) \mathcal{U}(t) \sigma_{\beta}(0)\right),
\end{align}
which can be graphically expressed as

\begin{align}
\vcenter{\hbox{\includegraphics[width=0.6\linewidth]{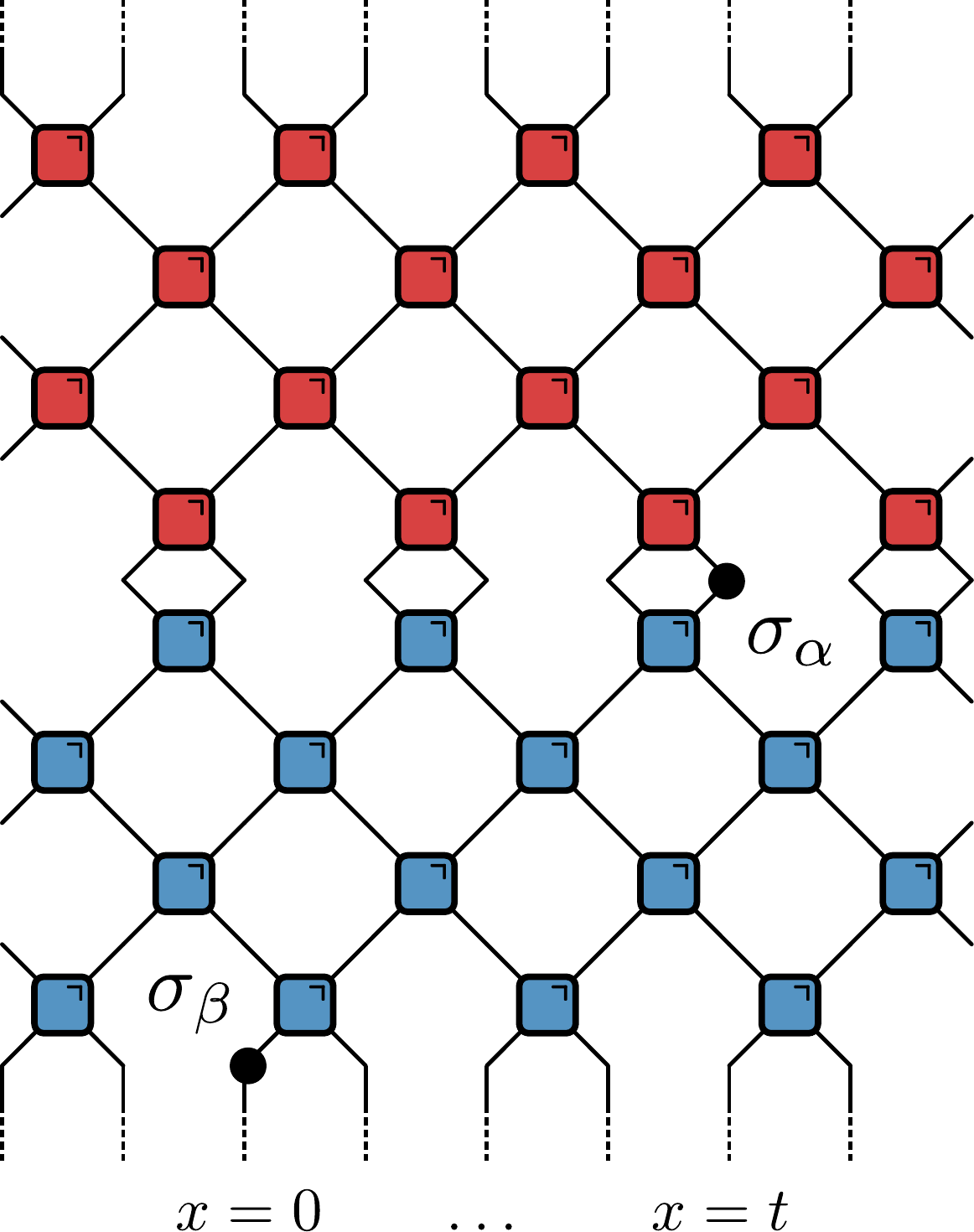}}} \label{eq:corr_1}\, ,
\end{align}
where the circuit has periodic boundary conditions in the $x$-direction and taking the trace corresponds to connecting the legs at the top and bottom. Here we also introduced a graphical notation for the (one-site) operators $\sigma_{\alpha,\beta}$ \footnote{While this diagram is explicitly symmetric under exchange of the indices, this is not necessarily the case for the operator itself, which should be taken into account when evaluating the diagrams.} as 
\begin{align}
\vcenter{\hbox{\includegraphics[scale=0.4]{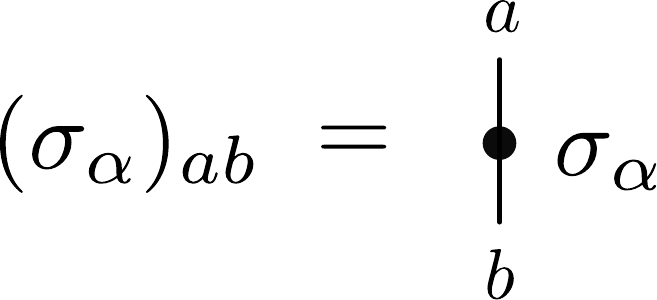}}} \label{eq:def_op}\, ,
\end{align}
where taking the trace can be graphically represented as
\begin{align}
\vcenter{\hbox{\includegraphics[scale=0.4]{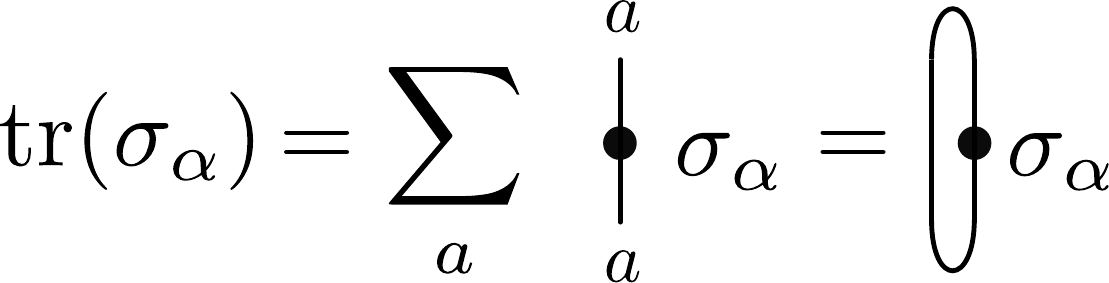}}} \label{eq:def_trace}\, .
\end{align}
The unitarity $\left(U^{\dagger} U \right)_{ab,cd} = \left(U U^{\dagger}\right)_{ab,cd} = \delta_{ac}\delta_{bd}$ similarly has a straightforward graphical representation
\begin{align}
\vcenter{\hbox{\includegraphics[width=0.6\columnwidth]{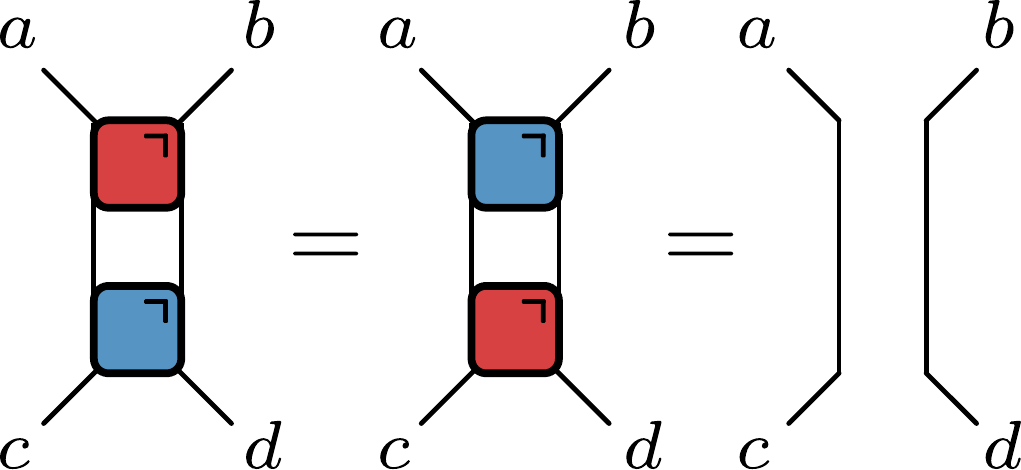}}} \label{eq:unitarity}\, .
\end{align}

Identifying all places where the unitarity of the underlying circuits can be used in this way, the correlator \eqref{eq:corr_1} simplifies to
\begin{align}
\vcenter{\hbox{\includegraphics[width=0.7\linewidth]{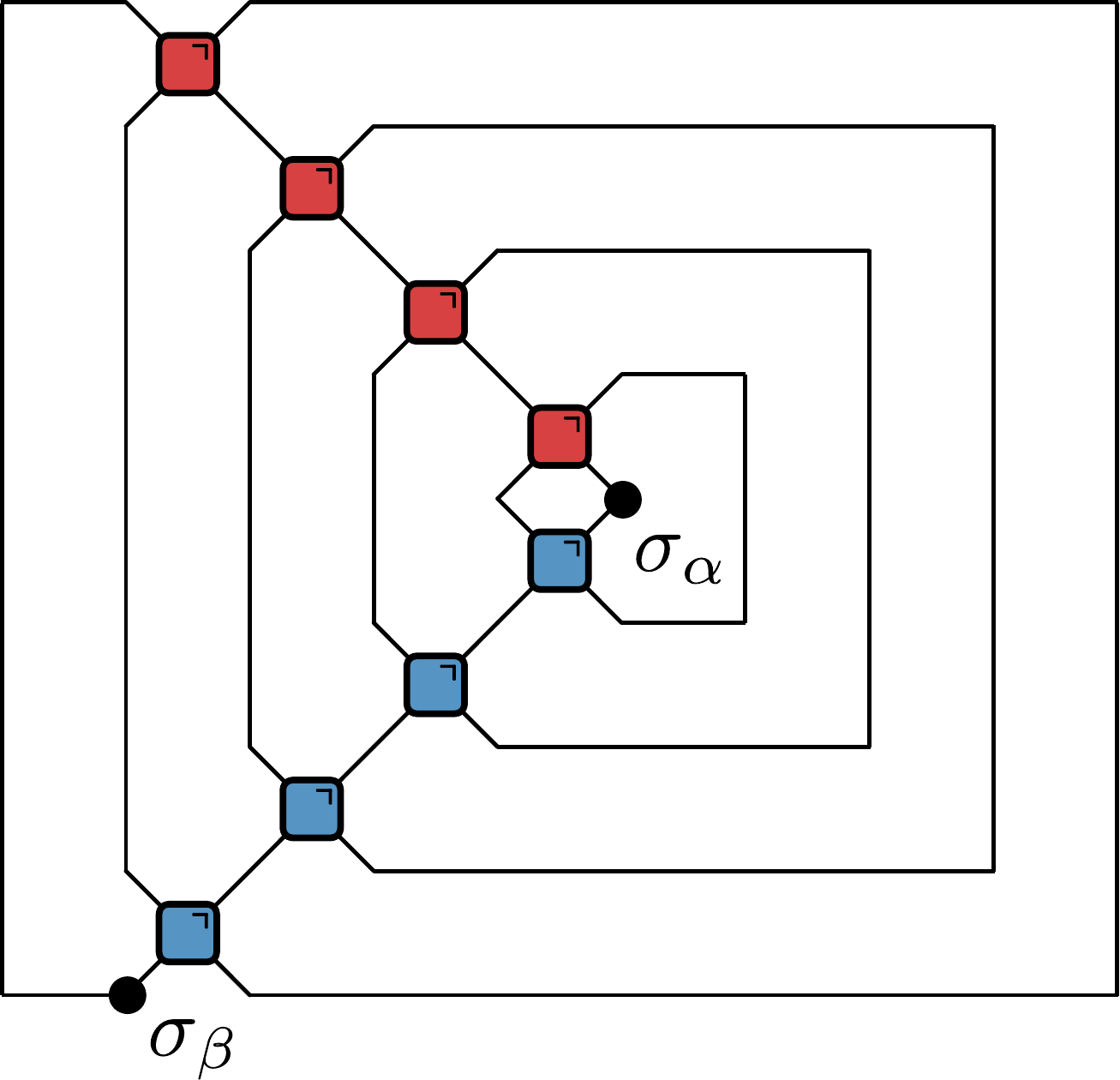}}} \label{eq:corr_2}\, \, .
\end{align}
The missing prefactor can be easily obtained as $1/q^{t+1}$ by noting that this prefactor does not depend on the choice of $\sigma_{\alpha,\beta}$ and the correlator simplifies to $\langle \mathbbm{1}\rangle = 1$ for $\sigma_{\alpha} = \sigma_{\beta} = \mathbbm{1}$, where the above diagram simplifies to $q^{t+1}$. The diagram in Eq.~\eqref{eq:corr_2} can be deformed to a more compact notation, returning $\langle \sigma_{\alpha}(t,t) \sigma_{\beta}(0,0) \rangle$
\begin{align}
\vcenter{\hbox{\includegraphics[width=0.8\linewidth]{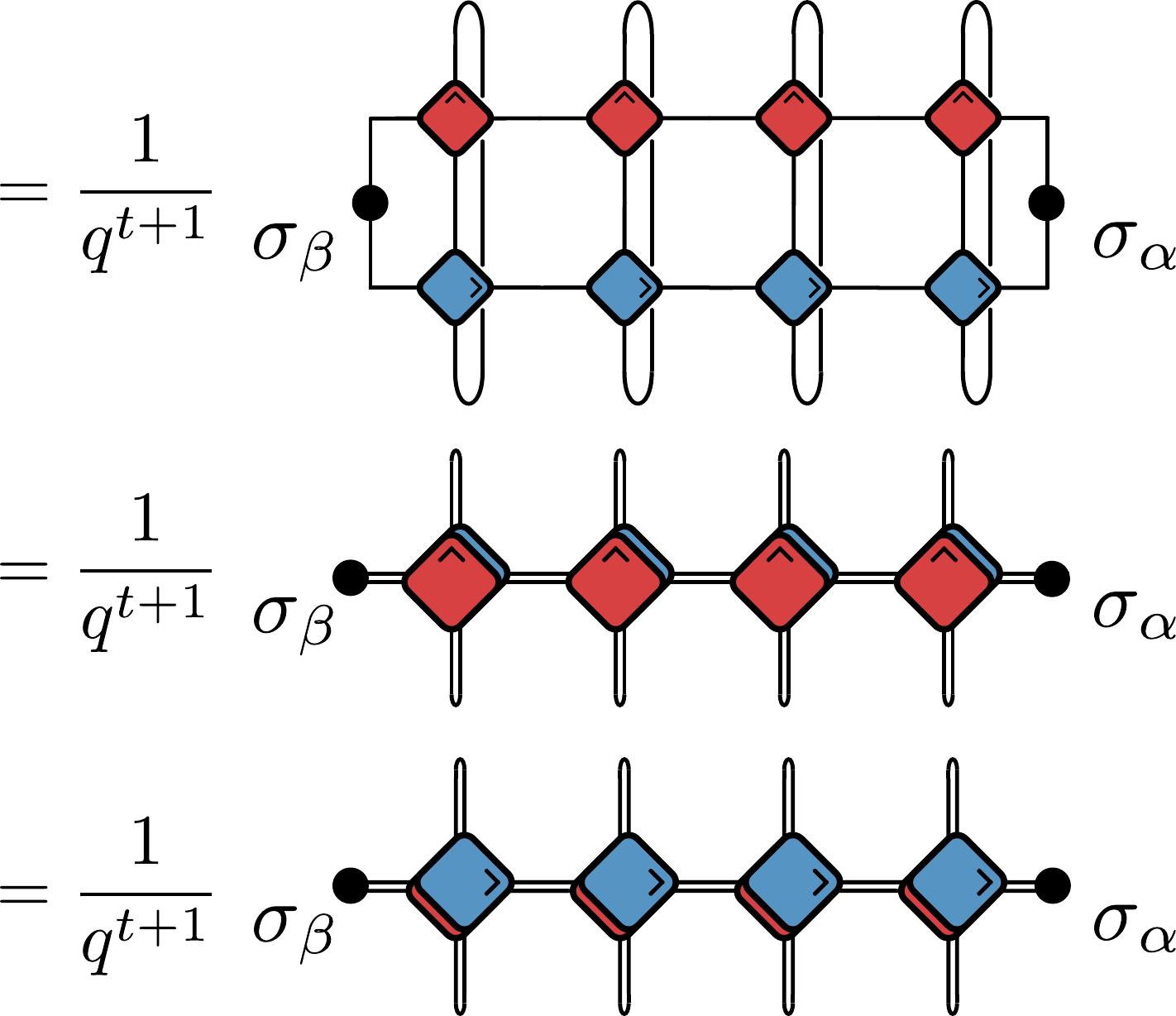}}} \label{eq:corr_3}\, ,
\end{align}
where we have introduced `folded' representations of the unitaries $U^* \otimes U$ and $U \otimes U^*$ as
\begin{align}
\vcenter{\hbox{\includegraphics[width=0.8\linewidth]{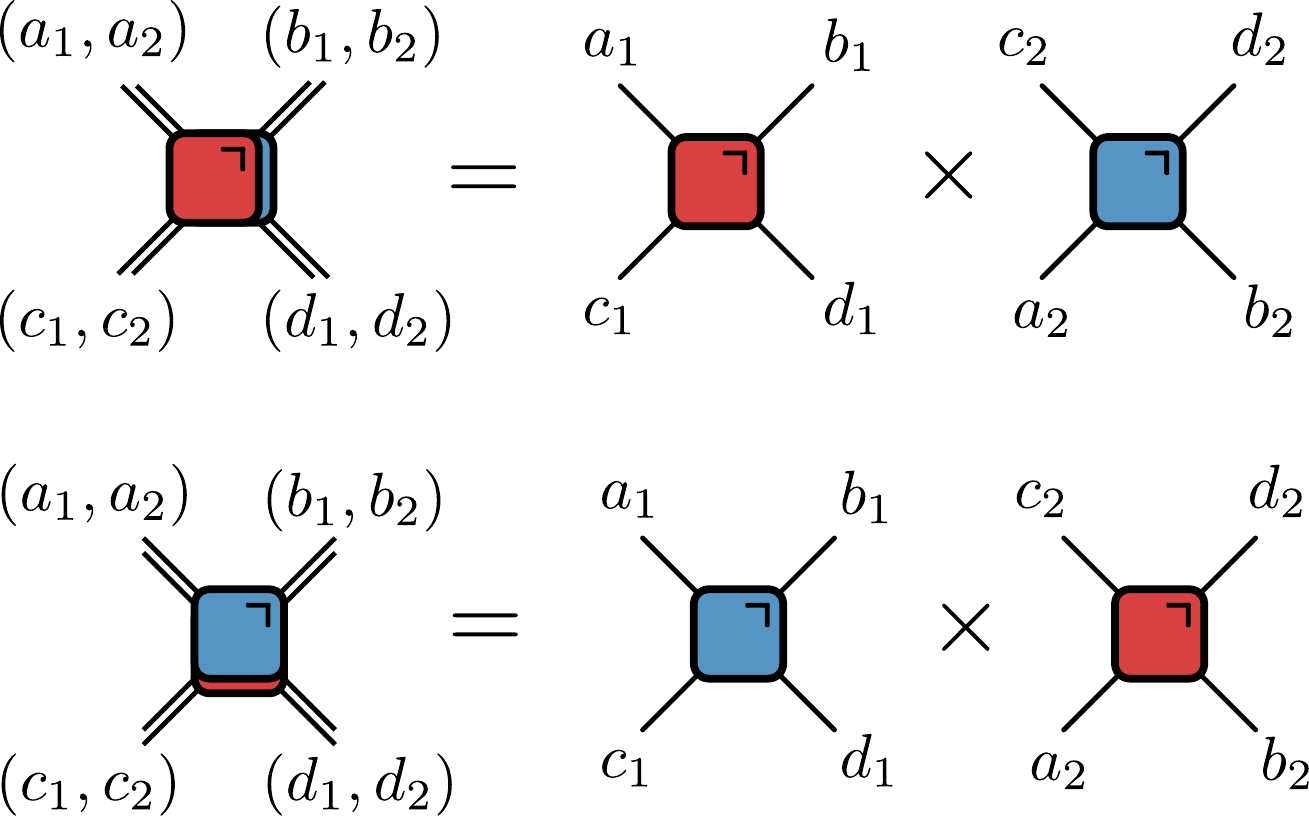}}} \label{eq:def_foldedU}.
\end{align}
The final expression for the correlator can be interpreted as the $t$-times repeated action of a linear map $\mathcal{M}_{\pm} \in \mathbb{C}^{q^2 \times q^2}$ acting on either $\sigma_{\alpha}$ or $\sigma_{\beta}$, subsequently traced out with the other operator, as
\begin{align}\label{eq:CorrChannels}
\langle \sigma_{\alpha}(t,t) \sigma_{\beta}(0,0) \rangle &= \tr \left[\sigma_{\beta}\mathcal{M}_{-}^t(\sigma_{\alpha})\right] / q \\
&=  \tr \left[ \sigma_{\alpha}\mathcal{M}_{+}^{t}(\sigma_{\beta})\right] / q,
\end{align}
where $\mathcal{M}_{\pm}$ are hermitian conjugate and defined as
\begin{align}\label{def:channels}
&\mathcal{M}_{+}(\sigma) = \tr_{1}\left[U(\sigma \otimes \mathbbm{1})U^{\dagger}\right]/q, \\
&\mathcal{M}_{-}(\sigma) = \tr_{2}\left[U^{\dagger}(\mathbbm{1} \otimes \sigma )U\right]/q.
\end{align}
Defined in this way, $\mathcal{M}_{\pm}$ is a completely positive and trace-preserving map, such that it acts as a quantum channel. From the unitarity it also immediately follows that $\mathcal{M}_{\pm}(\mathbbm{1}) = \mathbbm{1}$, such that these channels are furthermore unital.

These can also be graphically represented as
\begin{align}
\vcenter{\hbox{\includegraphics[width=0.8\linewidth]{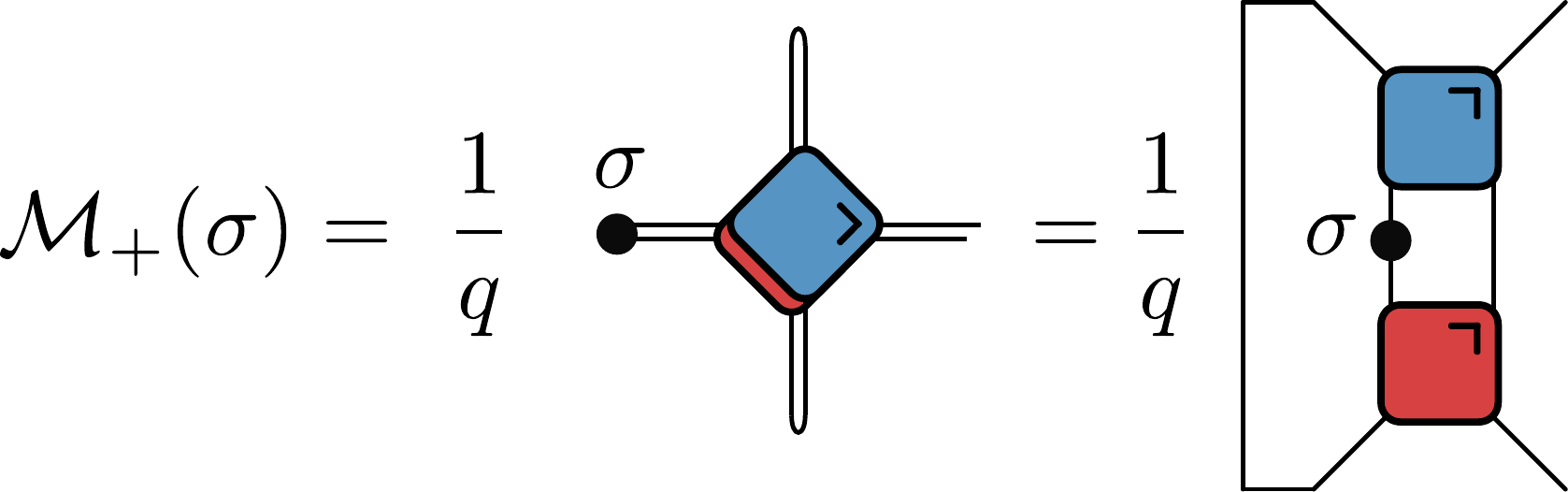}}}\, , \\ \nonumber \\
\vcenter{\hbox{\includegraphics[width=0.8\linewidth]{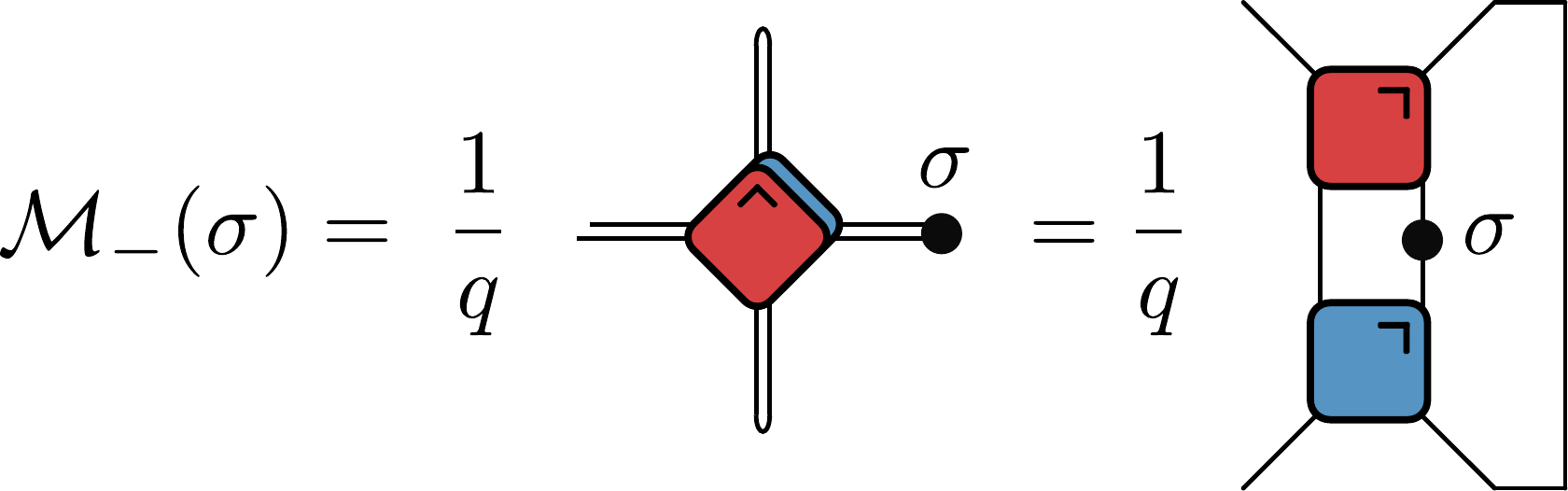}}}\, .
\end{align}

\begin{figure}[htb!]
\includegraphics[width=\columnwidth]{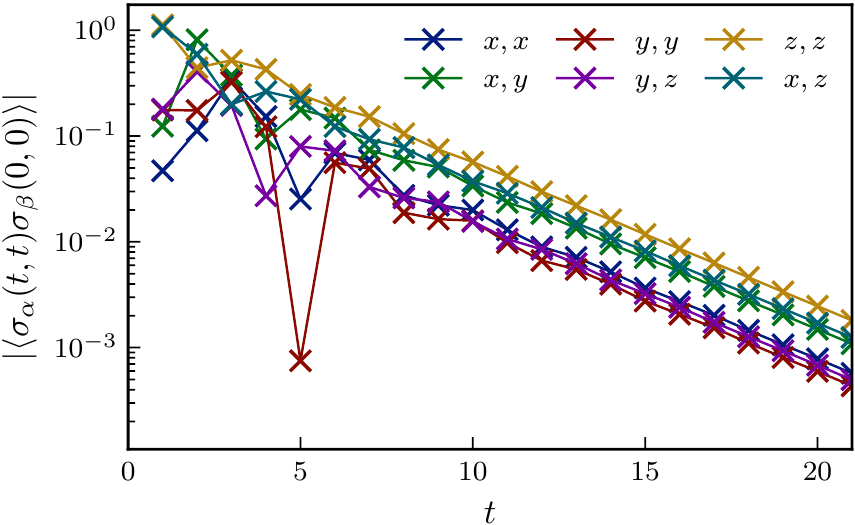}
\caption{Evolution of $\langle \sigma_{\alpha}(t,t) \sigma_{\beta}(0,0) \rangle$ for $t>0$, with $q=2$ and $\sigma_{\alpha}$ and $\sigma_{\beta}$ Pauli matrices, where the legend denotes $\alpha,\beta$. Exponential decay can be clearly observed after an initial transient regime, where all correlation functions decay at the same rate. The two-qubit gate $U$ is parametrized as in Appendix \ref{app:dualunit}. \label{fig:corr}}
\vspace{-1\baselineskip}
\end{figure}

This allows for a straightforward evaluation of the correlation functions on the light cone at long times, as illustrated in Fig. \ref{fig:corr}, where the long-time behaviour will generally exhibit exponential decay dominated by the eigenoperators of $\mathcal{M}_{\pm}$ with largest eigenvalue. While this was already pointed out for dual-unitary circuits in Ref.~\cite{bertini_exact_2019}, where all correlators that do not lie on the light cone vanish, this construction of correlation functions in terms of quantum channels holds for general unitary circuits.

\section{Out of time order correlators}\label{sec:OTOC}

In the following, we will consider out-of-time-order correlators (OTOCs) for unitary quantum circuits \cite{Larkin:1969aa,Bohrdt:2017aa,syzranov_out--time-order_2018,swingle_unscrambling_2018,von_keyserlingk_operator_2018,nahum_operator_2018,rakovszky_diffusive_2018,zhou_entanglement_2019}. These present a natural extension of the correlation functions~\eqref{eq:def_corr} and are defined as
\begin{equation}
C_{\alpha \beta}(x,t) = \langle \sigma_{\alpha}(0,t) \sigma_{\beta}(x,0) \sigma_{\alpha}(0,t) \sigma_{\beta}(x,0) \rangle.
\end{equation}
Whereas correlation functions are a measure for how excitations in a system relax towards equilibrium, OTOCs are a measure for chaos and the scrambling of quantum information. Their name follows from the fact that they contain two copies of both $\mathcal{U}$ and $\mathcal{U}^{\dagger}$, unlike the correlation functions where a single copy of each is present. 

As shown in Appendix \ref{app:derivationOTOCdiag}, explicitly writing out the OTOC and making use of the unitarity leads to a diagram for the OTOC that consists only of the gates lying in the intersection of the light cones of $\sigma_{\alpha}$ and $\sigma_{\beta}$. Again recasting these diagrams in a folded version leads to two possible expressions $C_{\alpha \beta}(x,t) = C^{\pm}_{\alpha \beta}(x,t)$,
\begin{align}
\vcenter{\hbox{\includegraphics[width=1.\linewidth]{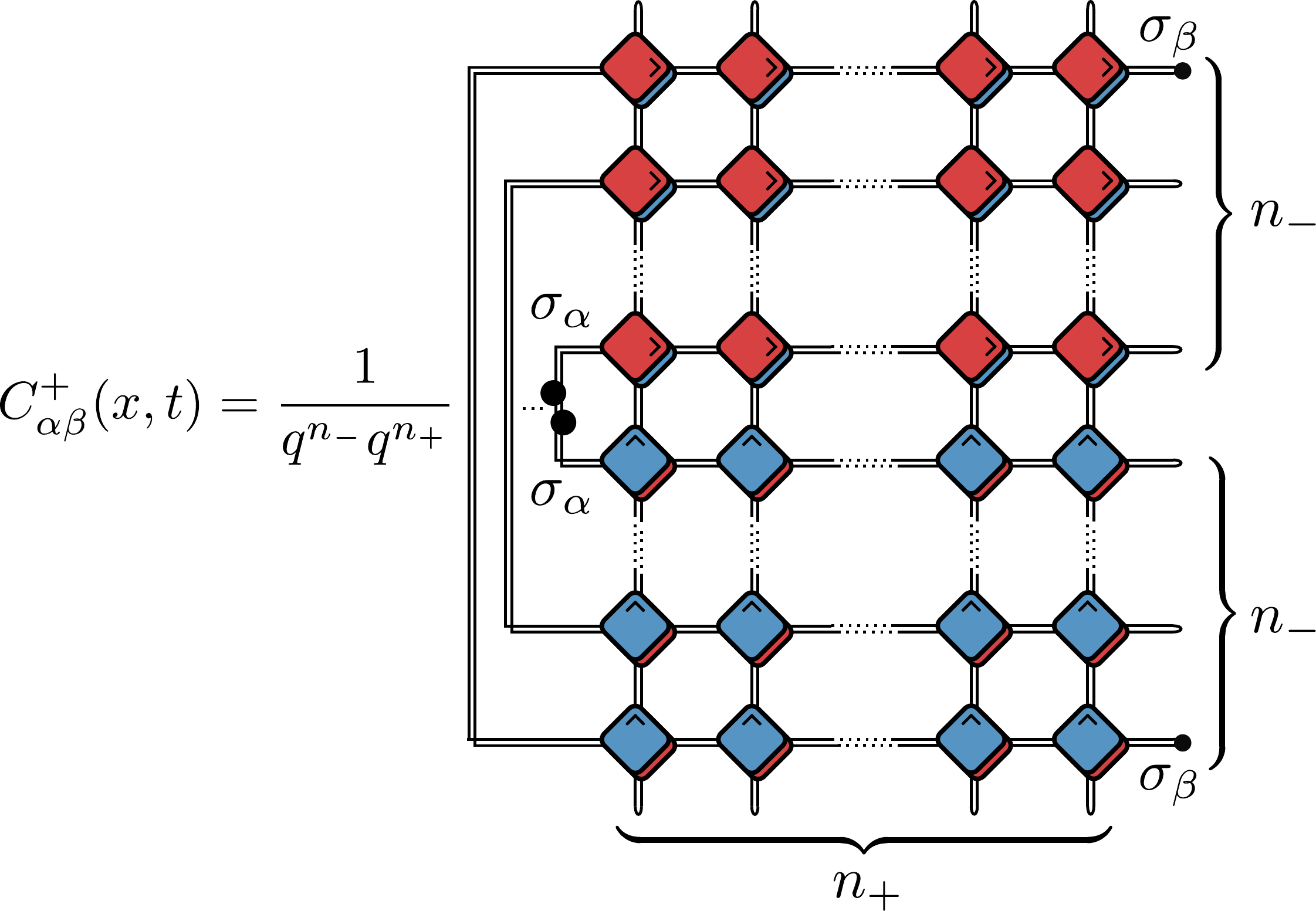}}} \label{eq:diag_def_LTR_even}
\end{align}
for $(x-t)$ even, where we define $n_{+} = (t+x)/2$ and $n_{-} = (t-x+2)/2$, and
\begin{align}
\vcenter{\hbox{\includegraphics[width=1.\linewidth]{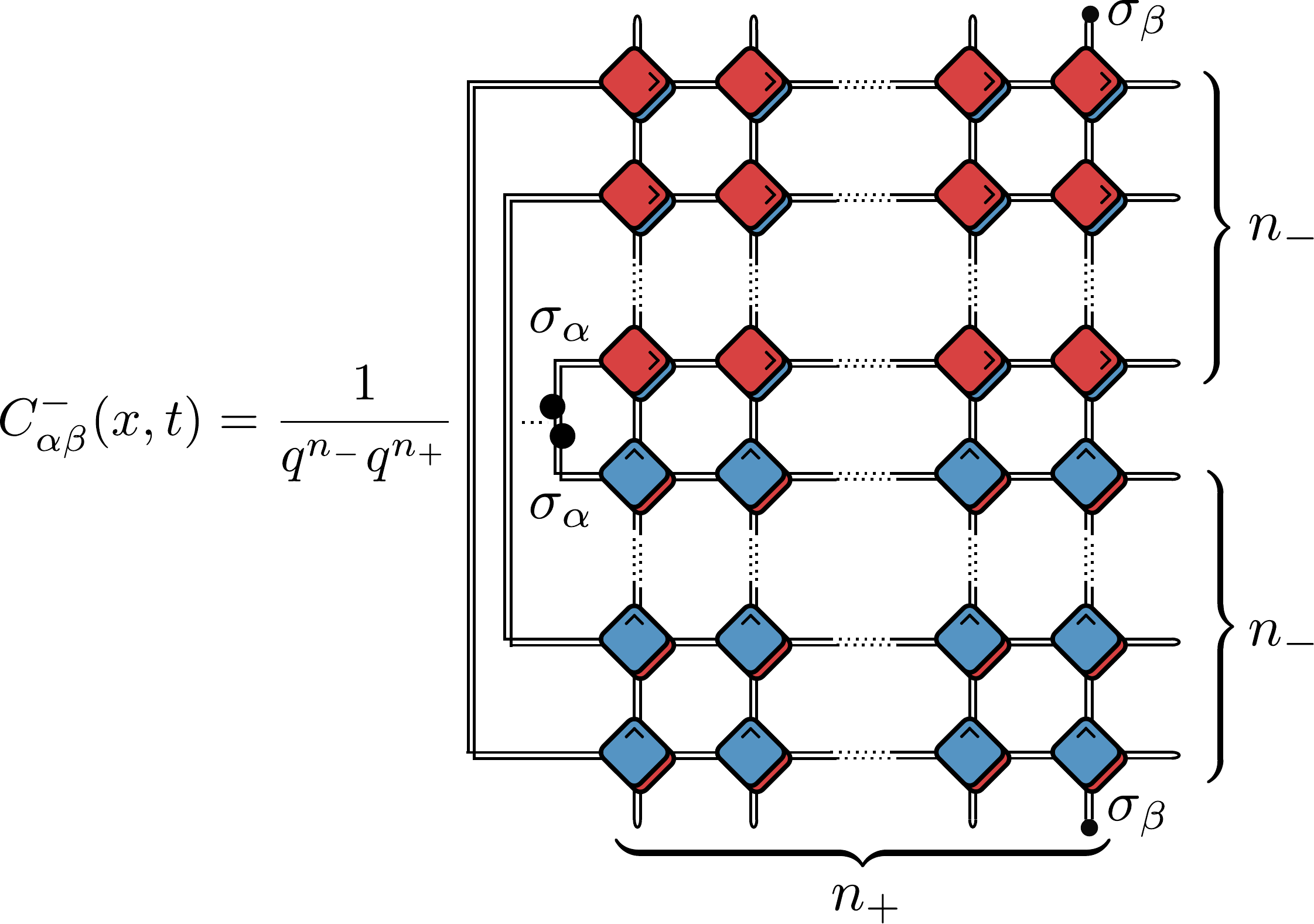}}} \label{eq:diag_def_LTR_odd}
\end{align}
for $(x-t)$ odd, with now $n_{+} = (t+x+1)/2$ and $n_{-} = (t-x+1)/2$. At finite times, the numerical evaluation of such diagrams is typically exponentially hard, and analytic results for OTOCs generally relie on either randomness in the underlying unitaries or cluster expansions in a fixed realization of a circuit. Here, we will be interested in the profile of the OTOC at long times and at fixed distances of $\sigma_{\beta}$ from the light cone of $\sigma_{\alpha}$, where analytic results can be obtained. Considering the right edge of the light cone (results are similar for the left edge), this corresponds to taking the limit $n_{+} \to \infty$ while keeping $n_{-} = n$ fixed. 

With this limit in mind, the expressions $C^{\pm}_{\alpha \beta }(x,t)$ can be reinterpreted as 
\begin{align}
&C^{+}_{\alpha \beta }(x,t) = \left(L(\sigma_{\alpha})\right|\left(T_{n_-}\right)^{n_+}\left|R^{-}(\sigma_{\beta})\right), \nonumber\\
&C^{-}_{\alpha \beta }(x,t) = \left(L(\sigma_{\alpha})\right|\left(T_{n_-}\right)^{(n_+-1)}\left|R^{+}(\sigma_{\beta})\right),
\end{align}
where all information about the long-time behaviour of the OTOC is encoded in the same column transfer matrix $T_n \in \mathbbm{C}^{q^{4n} \times q^{4n}}$ (which we here rotate by 90 degrees for ease of notation)
\begin{equation}
\vcenter{\hbox{
\includegraphics[width=0.8\linewidth]{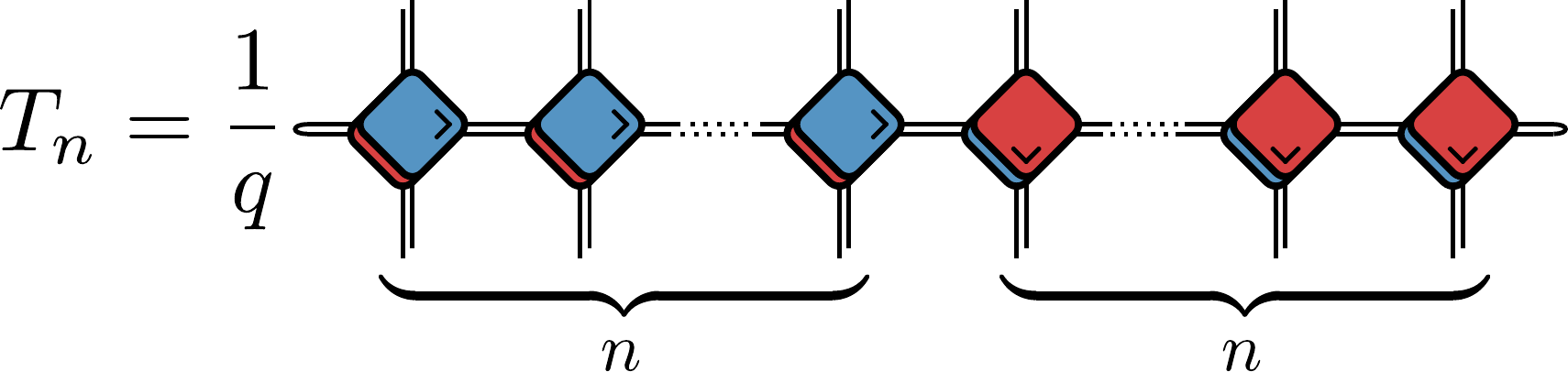}}} \label{eq:def_Tmat},
\end{equation}
and the left boundary  $(L_n(\sigma_{\alpha})|\in \mathbbm{C}^{q^{4n}}$ (similarly rotated) is given by
\begin{align}
\vcenter{\hbox{\includegraphics[width=0.8\linewidth]{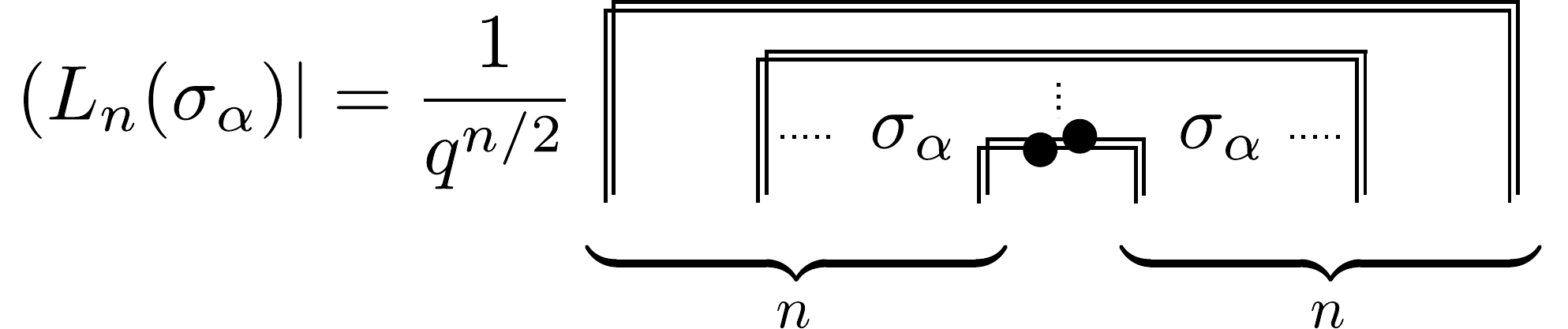}}} \label{eq:diag_bound_L}.
\end{align}
The right boundary $|R_n^{\pm}(\sigma_{\beta})) \in \mathbbm{C}^{q^{4n}}$ depends explicitly on the parity of $(x-t)$, leading to
\begin{align}
\vcenter{\hbox{\includegraphics[width=0.8\linewidth]{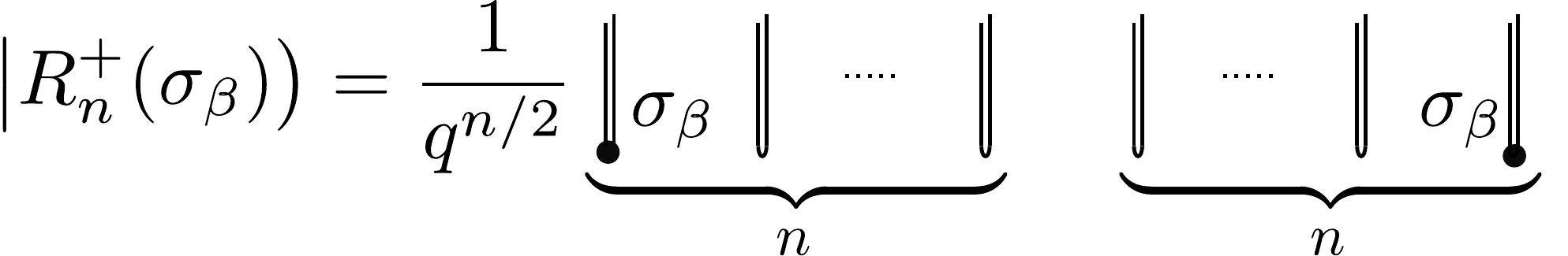}}} \label{eq:diag_bound_R_e},\\
\vcenter{\hbox{\includegraphics[width=0.8\linewidth]{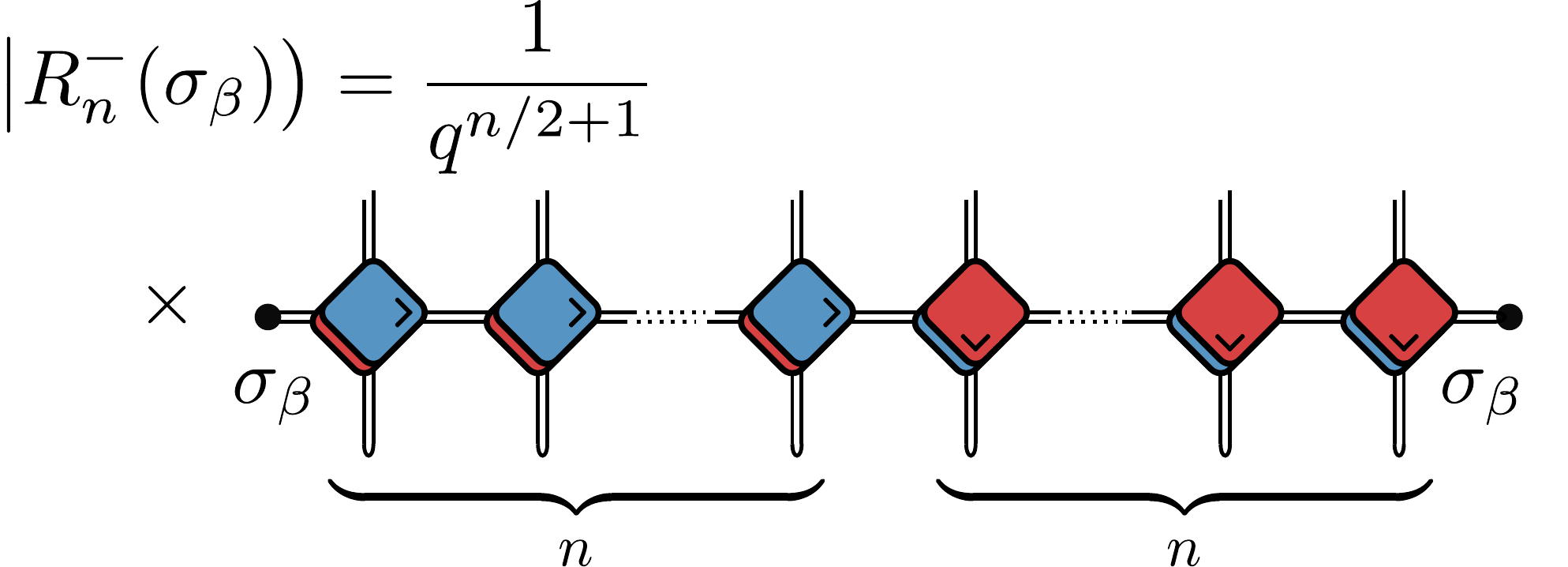}}} \label{eq:diag_bound_R_o}.
\end{align}
Since the transfer matrix is a contracting operator, all its eigenvalues $\lambda$ necessarily satisfy $|\lambda| \leq 1$, and the long-time behaviour of the OTOC at fixed $n_-=n$ will be fully determined by the eigenoperators of $T_n$ with maximal eigenvalue $|\lambda|=1$. Note that, since the transfer matrix is not necessarily hermitian, there is no guarantee that its left and right eigenstates/eigenoperators will be identical, and this will generally not be the case. 

This can already be illustrated when we assume no additional structure in the underlying circuits apart from their unitarity. In the folded representation, the conditions for unitarity can be rewritten as
\begin{equation}
\vcenter{\hbox{\includegraphics[width=0.8\linewidth]{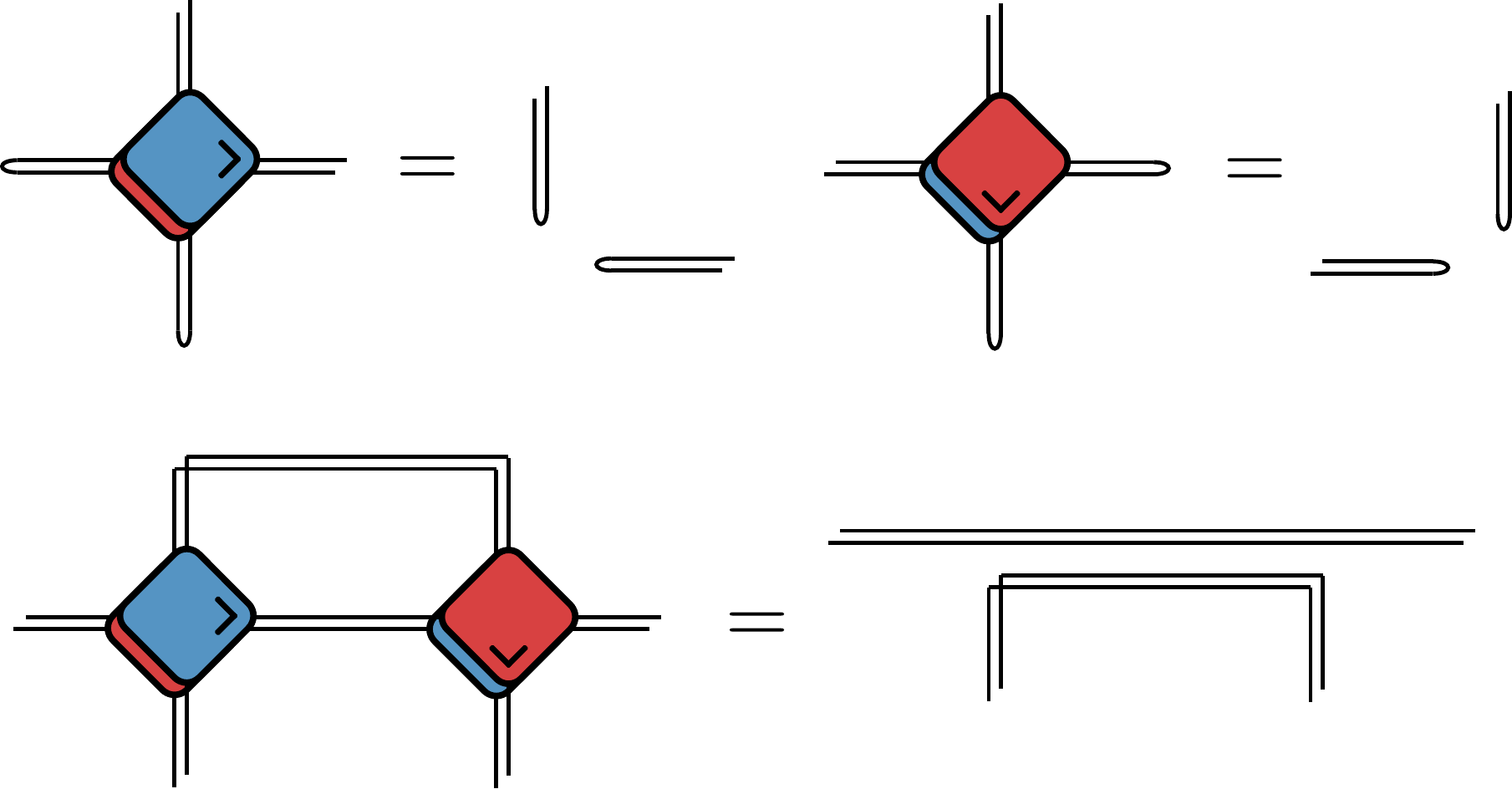}}} \label{eq:unitary},
\end{equation}
which can be used to construct a single right and left eigenoperator of the transfer matrix at arbitrary depth
\begin{align}
\vcenter{\hbox{\includegraphics[width=0.85\linewidth]{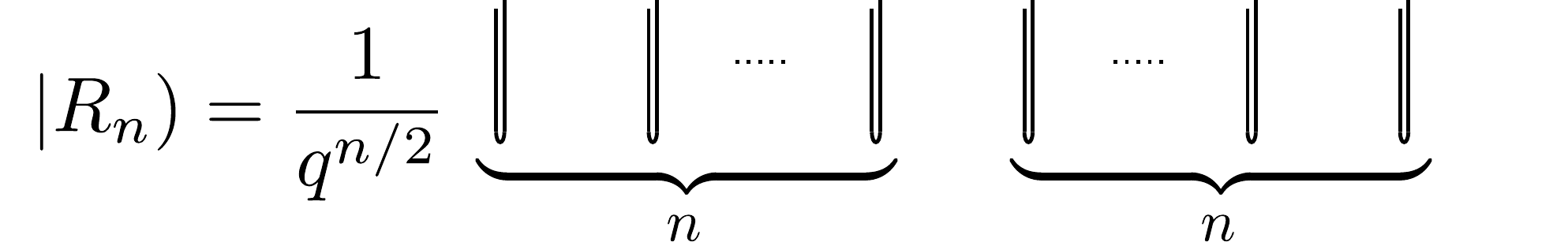}}},\\
\vcenter{\hbox{\includegraphics[width=0.85\linewidth]{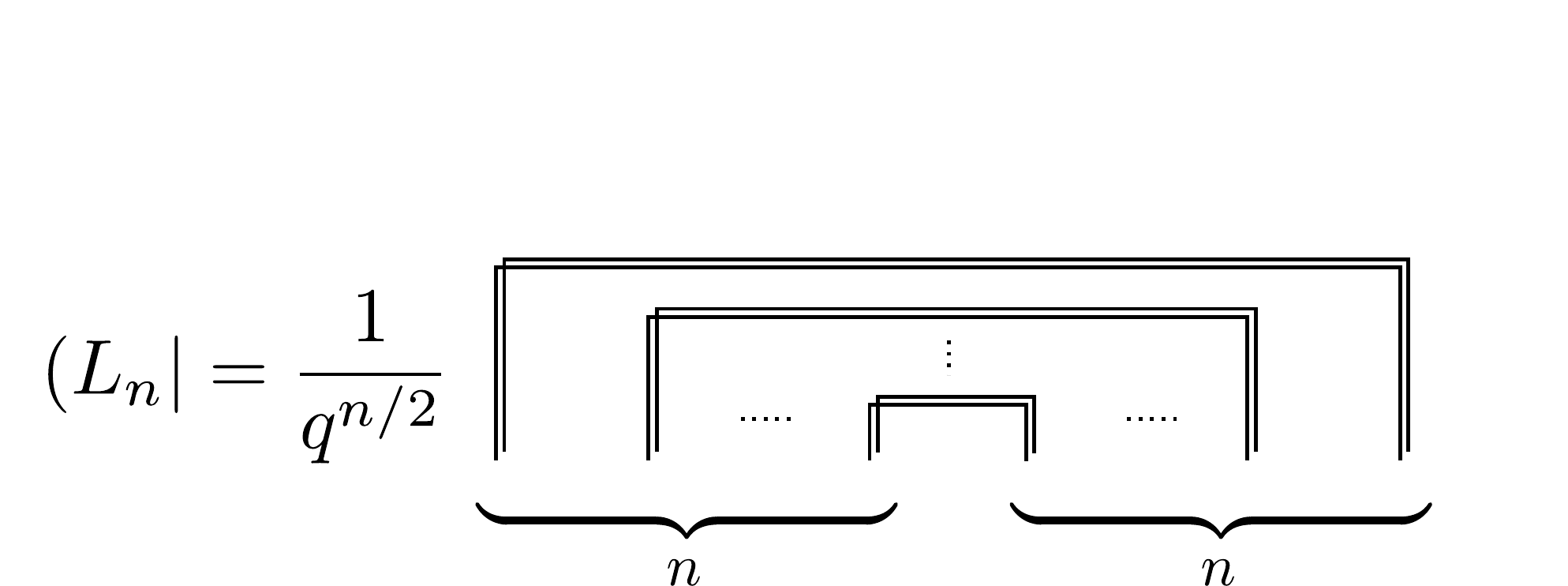}}}, 
\end{align}
satisfying $T_n |R_n) = R_n$ and $(L_n|T_n = (L_n|$, and normalized as $\left(L_n|R_n\right) = 1$, such that $\lim_{m \to \infty}(T_n)^m =  \left|R_n\right)\left(L_n\right|$. However, using this to evaluate the long-time value of the OTOC leads to
\begin{align}
&\lim_{m\to \infty} \left(L_n(\sigma_{\alpha})\right|\left(T_n\right)^m\left|R^{\pm}(\sigma_{\beta})\right) \nonumber\\
&\qquad= \left(L_n(\sigma_{\alpha})|R_n\right)\left(L_n|R_n^{\pm}(\sigma_{\beta})\right) = 1,
\end{align}
which is exactly the trivial value the OTOC takes outside of the light cone. This shows that the butterfly velocity satisfies $v_B < 1$ in generic unitary circuits without any additional structure, but provides no further information about the actual behaviour of the OTOC. 

A circuit with maximal butterfly velocity $v_B = 1$ necessitates additional unit-eigenvalue eigenoperators of the column transfer matrix $T_1$ for $x=t$, and we will provide exact results for two classes of unitary circuits where this is the case: \emph{dual-unitary circuits}, ranging from maximally chaotic to the kicked Ising model at both integrable and non-integrable points, and \emph{kicked XY circuits}. While these are not guaranteed to exhaust all classes of maximal-velocity circuits, dual-unitary circuits satisfy $v_B = 1$ at arbitrary $q$, and numerical investigations suggest that all maximal-velocity circuits for $q=2$ can be mapped to either dual-unitary circuits or kicked XY models.

For such maximal-velocity circuits, our approach consists of finding the accompanying left and right eigenoperators of the transfer matrix, constructing a dual basis out of these eigenoperators, and then calculating the long-time value by replacing the transfer matrix by the appropriate projector. In both cases, we calculate the nontrivial value the OTOC takes on the light cone at long times, and show that, up to parity effects, the profile of the OTOC either decays exponentially or remains constant inside the light cone. Dual-unitary circuits contain both chaotic and non-ergodic classes, which is directly reflected in the behaviour of the OTOC inside the light cone.

\subsection*{Notation}
In the following, we will make extensive use of different left and right eigenoperators of these transfer matrices, for which we introduce the following notation:
\begin{align}
&\vcenter{\hbox{\includegraphics[width=0.8\linewidth]{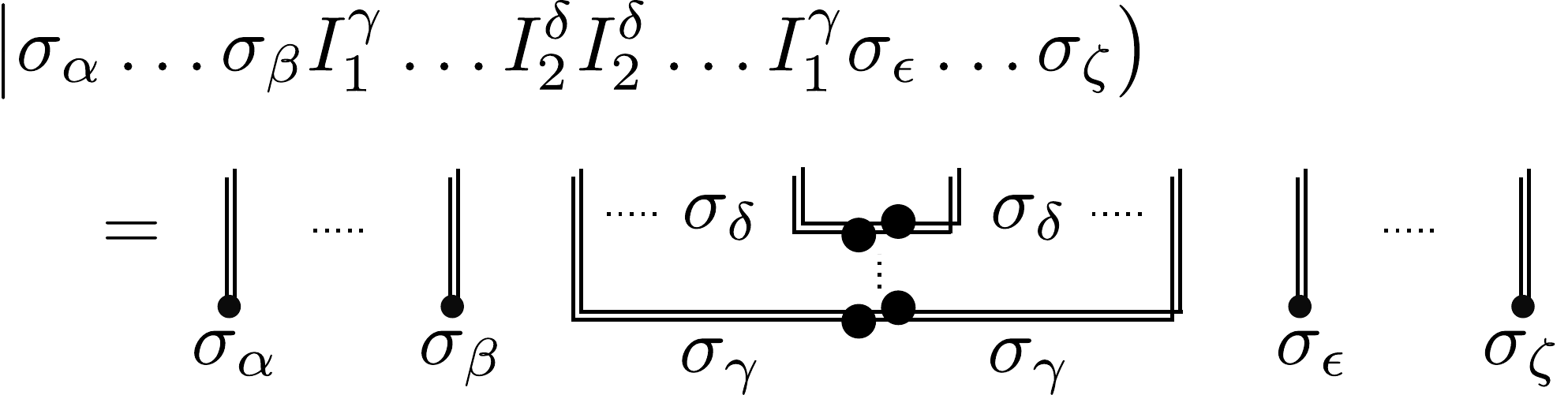}}}, \\
&\vcenter{\hbox{\includegraphics[width=0.8\linewidth]{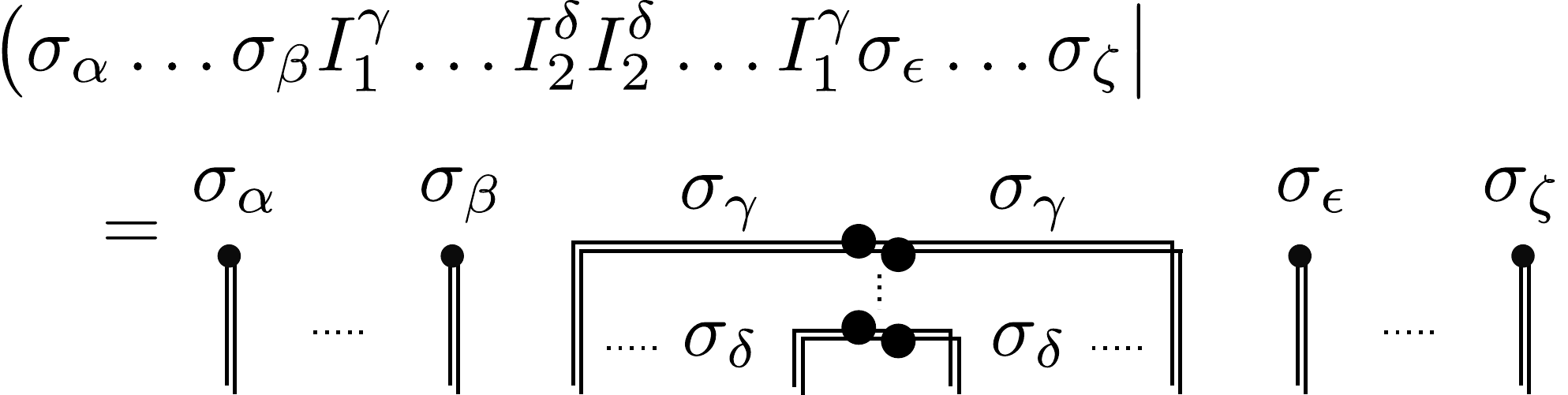}}}.
\end{align}
In order to lighten notation, we drop the subscript in $I_{\mathbbm{1}}$ and write $\mathbbm{1} = \circ$. In this notation, we already have 
\begin{align}
&|R_n) = \frac{1}{q^{n/2}} |\underbrace{\circ \circ \dots \circ \circ}_{2n}),\\
&(L_n| = \frac{1}{q^{n/2}}(I_1 I_2 \dots I_n I_n \dots I_2 I_1|,
\end{align}
and 
\begin{align}
&|R_n^+(\sigma_{\beta})) = \frac{1}{q^{n/2}} |\sigma_{\beta} \underbrace{\circ \circ \dots \circ \circ}_{2(n-1)} \sigma_{\beta} ),\\
&(L_n(\sigma_{\alpha})| = \frac{1}{q^{n/2}}(\underbrace{I_1 I_2 \dots I_{n-1}}_{n-1} I_n^{\alpha} I_n^{\alpha} \underbrace{I_{n-1} \dots I_2 I_1}_{n-1}|.
\end{align}
Overlaps can be evaluated as e.g. $(I_1 I_1 | \sigma_{\alpha} \sigma_{\beta}) = \tr(\sigma_{\alpha} \sigma_{\beta}) = q \delta_{\alpha \beta}$ \footnote{Note that the folding introduces an asymmetry between left/right and top/bottom operators, where the latter are the transpose of the former. Such ambiguities are removed when considering the unfolded diagram, such that this should not lead to any confusion.}.

\section{Dual-unitary circuits} \label{sec:dual}

Dual-unitary circuits are a class of unitary circuits that have recently gained increased attention, since they allow for exact calculations without the usual need to average over Haar-random unitary circuits. A unitary circuit $U$ with matrix elements $U_{ab,cd}$ is said to be dual-unitary if its dual $\tilde{U}_{ab,cd} = U_{db,ca}$ is also unitary \cite{bertini_exact_2019,gopalakrishnan_unitary_2019}. This has a clear interpretation: $U$ determines the evolution in time, which is guaranteed to be unitary, and $\tilde{U}$ determines the evolution in space, which is generally not unitary. Graphically, this can be represented as
\begin{align}
\vcenter{\hbox{\includegraphics[width=0.8\linewidth]{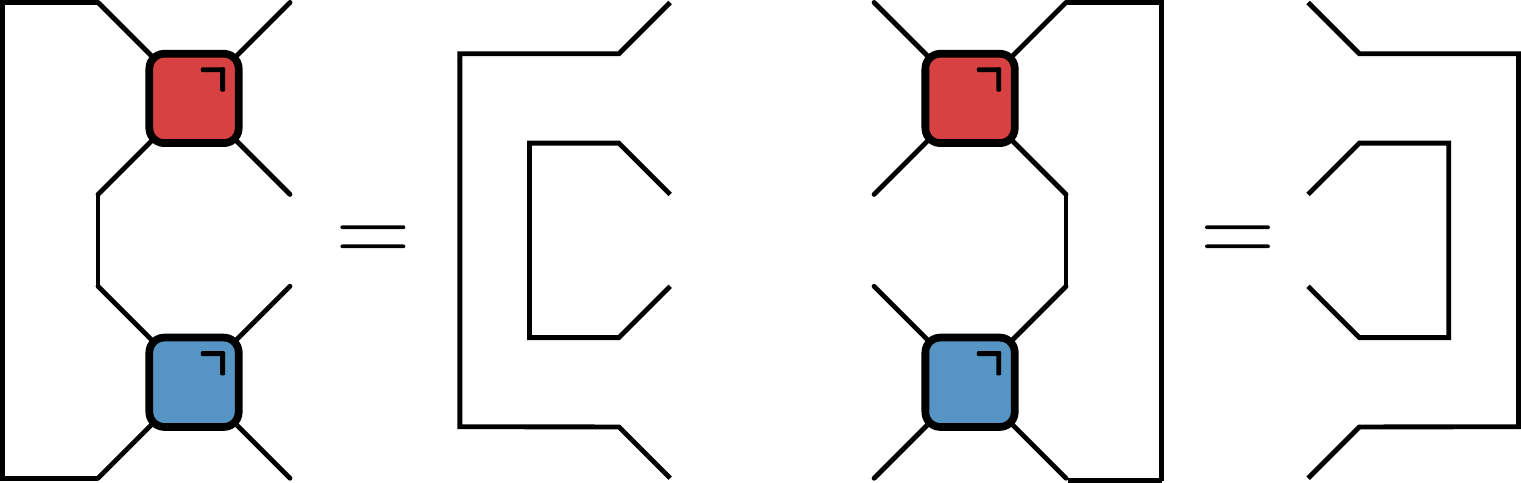}}} \label{eq:diag_def_dualunitary}\, ,
\end{align}
or in the folded representation as
\begin{align}
\vcenter{\hbox{\includegraphics[width=0.8\linewidth]{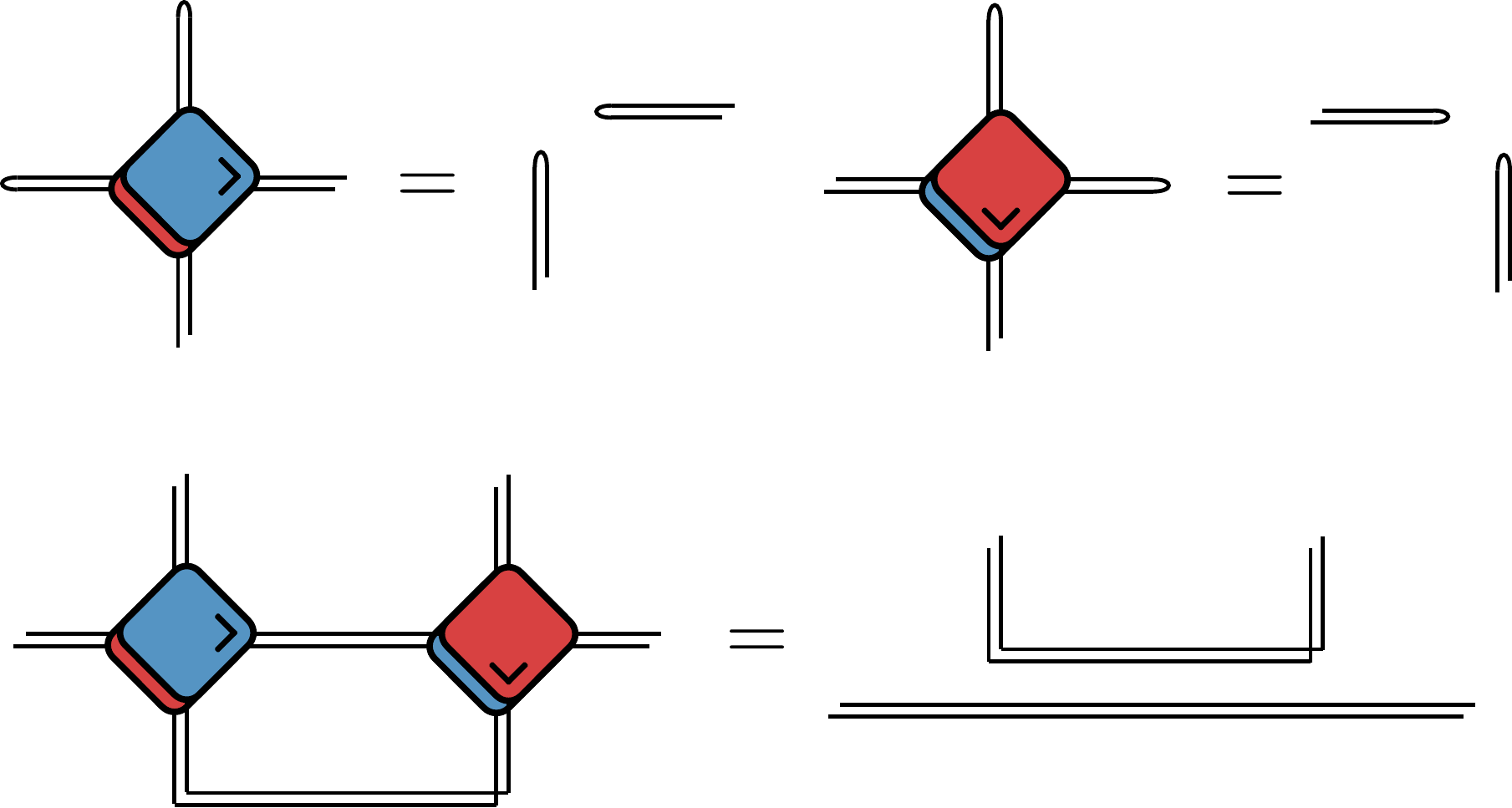}}} \label{eq:dualunitary}\, .
\end{align}
The resulting duality between space and time guarantees that all dynamical correlation functions $\langle \sigma_{\alpha}(x,t) \sigma_{\beta}(0,0) \rangle$ vanish unless the operators lie on the edges of a lightcone spreading at speed 1, which can then be expressed in terms of quantum channels acting on either $\sigma_{\alpha}$ or $\sigma_{\beta}$ (see Section \ref{sec:CorrelationFunctions} and Ref.~\cite{bertini_exact_2019}). This also allows for the exact calculation of operator entanglement, where the entanglement velocities are also maximal $v_E = 1$ \cite{bertini_operator_2019,gopalakrishnan_unitary_2019}. Note that the evolution of these operator entanglements is governed by the exact same transfer matrix as in Eq. \eqref{eq:def_Tmat}, albeit with different boundary conditions \cite{bertini_operator_2019}. 

Combining Eqs. \eqref{eq:unitary} and \eqref{eq:dualunitary}, an independent set of $n+1$ simultaneous left and right eigenoperators of $T_n$ can be constructed as
\begin{align}\label{eq:def_enk}
|e_{n,k}) = |\underbrace{\circ \dots \circ}_{n-k} I_1 I_2 \dots I_k I_k \dots I_2 I_1 \underbrace{\circ \dots \circ}_{n-k}) \\
(e_{n,k}| = (\underbrace{\circ \dots \circ}_{n-k} I_1 I_2 \dots I_k I_k \dots I_2 I_1 \underbrace{\circ \dots \circ}_{n-k}| 
\end{align}
for $k=0 \dots n$. Their orthonormal counterparts are given by
\begin{align}
&|\tilde{e}_{n,0}) = \frac{1}{q^n}|e_{n,0}), \\
&|\tilde{e}_{n,k \neq 0}) = \frac{1}{q^n}\frac{1}{\sqrt{q^2-1}} \left(q |e_{n,k})-|e_{n,k-1})\right),
\end{align}
and similar for $(\tilde{e}_{n,k}|$, leading to $(\tilde{e}_{n,i}|\tilde{e}_{n,j}) = \delta_{ij}$ (again following \cite{bertini_operator_2019}). Because of the dual-unitarity, the left eigenoperators are simply the transpose of the right eigenvectors, which is generally not the case.

\subsection{Maximally chaotic}
For maximally chaotic dual-unitary models, the set of eigenoperators \eqref{eq:def_enk} by definition exhausts all possible eigenoperators \cite{bertini_operator_2019}, such that the long-time value of the OTOC can be obtained from the appropriate projector constructed out of the eigenoperators. We will explicitly distinguish the even and odd cases, starting from the even case
\begin{align}
&\lim_{m\to \infty} \left(L_n(\sigma_{\alpha})\right|\left(T_n\right)^m\left|R^{\pm}(\sigma_{\beta})\right) \nonumber\\
&\qquad= \sum_{k=0}^n \left(L_n(\sigma_{\alpha})|\tilde{e}_{n,k}\right)\left(\tilde{e}_{n,k}|R_n^{\pm}(\sigma_{\beta})\right),
\end{align}
where the necessary overlaps can easily be evaluated as
\begin{align}
&\left(L_n(\sigma_{\alpha})|\tilde{e}_{n,0}\right) = \frac{1}{q^{n/2}}, \nonumber\\
&\left(L_n(\sigma_{\alpha})|\tilde{e}_{n,1}\right) = -\frac{1}{\sqrt{q^2-1}}\frac{1}{q^{n/2}}, \nonumber\\
&\left(L_n(\sigma_{\alpha})|\tilde{e}_{n,k>1}\right)=0,
\end{align}
where the overlaps vanish for $k>1$ since $\tr(\sigma_{\alpha})=0$, and
\begin{align}
&\left(\tilde{e}_{n,n}|R_n^{+}(\sigma_{\beta})\right) = \frac{q}{\sqrt{q^2-1}}\frac{1}{q^{n/2}}, \nonumber\\
&\left(\tilde{e}_{n,k<n}|R_n^{+}(\sigma_{\beta})\right) = 0,
\end{align}
again from $\tr(\sigma_{\beta})=0$. The only possible non-zero value for the OTOC at long times is when $n=n_{-}=1$ and subsequently $x=t$, leading to
\begin{align}
\lim_{(x+t) \to \infty }C^{+}_{\alpha \beta}(x,t) = 
\begin{cases}
-\frac{1}{q^2-1} \qquad &\text{if} \quad x=t,\\
0 \qquad &\text{if} \quad x \neq t.
\end{cases}
\end{align}
The value of the light-cone has a simple interpretation by assuming that, at long times, $\sigma_{\alpha}(0,t) \sigma_{\beta}(x,0)$ is essentially random and contains all $q^2-1$ traceless basis operators with equal amplitude $i/\sqrt{q^2-1}$. 

This can be contrasted with the expected behaviour for Haar-random unitary circuits, considering $n=1$ for simplicity. The matrix elements of the unfolded transfer matrix $T_1$ can be represented as
\begin{align}
\vcenter{\hbox{\includegraphics[width=0.55\linewidth]{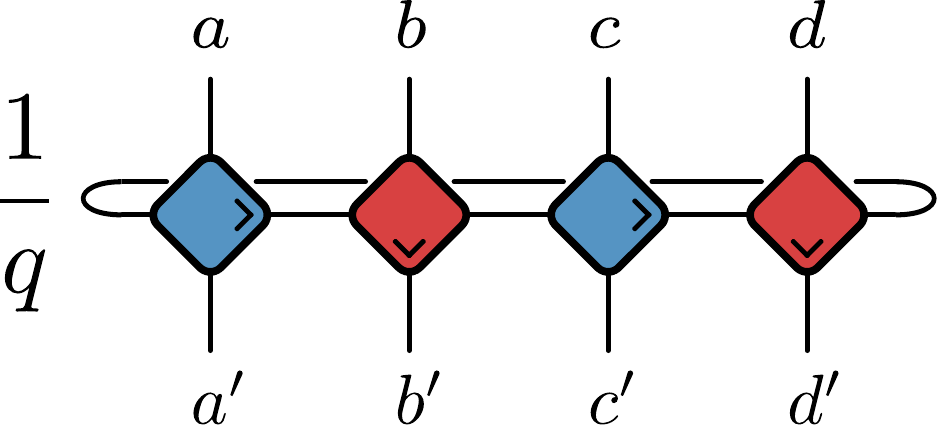}}}\ , \label{eq:Haar_1}
\end{align}
and in the long-time limit $T_1$ can be replaced by the projector $|\tilde{e}_{1,1})(\tilde{e}_{1,1}|+|\tilde{e}_{1,0})(\tilde{e}_{1,0}|$, where the matrix elements can be explicitly evaluated as
\begin{align}
&\frac{1}{q^2-1}\Big[\delta_{ab}\delta_{cd} \times \delta_{a'b'} \delta_{c'd'} + \delta_{ad}\delta_{bc} \times \delta_{a'd'} \delta_{b'c'} \nonumber\\
&- \frac{1}{q}\left(\delta_{ab}\delta_{cd} \times \delta_{a'd'} \delta_{b'c'}+\delta_{ad}\delta_{bc} \times \delta_{a'b'} \delta_{c'd'}\right)\Big].
\end{align}
Remarkably, this is exactly the expression that is obtained by taking the Haar average of random one-site unitary matrices $u_{aa'} u^*_{bb'}u_{cc'} u^*_{dd'}$, $u \in \mathbb{U}_q$, (see e.g. Ref.~\cite{nahum_operator_2018})
\begin{align}
\vcenter{\hbox{\includegraphics[width=0.6\linewidth]{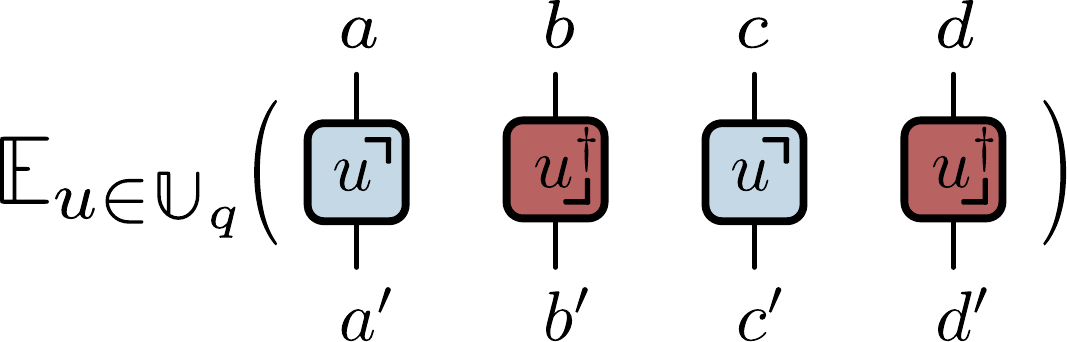}}}\, , \label{eq:Haar_2}
\end{align}
such that the long-time limit of evolution using two-site dual-unitary gates is here equivalent to the evolution using Haar-random one-site unitary gates with same dimension of the local Hilbert space. An additional observation is that a dual-unitary gate where a random one-site unitary is added to each leg remains dual-unitary. Constructing the transfer matrix for these unitaries returns the usual transfer matrix with an additional prefactor of the form \eqref{eq:Haar_2}. Averaging over the additional Haar-random one-site unitaries then returns as prefactor the projector by Eq.~\eqref{eq:Haar_1}, such that taking the Haar-average of the transfer matrix over one-site unitaries again results in the same projector. Since such unitaries can give rise to a basis rotation of the local operators, this provides an alternative argument for why all traceless basis operators should have equal amplitudes in the final OTOC value. However, it is worthwhile to note again that the result for the OTOC does not depend on any randomness in the circuits and holds for any circuit built out of dual-unitary circuits.

Returning to the calculation of the OTOCs, the right boundary for odd parity is more involved, but from the left boundary we see that we only need $(\tilde{e}_{n,0}|R_n^{-}(\sigma_{\beta}))$ and $(\tilde{e}_{n,1}|R_n^{-}(\sigma_{\beta}))$. This leads to
\begin{align}
&\sum_{k=0,1}\left(L_n(\sigma_{\alpha})|\tilde{e}_{n,k}\right) (\tilde{e}_{n,k}|R_n^{-}(\sigma_{\beta})) \nonumber\\
&\quad\quad= \frac{1}{q^2-1}\left(q^2\mathcal{M}_n(\sigma_{\beta})-\mathcal{M}_{n-1}(\sigma_{\beta})\right),
\end{align}
in which $\mathcal{M}_n(\sigma_{\beta})$ is given by
\begin{align}
\vcenter{\hbox{\includegraphics[width=0.85\linewidth]{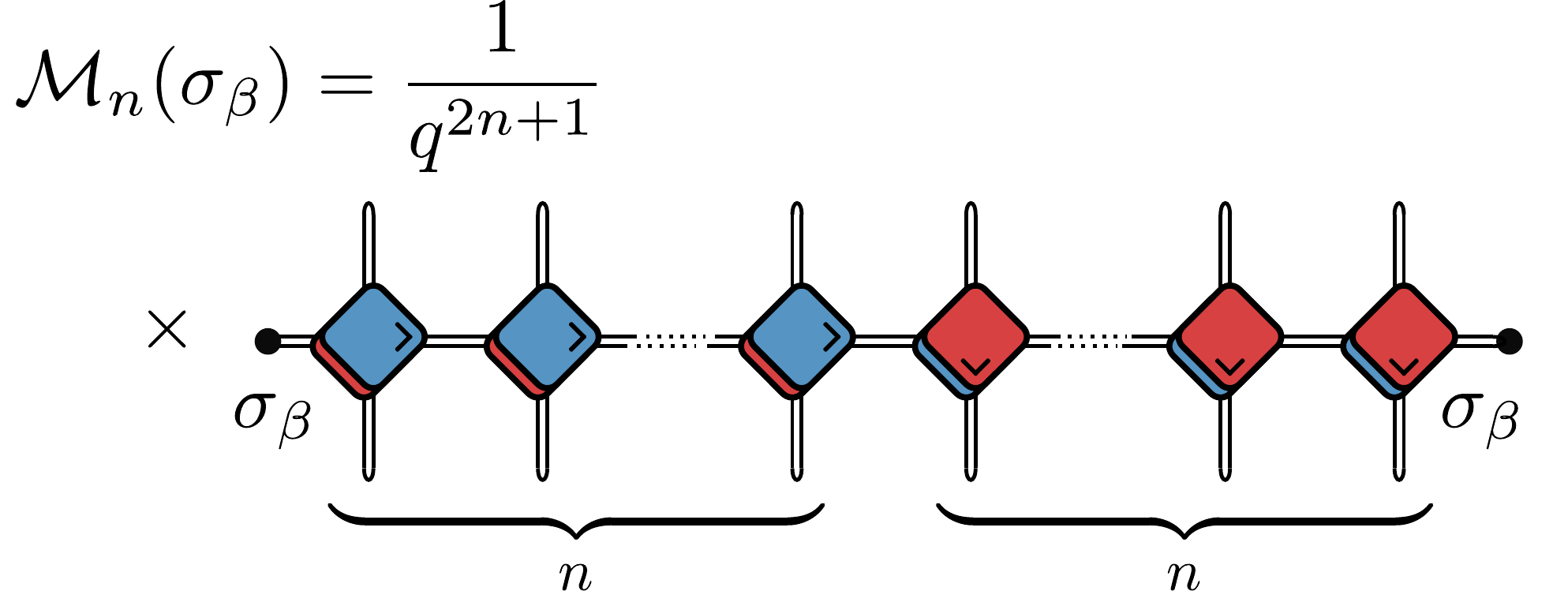}}}\, . \label{eq:diag_def_Mn}
\end{align}
Plugging this in the expression for the OTOC returns
\begin{align}\label{eq:OTOC_ss_o}
&\lim_{(x+t) \to \infty }C^{-}_{\alpha \beta}(x,t) \nonumber\\
&\qquad = \frac{q^2\mathcal{M}_{(t-x+1)/2}(\sigma_{\beta})-\mathcal{M}_{(t-x-1)/2}(\sigma_{\beta})}{q^2-1}.
\end{align}
The behaviour of $\mathcal{M}_n$ can immediately be linked to the dynamical correlations on the light cone, since
\begin{align}
\mathcal{M}_n(\sigma_\beta) &= \tr\left[\sigma_{\beta}\mathcal{M}_{-}^n(\mathcal{M}_{+}^n(\sigma_{\beta}))\right]/q \nonumber \\
&=\tr\left[\mathcal{M}_{+}^n(\sigma_{\beta})^{\dagger} \mathcal{M}_{+}^n(\sigma_{\beta})\right]/q,
\end{align}
with $\mathcal{M}_{\pm}$ defined as previously \eqref{def:channels}. So not only do these quantum channels fully determine the two-point correlation functions on the light cone, the only non-zero correlations in dual-unitary circuits, they determine the decay of the OTOC inside the light cone in such circuits. This explicitly connects both the decay rate and the steady-state values of the OTOC with those of the correlation functions. Since dual-unitary circuits have been classified in terms of the increasing level of ergodicity encoded in the eigenvalues of $\mathcal{M}_{\pm}$ (see Ref.~\cite{bertini_exact_2019}), this is immediately reflected in the OTOC behaviour (for $(x-t)$ odd):
\begin{enumerate}
\item  \emph{Non-interacting}: All $2(q^2-1)$ nontrivial eigenvalues of $\mathcal{M}_{\pm}$ are equal to 1. All dynamical correlations remain constant, and the OTOC similarly remains constant and equal to one both inside and outside the light cone.
\item \emph{Non-ergodic}: There exist more than zero but less than $2(q^2-1)$ nontrivial eigenvalues equal to 1. Some dynamical correlations remain constant, and the OTOC similarly decays exponentially to a constant value inside the light cone since, for some $\sigma$, $\lim_{t\to \infty} \mathcal{M}_{+}^{t}(\sigma)$ converges to a non-zero operator. The limiting value of the OTOC then simply equals the norm of this operator up to a factor $q$. The decay rate for the OTOC is twice that of the corresponding dynamical correlation.
\item \emph{Ergodic and non-mixing}: All nontrivial eigenvalues are different from 1, but there exists at least one eigenvalue with unit modulus. All time-averaged dynamical correlations vanish at large times, but $\mathcal{M}_{+}^{t}(\sigma)$ keeps oscillating. The time-averaged OTOC similarly keeps oscillating, but around a value that is larger than zero and (generally) smaller than one.
\item \emph{Ergodic and mixing}: All nontrivial eigenvalues are within the unit disc and all dynamical correlations decay to zero since $\lim_{t\to \infty} \mathcal{M}_{+}^{t}(\sigma)$ vanishes for all initial $\sigma$. The OTOC similarly exponentially decays to zero inside the light cone, where the decay rate is again twice that of the corresponding dynamical correlation.
\end{enumerate}

\begin{figure}[htb!]
\includegraphics[width=\columnwidth]{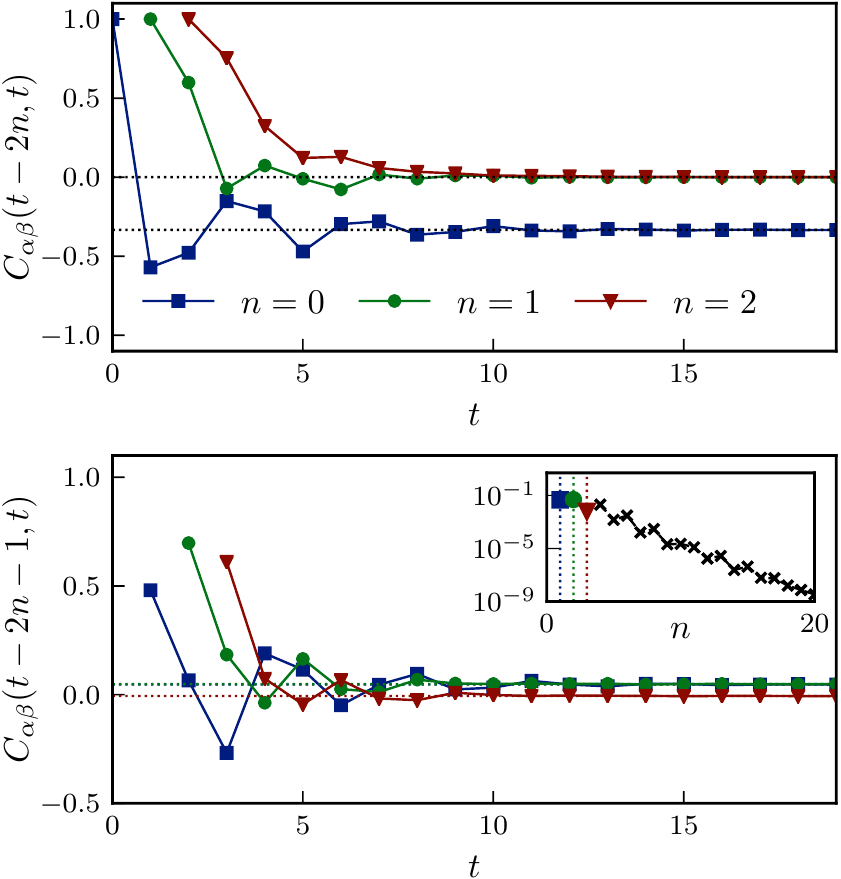}
\caption{$C^{+}_{\alpha\beta}(x,t)$ (top) and $C^{-}_{\alpha\beta}(x,t)$ (bottom) for $\sigma_{\alpha} = \sigma_{\beta} =\sigma_x$ and a random dual-unitary circuit (see Appendix \ref{app:dualunit}). Only the values inside and on the light cone are shown. In the top plot $(t-x)$ is even, and the long-time values are given by $-1/(q^2-1)=-1/3$ for $x=t$ and $0$ otherwise. The bottom plot denotes the evolution for $(t-x)$ odd, where the inset details the (logarithm of the) long-time values of $\lim_{t\to \infty}|C_{\alpha \beta}(t-2n-1,t)|$ from Eq.~\eqref{eq:OTOC_ss_o} for a larger range of $n$, where the exponential decay can be clearly observed.
 \label{fig:maxchaos}}
\vspace{-1\baselineskip}
\end{figure}

This is illustrated in Fig.~\ref{fig:maxchaos} for an ergodic and non-mixing dual-unitarity circuit, showing both the OTOC at finite times for $n_{-}=1,2,3$ and their steady-state value for a large range of $n_{-}$, illustrating the exponentially-decaying profile of the OTOC inside the light cone.

\subsection{Kicked Ising Model}
The previous calculation explicitly assumed no other eigenoperators with eigenvalue 1 other than the ones from Eq. \eqref{eq:def_enk}. However, within the class of dual-unitary circuits with $q=2$ there exists a subclass that are equivalent to the Kicked Ising Model (KIM) at the self-dual point, given by
\begin{align}
\vcenter{\hbox{\includegraphics[width=0.45\linewidth]{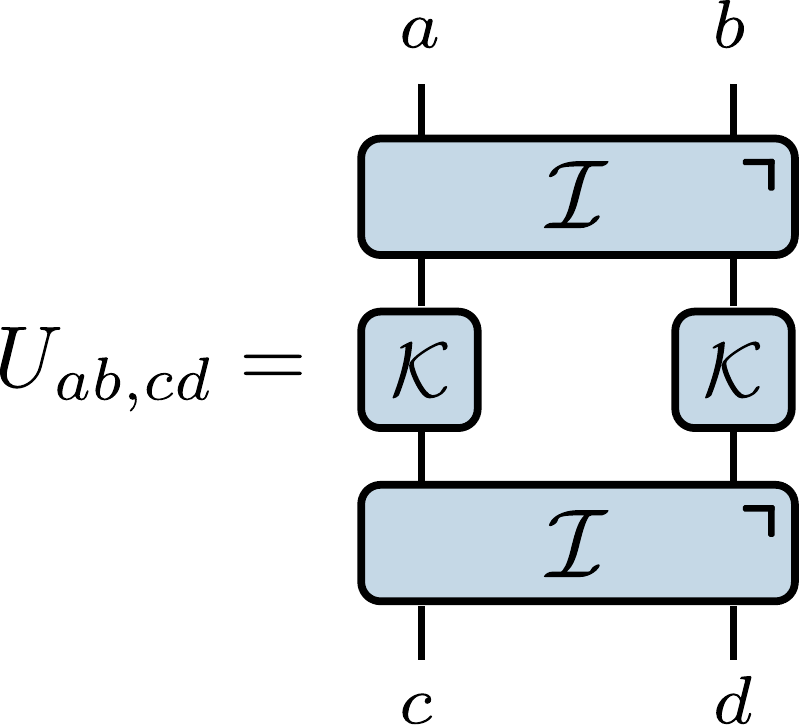}}}\ , \label{eq:diag_def_U_KIM}
\end{align}
defined in terms of two-qubit ($\mathcal{I}$) and one-qubit ($\mathcal{K}$) gates
\begin{align}
&\mathcal{I} = \exp\left[-i J \sigma_z \otimes \sigma_z\right] \nonumber\\
&\qquad\times \exp\left[- i (h_1 (\sigma_z  \otimes \mathbbm{1}) + h_2 (\mathbbm{1} \otimes \sigma_z))/2\right], \\
&\mathcal{K} =  \exp\left[i b \sigma_x\right],
\end{align}
where dual-unitarity fixes $|J|=|b| = \pi/4$ and $h_1,h_2 \in \mathbbm{R}$ can be chosen freely. Taking $J=b=\pi/4$, the matrix elements of this gate are given by
\begin{align}\label{eq:paramKIM}
U_{ab,cd} =& -\frac{i}{2} \exp \left[i\frac{\pi}{4}(a-d)(c-b)\right] \nonumber\\
&\qquad \times \exp \left[-i\frac{h_1}{2}(a+c)-i\frac{h_2}{2}(b+d)\right], 
\end{align}
with $a,b,c,d \in \{-1,1\}$. As also noted in Ref. \cite{bertini_entanglement_2019}, these gates exhibit an additional symmetry that allows for the construction of additional eigenoperators with eigenvalue one as
\begin{equation}
|z_{n,k}) = |\underbrace{\circ \dots \circ}_{n-k} \sigma_z \underbrace{I_1 \dots I_{k-1} I_{k-1} \dots I_1}_{2(k-1)} \sigma_z \underbrace{\circ \dots \circ}_{n-k}),
\end{equation}
with $k = 1 \dots n$, which can again be orthonormalized as 
\begin{equation}
|\tilde{e}_{n,n+k}) =  \frac{1}{2^n}\left(\sqrt{\frac{3}{2}} |z_{n,k}) - \sqrt{\frac{2}{3}}|e_{n,k}) + \sqrt{\frac{1}{6}}|e_{n,k-1})\right),
\end{equation}
and similar for $(\tilde{e}_{n,n+k}|$. The necessary overlaps with the left boundary now explicitly depend on the choice of operators. Writing $\sigma_{\alpha} = \alpha_x \sigma_x + \alpha_y \sigma_y + \alpha_z \sigma_z$ and $\sigma_{\beta}= \beta_x \sigma_x + \beta_y \sigma_y + \beta_z \sigma_z$, with $\alpha_x^2+\alpha_y^2+\alpha_z^2=\beta_x^2+\beta_y^2+\beta_z^2=1$, the only relevant non-zero overlaps follow from
\begin{align}
&(L_n(\sigma_{\alpha}) | z_{n,1}) = 2^{n/2}\left(2\alpha_z^2-1\right), \\
&(z_{n,n} | R_n^+(\sigma_{\beta})) = 2^{n/2+1} \beta_z^2,
\end{align}
leading (for the left boundary) to
\begin{align}
&(L_n(\sigma_{\alpha}) | \tilde{e}_{n,n+1}) = \frac{1}{2^{n/2}} \left[\sqrt{6}\left(2\alpha_z^2-1\right) + \sqrt{\frac{1}{6}}\right], \\
&(L_n(\sigma_{\alpha}) | \tilde{e}_{n,k>n+1}) = 0,
\end{align}
and for the right boundary to
\begin{align}
&(\tilde{e}_{n,2n}|R_n^+(\sigma_{\beta})) = \frac{1}{2^{n/2}} \left[\sqrt{6}\ \beta_z^2-\sqrt{\frac{2}{3}}\right], \\
&(\tilde{e}_{n,n<k<2n}|R_n^+(\sigma_{\beta}))  = 0.
\end{align}
Considering even parity, this will only modify the steady-state value for $n=n_{-}=1$, where we find
\begin{align*}
\lim_{(x+t) \to \infty}  C^+_{\alpha \beta}(x,t) = 
\begin{cases}
3 \beta_z^2 \alpha_z^2-\alpha_z^2-\beta_z^2\ &\text{if} \ x=t,\\
0 \qquad &\text{if} \  x \neq t.
\end{cases}
\end{align*}
For the case of odd parity the profile will again be determined by $\mathcal{M}_{n}$, where the explicit parametrization of $U$ allows us to find analytic expressions for $C^{-}_{\alpha \beta}(x,t)$. The necessary additional overlap follows from
\begin{align}
(z_{n,1}|R^-_n(\sigma_{\beta})) = 0,
\end{align}
where we have evaluated the diagram using that, for the KIM,
\begin{align}
\vcenter{\hbox{\includegraphics[width=0.6\linewidth]{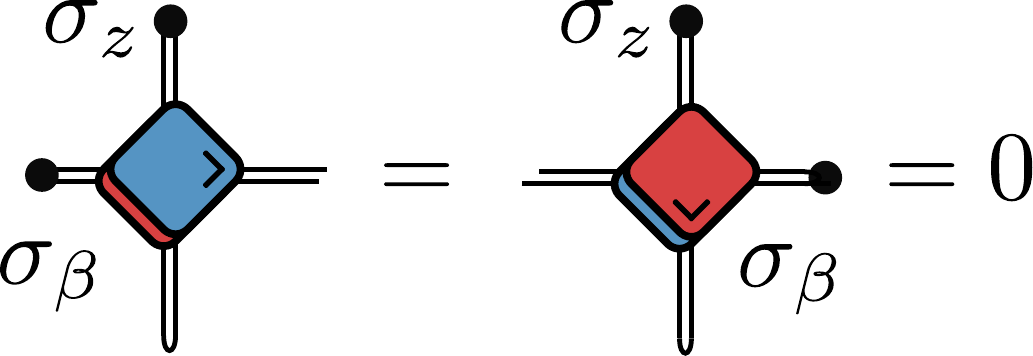}}}\, .
\end{align}
The final value for the OTOC is given by
\begin{align}
&\sum_{k=0}^{2n}\left(L_n(\sigma_{\alpha})|\tilde{e}_{n,k}\right) (\tilde{e}_{n,k}|R_n^{-}(\sigma_{\beta})) \nonumber\\
&\qquad=\left(1+\alpha_z^2\right)\mathcal{M}_n(\sigma_{\beta})- \alpha_z^2 \mathcal{M}_{n-1}(\sigma_{\beta}).
\end{align}

\begin{figure}[htb!]
\includegraphics[width=\columnwidth]{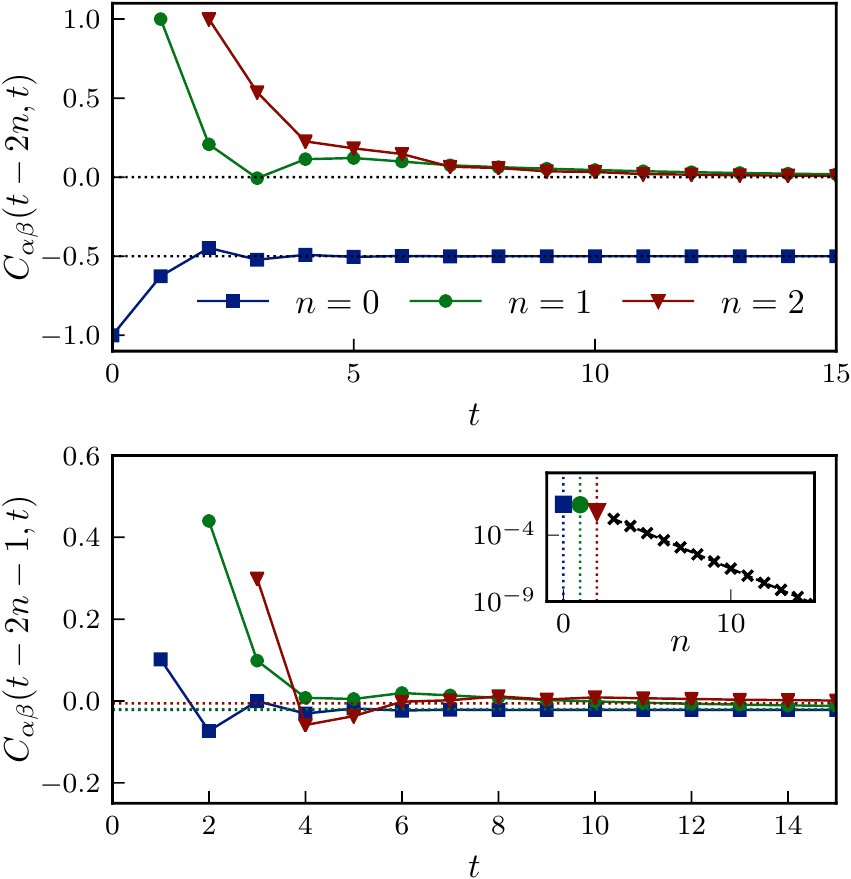}
\caption{Values of $C^{+}_{\alpha\beta}(x,t)$ (top) and $C^{-}_{\alpha\beta}(x,t)$ (bottom) for $\sigma_{\alpha} = \left(\sigma_x+\sigma_z\right)/\sqrt{2}$ and $\sigma_{\beta} =\sigma_y$ for evolution using the KIM with $h_1=0.4$ and $h_2=0.6$. For even $(t-x)$ the long-time values are given by $-1/2$ for $x=t$ and $0$ otherwise. The bottom plot denotes the evolution for odd $(t-x)$, where the inset details the (logarithm of the) long-time values of $\lim_{t\to \infty}|C_{\alpha \beta}(t-2n-1,t)|$ for a larger range of $n$, where the exponential decay $\propto \cos(h_1+h_2)^{2n}$ can be clearly observed.
 \label{fig:KIM}}
\vspace{-1\baselineskip}
\end{figure}

Using the explicit construction of the eigenoperators of $\mathcal{M}_{+}$ (see Appendix \ref{app:MapKIM}), we can evaluate
\begin{align}
\mathcal{M}_n(\sigma_{\beta}) = \left(\beta_x \cos(h_1)-\beta_y \sin(h_1)\right)^2 \cos(h_1+h_2)^{2(n-1)},
\end{align}
for $n \geq 1$ and and $\mathcal{M}_0(\sigma_{\beta})=1$. The final value for the OTOC at odd values of $(x-t)$ follows as
\begin{align}
&\lim_{(x+t) \to \infty }C^{-}_{\alpha \beta}(x,t) = \left(\beta_x \cos(h_1)-\beta_y \sin(h_1)\right)^2 \nonumber\\
&\times\cos(h_1+h_2)^{t-x-3}\left(\cos(h_1+h_2)^2-\alpha_z^2 \sin(h_1+h_2)^2\right),
\end{align}
for $x<t-1$, and 
\begin{align}
&\lim_{(x+t) \to \infty }C^{-}_{\alpha \beta}(t-1,t)\nonumber\\
&\qquad = (1+\alpha_z^2)\left(\beta_x \cos(h_1)-\beta_y \sin(h_1)\right)^2-\alpha_z^2.
\end{align}
This is illustrated in Fig.~\ref{fig:KIM}, both the transient regime for small values of $n$ and the long-time value for a larger range of $n$.

\begin{figure}[htb!]
\includegraphics[width=\columnwidth]{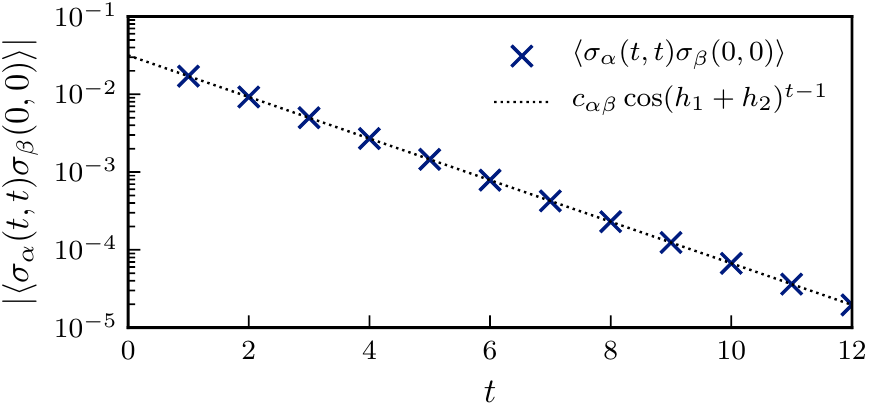}
\caption{Evolution of the correlation function on the light cone  $\langle \sigma_{\alpha}(t,t) \sigma_{\beta}(0,0) \rangle$ for the KIM and $\sigma_{\alpha,\beta}$ parametrized as in Fig.~\ref{fig:KIM}, showing exponential decay $\propto \cos(h_1+h_2)^t$ with prefactor $c_{\alpha \beta} = (\alpha_x\cos(h_2)-\alpha_y \sin(h_2))(\beta_x \cos(h_1)-\beta_y\sin(h_1))$.
 \label{fig:corrKIM}}
\vspace{-1\baselineskip}
\end{figure}

Having constructed $\mathcal{M}_{+}$, the correlation functions on the light cone also immediately follow (for $t>0$) as
\begin{align}
&\langle \sigma_{\alpha}(t,t) \sigma_{\beta}(0,0) \rangle = \cos(h_1+h_2)^{t-1} \nonumber \\
&\quad \times(\alpha_x\cos(h_2)-\alpha_y \sin(h_2))(\beta_x \cos(h_1)-\beta_y\sin(h_1)),
\end{align}
which is illustrated in Fig.~\ref{fig:corrKIM}.

\subsection{Kicked Ising Model at the integrable point}
The final dual-unitary circuit we will consider is the KIM model at the integrable point, where $h_1 = -h_2$. The time evolution governed by this circuit is easily seen to be equivalent to the circuit with $h_1 = h_2 = 0$, which is exactly the Trotterization of the Kicked Ising Model at the integrable point. As shown in the previous section, the decay rate of the KIM is set by $\cos(h_1+h_2)$, such that at these values the OTOC is naively not expected to decay. While this will turn out to be the case, the argument needs to take into account that the integrability is reflected in the fact that the transfer matrix supports an exponentially large number of eigenoperators with eigenvalue one. 

More specifically, any `product state' of the form
\begin{equation}
|\sigma_{\alpha_1} \sigma_{\alpha_2} \dots \sigma_{\alpha_{2n}})
\end{equation}
is an eigenoperator of the transfer matrix, with the eigenvalue either zero or one depending on $n_{y}+n_{z}$, the total combined number of $\sigma_{y}$ and $\sigma_{z}$ operators in this eigenoperator. If this is even, the state has eigenvalue one, otherwise the state has eigenvalue zero. The unitarity can be combined with a set of relations for $\sigma_{x}$, $\sigma_{y}$ and $\sigma_{z}$ that are satisfied precisely at the integrable point: $U (\mathbbm{1} \otimes \sigma_{\alpha}) = \sigma_{\alpha}\otimes \sigma_{\beta}$ and $U (\sigma_x \otimes \sigma_{\alpha}) = \sigma_{\alpha}\otimes \sigma_{\beta}$, for general $\sigma_{\alpha}$ and with $\sigma_{\beta}$ either $\sigma_x$ or $\mathbbm{1}$ (see Appendix \ref{app:IdentitiesKIM}), such that the action of the transfer matrix on such a product state results in an eigenvalue that is either proportional to $\tr(\sigma^x)$ if $n_{y}+n_{z}$ is odd, or $\tr(\mathbbm{1})$ for $n_{y}+n_{z}$ even. As such, the transfer matrix is effectively a projector and the OTOC immediately saturates to a constant value inside the light cone.

Since the transfer matrix is a projector, the OTOC diagram at arbitrary values of $n_{+}$ equals the diagram with $n_{+}=1$, which can be explicitly contracted using the identities \eqref{eq:diag_U_KIM_int} from Appendix \ref{app:IdentitiesKIM} to return
\begin{align}
&C^+(t,t) = 2\left[(\alpha_y\beta_y+\alpha_z\beta_z)^2+\alpha_x^2 \beta_x^2\right]-1,\\
&C^+(x<t,t) = 1,
\end{align}
and for odd parity,
\begin{align}
C^-(x,t) = \alpha_x^2+(1-\alpha_x)^2(2\beta_x^2-1),
\end{align}
where the limit $(x+t)\to \infty$ does not need to be taken because the OTOC does not depend on $(x+t)$ away from the light cone.

\begin{figure}[htb!]
\includegraphics[width=\columnwidth]{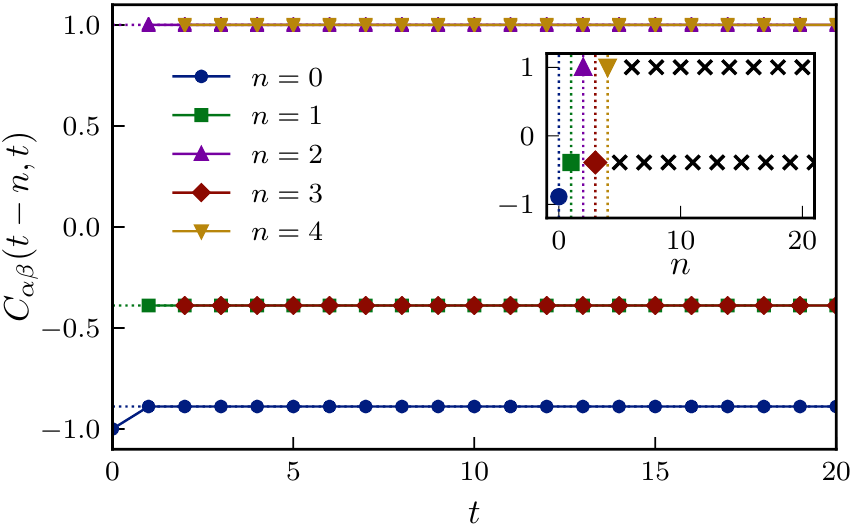}
\caption{Values of $C(x,t)$  for $x=t-n$, $\sigma_{\alpha} = \sigma_x/\sqrt{6} + \sigma_y/\sqrt{2} + \sigma_z/\sqrt{3}$, and $\sigma_{\beta} =\sigma_x/\sqrt{6} - \sigma_y/\sqrt{2} + \sigma_z/\sqrt{3}$ for evolution using the kicked XY model with $J_z = \pi/10$. Dotted lines represent the analytic results, and the inset details $\lim_{(x+t)\to \infty}C(t-n,t)$ for a larger range of $n$. The immediate saturation to a constant value can be clearly observed.
 \label{fig:KIM_int}}
\vspace{-1\baselineskip}
\end{figure}

The correlation functions on the light cone can also be explicitly evaluated from the known eigenoperators of the quantum channels (see Appendix \ref{app:IdentitiesKIM}) and exhibit the same behaviour, immediately saturating to a constant and non-zero value on the light cone as
\begin{align}
\langle \sigma_{\alpha}(t,t) \sigma_{\beta}(0,0) \rangle = 
\begin{cases}
\delta_{\alpha \beta} \qquad &\textrm{if} \quad t=0, \\ 
\alpha_x \beta_x \qquad &\textrm{if} \quad t>0.
\end{cases}
\end{align}

\section{Kicked XY Models}\label{sec:XY}

In this section, we will show how the results for the integrable KIM can be extended towards a closely related class of kicked XY models, highlighting that it is not the dual-unitarity that is responsible for the maximal butterfly velocity. We will consider circuits of the form
\begin{align}
\vcenter{\hbox{\includegraphics[scale=0.5]{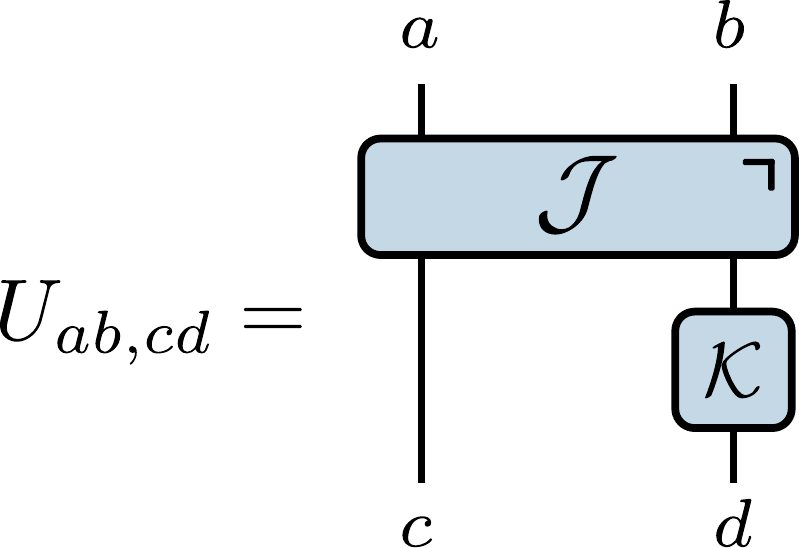}}}\ , \label{eq:diag_U_XY}
\end{align}
with one-qubit gate $\mathcal{K}=\exp\left[i \frac{\pi}{4}\sigma_x\right]$ and the two-qubit gate $\mathcal{J}=\mathcal{J}[J]$ part of a one-parameter family of unitary circuits
\begin{align}
&\mathcal{J}[J] = \exp\left[iJ \sigma_z \otimes \sigma_z\right] \exp\left[i \frac{\pi}{4}\sigma_y \otimes \sigma_y\right].
\end{align}
This circuit introduces an explicit anisotropy in the one-qubit operator that only acts on a single site. However, the building block for time evolution over two time steps is given by
\begin{align}
\vcenter{\hbox{\includegraphics[scale=0.5]{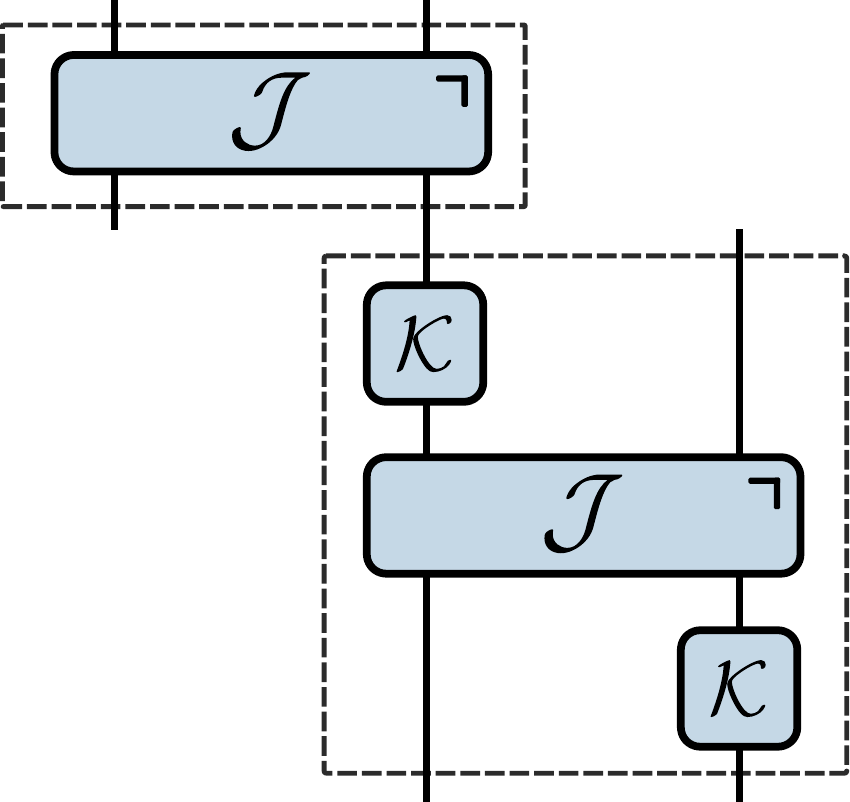}}}\ , \label{eq:diag_U_XY_comb}
\end{align}
both operators of which we can interpret as the Trotterization of an Ising model with local interactions $J \sigma_{z}\otimes \sigma_{z}+\frac{\pi}{4} \sigma_y \otimes \sigma_y$, where the transverse field alternates between $h_x=\pi/4$ and $0$. After a spin rotation, this can be interpreted as a kicked XY spin model with transverse field with magnetization strength $\pi/4$, which is why we refer to this model as a kicked XY model (following the name of e.g. Ref.~\cite{lieb_two_1961})). However, we will stick with the ZY parametrization because it highlights the similarities with the Kicked Ising Model. More specifically, at $J=\pi/4$ the model is dual-unitary and $-U(\mathcal{K}\otimes \mathbbm{1})$ equals the self-dual KIM unitary at the integrable point. In the following, we will consider the model away from the dual-unitary point.

Indeed, this model behaves in the same vein as the integrable self-dual KIM, in that it has a maximal butterfly velocity $v_{B} = 1$ and satisfies a (more restricted) set of identities. However, it also differs in some crucial ways: it is not dual-unitary (except for $|J|=\pi/4$), and the transfer matrix will no longer be a projector. As such, the OTOC in these models will again exhibit some transient dynamics, as in maximally chaotic dual-unitary circuits, before converging to a steady-state value,  where parity effects will turn out to be crucial.

In order to explicitly construct eigenoperators, we can make use of the relations
\begin{align}\label{eq:XY_rightIdentities}
&U(\mathbbm{1} \otimes \mathbbm{1})U^{\dagger} =\mathbbm{1} \otimes \mathbbm{1}, \quad U(\sigma_x \otimes \sigma_x)U^{\dagger} = \sigma_x \otimes \sigma_x , \nonumber\\
&U(\mathbbm{1} \otimes \sigma_y)U^{\dagger} = \sigma_y \otimes \sigma_x, \quad U(\sigma_x \otimes \sigma_z)U^{\dagger} =  \sigma_z \otimes \mathbbm{1}.
\end{align}
and 
\begin{align}\label{eq:XY_leftIdentities}
&U^{\dagger}(\mathbbm{1} \otimes \mathbbm{1})U =\mathbbm{1} \otimes \mathbbm{1}, \quad U^{\dagger}(\sigma_x \otimes \sigma_x)U = \sigma_x \otimes \sigma_x , \nonumber\\
& U^{\dagger}(\sigma_y \otimes \sigma_x)U =  \mathbbm{1} \otimes \sigma_y, \quad U^{\dagger}(\sigma_z \otimes \mathbbm{1})U = \sigma_x \otimes \sigma_z.
\end{align}
These are graphically represented in Appendix \ref{app:IdentitiesXY}.  Note that these are not all independent and the identities for $\mathbbm{1}\otimes \mathbbm{1}$ are a simple rewriting of unitarity, but written in this way they can be used to construct a set of eigenoperators with eigenvalue $1$ as product states. In this model left and right eigenoperators differ, so we will first focus on the construction of right eigenoperators using Eqs.~\eqref{eq:XY_rightIdentities}. While the number of eigenoperators that can be constructed in this way is exponentially large, a large part of these eigenoperators will be irrelevant for the calculation of the OTOC -- demanding a non-zero overlap between the eigenoperators and the left or right boundary fixes the eigenoperators to be symmetric w.r.t. space inversion. Defining
\begin{align}
&|\{r_1 r_2 \dots  r_n\}) \nonumber\\
&= |\sigma(0, r_1)\ \sigma(r_1,r_2) \dots \sigma(r_{n-1}, r_n) \nonumber \\
& \qquad \qquad \sigma(r_{n-1}, r_n) \dots \sigma(r_1, r_2)\ \sigma(0,r_1)) \\
&=\vcenter{\hbox{\includegraphics[width=0.95\linewidth]{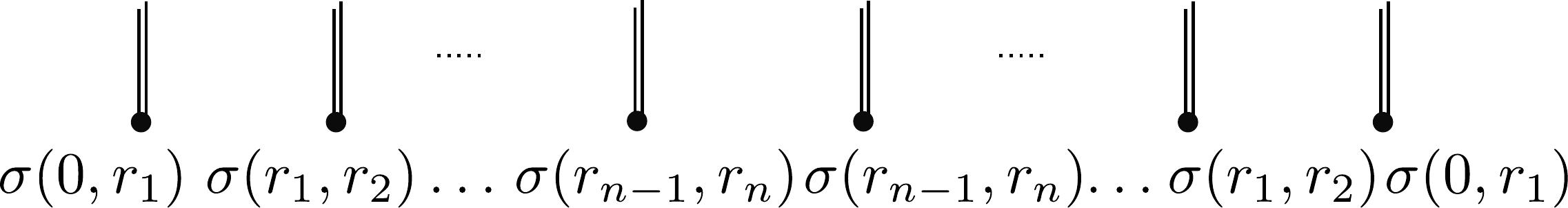}}} \nonumber
\end{align}
with $r_i \in \{0,1\}$ and $(r_{i-1},r_{i})$ determining the operator on leg $i$ and $2n+1-i$ as
\begin{align}
\sigma(0,0) = \mathbbm{1},\  \sigma(0,1) = \sigma_y,\ \sigma(1,0) = \sigma_z, \  \sigma(1,1) = \sigma_x,
\end{align}
with implicit $r_0=0$. It can easily be checked from Eqs.~\eqref{eq:XY_rightIdentities} that every choice of $\{r_1, r_2, \dots , r_n\}$ leads to an eigenoperator with eigenvalue $1$, which is graphically illustrated in Appendix \ref{app:IdentitiesXY}. Numerically, it can be checked that the resulting set of $2^n$ eigenoperators seems to exhaust all eigenoperators with a non-vanishing overlap with the left and right boundaries for small $n$, and we conjecture that this holds for arbitrary $n$.

The same procedure can be followed for the left eigenoperators, using the same symmetry constraint and Eqs.~\eqref{eq:XY_leftIdentities}. Any eigenoperator is now denoted as $(\{l_1 l_2 \dots l_{n}\}|$ with $l_i \in \{0,1\}$, and can be constructed as
\begin{align}
&(\{l_1 l_2 \dots l_n\}| \nonumber\\
&=(I_n(l_{n-1},l_n)\dots I_2(l_1,l_2)I_1(0,l_1)\nonumber\\
&\qquad \qquad I_1(0,l_1)I_2(l_2,l_1)\dots I_n(l_{n-1},l_n)| \\
&=\vcenter{\hbox{\includegraphics[width=0.95\linewidth]{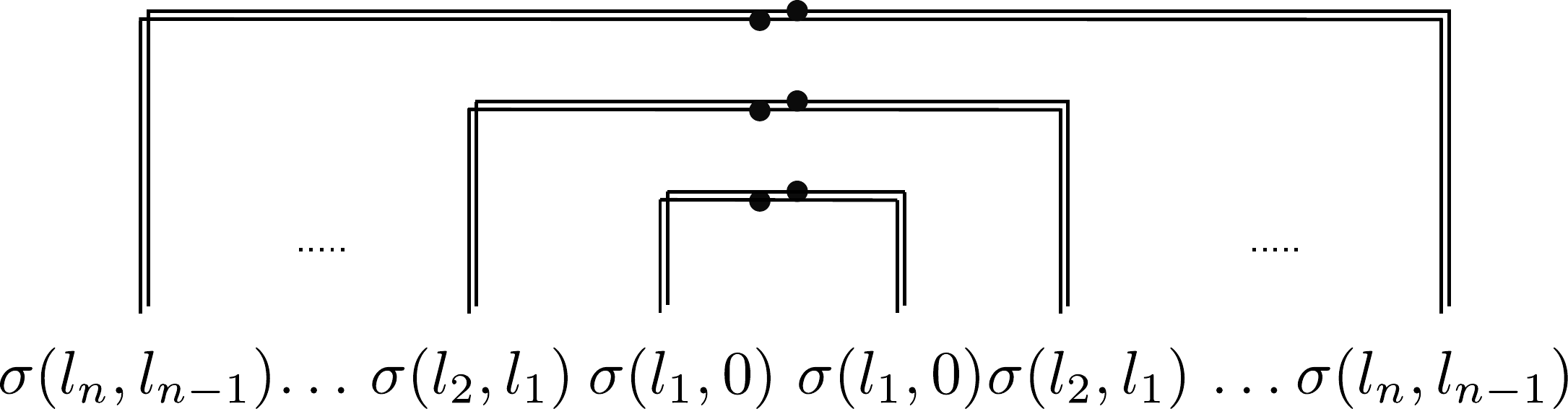}}} \nonumber
\end{align}
with $I(l_{i},l_{i+1}) = I^{\sigma(l_{i+1},l_i)}$ given by
\begin{align}
I(0,0) = I, \ I(0,1) = I^z, \ I(1,0) = I^{y}, \ I(1,1) = I^{x}.
\end{align}
The pair of coefficients $(l_{i-1}, l_{i})$ now determines the operators connecting leg $(n+1-i)$ and $(n+i)$. For small $n$ these operators again seem to exhaust all left eigenoperators with eigenvalue $1$, leading to a set of $2^n$ left and right eigenoperators of $T_n$. However, these do not yet form a dual basis. The overlap between a left and right eigenoperator is generally non-zero and can be obtained as
\begin{align}\
&(\{l_1, l_2 \dots l_n\}| \{r_1, r_2 \dots r_n\}) \nonumber\\
&\qquad= 2^n (-1)^{r_1 \cdot l_{n-2}+r_2 \cdot l_{n-3}+\dots + r_{n-3} \cdot l_2  + r_{n-2} \cdot l_{1}} \nonumber\\
&\qquad\qquad \times (-1)^{r_1 \cdot l_n + r_2 \cdot l_{n-1}+ \dots + r_{n-1}\cdot l_2+ r_{n}\cdot l_1} \label{eq:overlap_XY}
\end{align}
since the overlap consists of the product of the overlaps on leg $i$ and $2n+1-i$. Considering leg $i$, the right operator follows from $(r_{i-1},r_i)$ and the left one from $(l_{n-i},l_{n-i+1})$, leading to a factor
\begin{align}
&\tr\left(\sigma(r_{i-1},r_{i})\sigma(l_{n-i+1},l_{n-i})\sigma(r_{i-1},r_{i})\sigma(l_{n-i+1},l_{n-i})\right) \nonumber\\
&\qquad= 2 (-1)^{r_{i-1}\cdot l_{n-i}+r_{i} \cdot l_{n-i+1}},
\end{align}
following from the explicit definition of these operators, with again implicit $r_0=l_0=0$. The overlap matrix has the property that it is an orthonormal matrix (up to normalization, see Appendix \ref{app:IdentitiesXY}), such that we can construct a properly orthonormalized dual basis by choosing $\{l_1 \dots l_{n}\}$ as labels and writing
\begin{align}
&|R(\{l_1 \dots l_n\}) = \frac{1}{2^{2n}} \nonumber\\
&\qquad \times\sum_{r_1 \dots r_n} |\{r_1 \dots r_n\})(\{l_1 \dots l_n\}|\{r_1 \dots r_n\}), \nonumber\\
&(L(\{l_1 \dots l_n\})| = \frac{1}{2^n}(\{l_1 \dots l_n\}|,
\end{align}
satisfying
\begin{align}
(L(\{l_1 \dots l_n\})|R(\{l_1' \dots l_n'\})= \delta_{l_{1},l_{1}'}\delta_{l_{2},l_{2}'}\dots \delta_{l_{n},l_{n}'}.
\end{align}
The long-time value of the OTOC can now be evaluated using the usual construction,
\begin{align}
&\lim_{m\to \infty} \left(L_n(\sigma_{\alpha})\right|\left(T_n\right)^m\left|R^{\pm}(\sigma_{\beta})\right) \nonumber\\
&=\sum_{l_1 \dots l_n} \left(L_n(\sigma_{\alpha})|R(\{l_1  \dots l_n\}\right)\left(L(\{l_1  \dots l_n\})|R^{\pm}(\sigma_{\beta})\right) \nonumber\\
&=\frac{1}{2^{3n}}\sum_{\substack{r_1 \dots r_n \\ l_1 \dots l_n}}\left(L_n(\sigma_{\alpha})|\{r_1 \dots r_n\}\right)(\{l_1 \dots l_n\}|R^{\pm}(\sigma_{\beta}))\nonumber\\
&\qquad \qquad \qquad \qquad \times \left(\{l_1 \dots l_n\}|\{r_1 \dots r_n\}\right).
\end{align}
The overlap between the left and right boundaries can be evaluated in a similar way as the overlaps between eigenoperators, where it is important to note that these will only depend on the operators on either the outer (right boundary) or inner (left boundary) legs, and hence on $(l_{n-1},l_{n})$ and $(r_{n-1},r_n)$, leading to
\begin{align}
&\left(L_n(\sigma_{\alpha})|\{r_1 \dots r_n\}\right)   \nonumber\\
&\qquad=2^{n/2-1}\tr\left(\sigma_{\alpha}\sigma(r_{n-1},r_{n})\sigma_{\alpha}\sigma(r_{n-1}, r_{n})\right), \\
&\left(\{l_1 \dots l_n\}|R_n^+(\sigma_{\beta})\right)   \nonumber\\
&\qquad=2^{n/2-1}\tr\left(\sigma_{\beta}\sigma(l_{n},l_{n-1})\sigma_{\beta}\sigma(l_{n},l_{n-1})\right), \\
&\left(\{l_1 \dots l_n\}|R_n^-(\sigma_{\beta})\right)   \nonumber\\
&\qquad=2^{n/2-1}\tr\left((1-l_n)\mathbbm{1}+l_n \sigma_{\beta}\sigma_x\sigma_{\beta}\sigma_x\right), 
\end{align}
where the overlap with $|R_n^-(\sigma_{\beta}))$ can be simplified since all contractions can be evaluated, and the final result will depend on the operator on the horizonal line (see Appendix \ref{app:IdentitiesXY}), which is $\mathbbm{1}$ for $l_n=0$ and $\sigma_x$ for $l_n=1$.

Expressing the overlaps as a phase following Eq.~\eqref{eq:overlap_XY}, the summation over $l_{1}\dots l_{n-2}$ can be evaluated to return $2^{n-2} \delta_{r_1, r_3}\delta_{r_2,r_4}\dots \delta_{r_{n-2},r_n}$. The only remaining summations run over $(r_{n-1}, r_{n})$ and $(l_{n-1}, l_{n})$, where the remaining phase $(-1)^{r_1\cdot l_n + r_2 \cdot l_{n-1}}$ will either return $(-1)^{r_{n-1} \cdot l_n + r_{n}\cdot l_{n-1}}$ for $n$ even, or  $(-1)^{r_{n} \cdot l_n + r_{n-1}\cdot l_{n-1}}$ for $n$ odd. For $n$ even, the final value follows as
\begin{align}
&\frac{1}{4} \sum_{\substack{r_{n-1}, r_{n} \\ l_{n-1}, l_{n}}} (-1)^{r_{n-1}\cdot l_{n}+r_n\cdot l_{n-1}} \nonumber\\
&\qquad \qquad \times\tr\left(\sigma_{\alpha}\sigma(r_{n-1}, r_{n})\sigma_{\alpha}\sigma(r_{n-1}, r_{n})\right) \nonumber\\
&\qquad \qquad \times\tr\left(\sigma_{\beta}\sigma(l_n,l_{n-1})\sigma_{\beta}\sigma(l_n,l_{n-1})\right),
\end{align}
while for $n$ odd this follows as
\begin{align}
&\frac{1}{4} \sum_{\substack{r_{n-1}, r_{n} \\ l_{n-1}, l_{n}}} (-1)^{r_{n-1}\cdot l_{n-1}+r_n\cdot l_{n}} \nonumber\\
&\qquad \qquad \times \tr\left(\sigma_{\alpha}\sigma(r_{n-1}, r_{n})\sigma_{\alpha}\sigma(r_{n-1}, r_{n})\right)\nonumber\\
&\qquad \qquad \times \tr\left(\sigma_{\beta}\sigma(l_n,l_{n-1})\sigma_{\beta}\sigma(l_n,l_{n-1})\right).
\end{align}
These already highlight how the OTOC will not decay within the light cone, since the final value only depends on the parity of $n$ rather than its explicit value. The summations can be explicitly evaluated to obtain the long-time value of the OTOC, which will lead to two possible values for $C^-_{\alpha \beta}(x,t)$ and three possible final values for $C^+_{\alpha \beta}(x,t)$: $n=n_{-}$ either even or odd, and the case $n=1$ needs to be treated separately because there is no summation over $l_0=r_0=0$. 

\begin{figure}[htb!]
\includegraphics[width=\columnwidth]{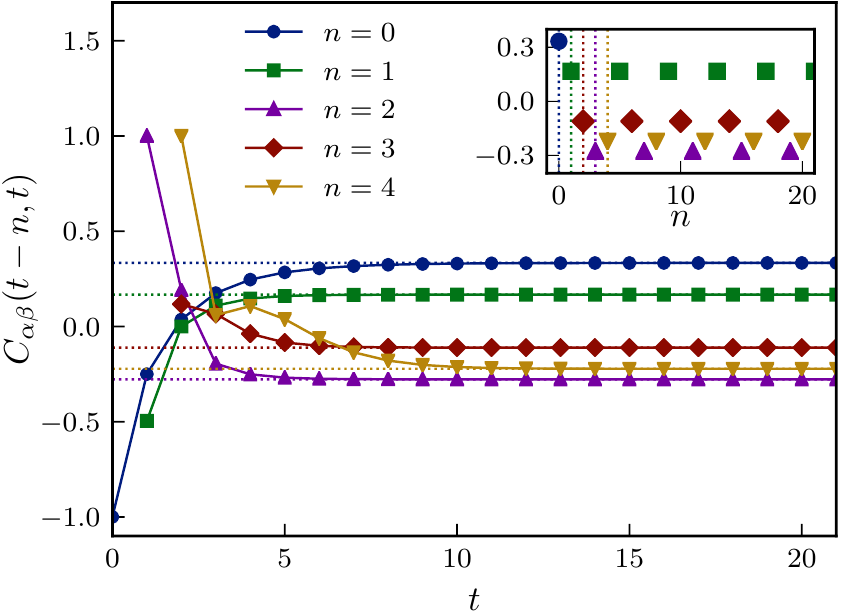}
\caption{Values of $C_{\alpha\beta}(x,t)$ for $x=t-n$, $\sigma_{\alpha} = \sigma_x/\sqrt{6} + \sigma_y/\sqrt{2} + \sigma_z/\sqrt{3}$, and $\sigma_{\beta} =\sigma_x/\sqrt{6} - \sigma_y/\sqrt{2} + \sigma_z/\sqrt{3}$ for evolution using the kicked XY model with $J_z = \pi/10$. Dotted lines represent the analytic results for $\lim (x+t)\to \infty$, and the inset details $\lim_{(x+t)\to \infty}C(t-n,t)$ for a larger range of $n$.
 \label{fig:XY}}
\vspace{-1\baselineskip}
\end{figure}

The resulting long-time values of the OTOC $\lim_{(x+t)\to\infty}C_{\alpha \beta}(x,t)$ then follow as
\begin{align}
\begin{cases}
\alpha_y^2 + (1-\alpha_y^2)(2\beta_z^2-1) \  &\text{if} \ x=t,\\
2\left(\beta_x^2 \alpha_x^2+\beta_y^2\alpha_y^2+\beta_z^2 \alpha_z^2\right)-1 \  &\text{if} \ (t-x)\in 4\mathbbm{N}, \\
\alpha_y^2+(1-\alpha_y^2)(2 \beta_x^2-1) \  &\text{if} \  (t-x)\in 4\mathbbm{N}+1, \\
2\left(\beta_x^2 \alpha_x^2+\beta_y^2\alpha_z^2+\beta_z^2 \alpha_y^2\right)-1 \  &\text{if} \ (t-x)\in 4\mathbbm{N}+2, \\
\alpha_z^2+(1-\alpha_z^2)(2\beta_x^2-1) \  &\text{if} \  (t-x)\in 4\mathbbm{N}+3, 
\end{cases}
\end{align}
This is illustrated in Fig. \ref{fig:XY}, where the 5 possible limiting values can be clearly observed after an initial transient regime.

The correlation functions on the light cone can similarly be evaluated by constructing the eigenoperators of the quantum channels $\mathcal{M}_{\pm}$, as is done in Appendix \ref{app:IdentitiesXY} and illustrated in Fig.~\ref{fig:corrXY}. Unlike the integrable KIM, these now decay exponentially to a zero value as
\begin{align}
\langle \sigma_{\alpha}(t,t) \sigma_{\beta}(0,0) \rangle = 
\begin{cases}
\delta_{\alpha \beta} \quad &\textrm{if} \quad t=0, \\ 
\alpha_x \beta_x \sin(2J)^t \quad &\textrm{if} \quad t>0.
\end{cases}
\end{align}
While the OTOC decays to non-zero values inside the light cone, the correlation functions in this model decay exponentially to zero as $\sin(2J)^t$. These do not decay for $J=\pm \pi/4$, which is exactly when the model becomes dual-unitary and returns the correlation functions of the self-dual integrable KIM.

\begin{figure}[htb!]
\includegraphics[width=\columnwidth]{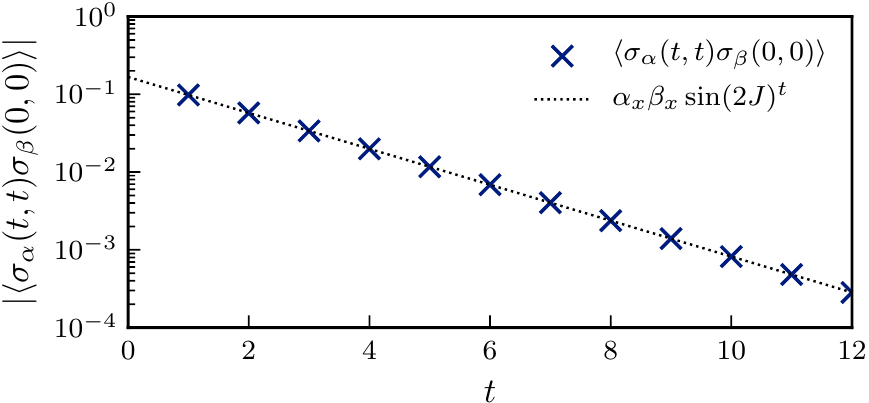}
\caption{Evolution of the correlation function on the light cone  $\langle \sigma_{\alpha}(t,t) \sigma_{\beta}(0,0) \rangle$ for the kicked XY model and $\sigma_{\alpha,\beta}$ parametrized as in Fig.~\ref{fig:XY}.
 \label{fig:corrXY}}
\vspace{-1\baselineskip}
\end{figure}

\section{Conclusions}\label{sec:conclude}
We have provided analytical results for the long-time behaviour of out-of-time-order correlators (OTOCs) in maximal velocity circuits (MVCs). Representing general OTOCs in a transfer matrix formalism, a maximal butterfly velocity $v_B = 1$ implies the existence of non-trivial eigenoperators of the transfer matrix with eigenvalue one. This provides both a criterion for maximal velocity circuits and a way of evaluating the long-time limit of the OTOCs using the resulting eigenoperators, as was done for two classes of MVCs: dual-unitary models and kicked XY models. These did not require the usual averaging over random local unitaries in analytically-tractable chaotic systems but rather hold for any realization of the quantum circuit (including, but not restricted to, Floquet models).

The resulting behaviour for the OTOCs in ergodic MVCs differs from that in generic unitary circuits not only in the sense that $v_B=1$, but also in the absence of a diffusively-broadening front: at long times the OTOC takes a maximal value on the light cone and decays exponentially away from the light cone, consistent with recent numerical observations \cite{von_keyserlingk_operator_2018,zhou_entanglement_2019}. Furthermore, this exponential decay of the OTOC is governed by the same quantum channels that fully determine the correlation functions, connecting the scrambling of quantum information with the relaxation of excitations towards equilibrium. This was observed both in maximally chaotic dual-unitary circuits and non-integrable kicked Ising models at the self-dual point. Apart from ergodic models, these MVCs also contain non-ergodic integrable classes (the self-dual kicked Ising model at the integrable point and kicked XY models), where no such exponential decay is observed. Rather, the OTOC immediately saturates to a constant value inside the light cone, whereas the correlation functions on the light cone can either exhibit a similar saturation or decay to zero.

We close with a natural question that merits further study: is it possible to completely characterize the set of MVCs, starting with the case $q=2$? A necessary condition is that the transfer matrix $T_1$ has at least one additional unit eigenvalue eigenoperator. Using the explicit parametrization for $q=2$ unitary gates (following, e.g., Refs.~\cite{bertini_exact_2019,kraus_optimal_2001,vatan_optimal_2004}), our numerical investigations suggest that in this case all models for which such an additional eigenoperator exists are either dual unitary or gauge-equivalent to the kicked XY models. However, for $q>2$ the problem is much more involved. Numerically, both the construction and diagonalization of the transfer matrix grows exponentially harder with increasing dimension of the local Hilbert space. Theoretically, dual-unitary circuits can be constructed at arbitrary $q$, but for $q=2$ it was already shown that these are only a subclass of all possible MVCs. While dual-unitary circuits for larger $q>2$ have been constructed building on complex Hadamard matrices \cite{gutkin_local_2020}, there is no guarantee that these exhaust all dual-unitary models. Even more, the full classification of complex Hadamard matrices itself remains an open problem \cite{tadej_concise_2006}.  Still, if such additional eigenoperators of $T_1$ with unit eigenvalue are known, our proposed construction can be straightforwardly extended to calculate the OTOCs in general MVCs.

\section*{Acknowledgements} 
We gratefully acknowledge support from EPSRC Grant No. EP/P034616/1. 
\vspace{\baselineskip}

\appendix
\newpage
\section{Explicit derivation of the OTOC diagram}
\label{app:derivationOTOCdiag}
Starting from the explicit definition of 
\begin{equation}
C_{\alpha \beta}(x,t) = \langle \sigma_{\alpha}(0,t) \sigma_{\beta}(x,0) \sigma_{\alpha}(0,t) \sigma_{\beta}(x,0) \rangle,
\end{equation}
the diagrams for the OTOC follow from the unitarity of the circuit, leading to a final diagram that consists of the intersection of the light cones of $\sigma_{\alpha}$ and $\sigma_{\beta}$, where the top and bottom legs along the vertical axis are connected. At each step, the proportionality factors are given by powers of $q$ following from tracing out local degrees of freedom, independent of the choice of $\sigma_{\alpha, \beta}$, such that the final prefactor can easily be obtained from $C_{00}(x,t) = \tr(\mathcal{U}^{\dag}(t) \mathcal{U}(t) \mathcal{U}^{\dag}(t) \mathcal{U}(t) )/\tr(\mathbbm{1}) =  \tr(\mathbbm{1})/\tr(\mathbbm{1}) = 1$.
\onecolumngrid
\begin{align}
\vcenter{\hbox{
\includegraphics[width=0.9\linewidth]{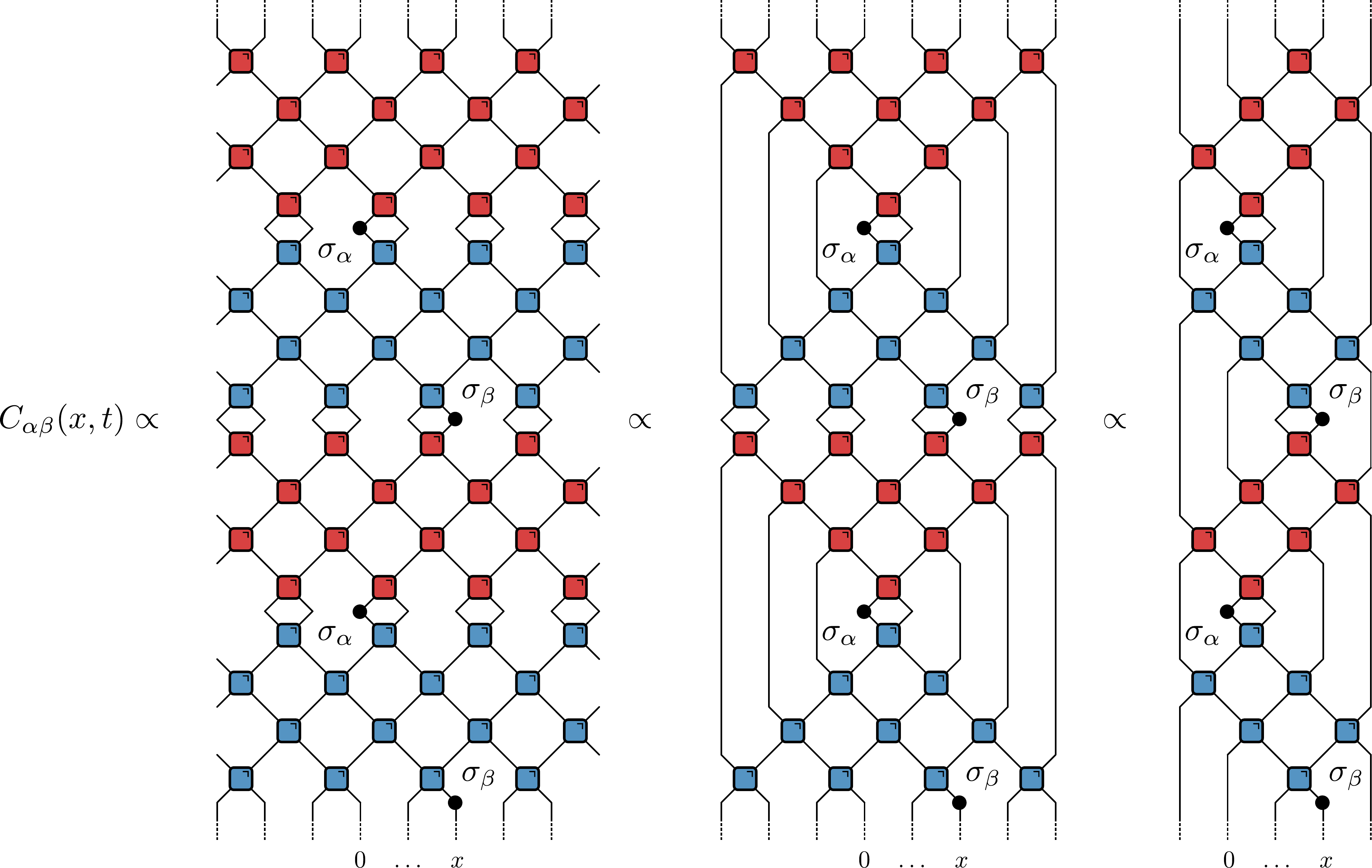}}} \label{eq:OTOC_tot}
\end{align}
\twocolumngrid

\section{Parametrization of dual-unitary matrices}
\label{app:dualunit}
For a local two-dimensional Hilbert space, any two-qubit unitary gate can be parametrized as
\begin{align}
U = e^{i \phi} (u_{+} \otimes u_{-}) V[J_x,J_y,J_z] (v_{-} \otimes v_{+}),
\end{align}
where $\phi, J_{x}, J_{y}, J_{z} \in \mathbbm{R}$, and $u_{\pm}, v_{\pm} \in SU(2)$ are one-qubit special-unitary matrices \cite{kraus_optimal_2001,vatan_optimal_2004}. All entanglement is generated by the two-qubit unitary
\begin{align}
&V[J_x,J_y,J_z] \nonumber\\
&\  = \exp \left[-i(J_x \sigma_x \otimes \sigma_x+J_y \sigma_y \otimes \sigma_y+J_z \sigma_z \otimes \sigma_z) \right].
\end{align}
As shown in Ref.~\cite{bertini_exact_2019}, dual-unitarity fixes two of the parameters in $V[J_x,J_y,J_z]$ as $J_x = J_y = \frac{\pi}{4}$, leaving $J_z$, as well as $\phi$ and $u_{\pm}, v_{\pm}$, free variables. Any permutation of $J_x, J_y, J_z$ works equally well, but can be brought in this parametrization through the $SU(2)$ rotations $u_{\pm}, v_{\pm}$. These one-qubit operators can be parametrized as $\exp\left[-i (n_x \sigma_x + n_y \sigma_y + n_z \sigma_z) \right]$, with $n_x, n_y, n_z \in \mathbbm{R}$. For the figures in the main text, random dual-unitary circuits were generated by choosing all parameters within these parametrizations randomly.

Note that no such full parametrizations exist for $q>2$, although Rather \emph{et al.} recently outlined a method for the generation of operators that are arbitrarily close to being dual-unitary \cite{rather_creating_2019} and Gutkin \emph{et al.} showed how it was possible to construct dual-unitary kicked models for arbitrary $q$ based on complex Hadamard matrices \cite{gutkin_local_2020}.

\section{Eigenvalues and eigenoperators for the KIM channel}
\label{app:MapKIM}

In this Appendix, we explicitly construct the quantum channel following from the KIM and its left and right eigenoperators (similar results were presented in Ref.~\cite{gutkin_local_2020} and are included here for completeness). Starting from the matrix elements of $U$, given by
\begin{align}
U_{ab,cd} =& -\frac{i}{2} \exp \left[i\frac{\pi}{4}(a-d)(c-b)\right] \nonumber\\
&\qquad \times \exp \left[-i\frac{h_1}{2}(a+c)-i\frac{h_2}{2}(b+d)\right],
\end{align}
with $a,b,c,d \in \{-1,1\}$, the linear map $\mathcal{M}_n$ can be found through an explicit construction of $\mathcal{M}_{\pm}$. Since $\mathcal{M}_{+}(\sigma) = \tr_1(U(\sigma \otimes \mathbbm{1})U^{\dagger})$, the matrix elements of $\mathcal{M}_{+}$ follow from
\begin{align}
\vcenter{\hbox{\includegraphics[width=0.75\linewidth]{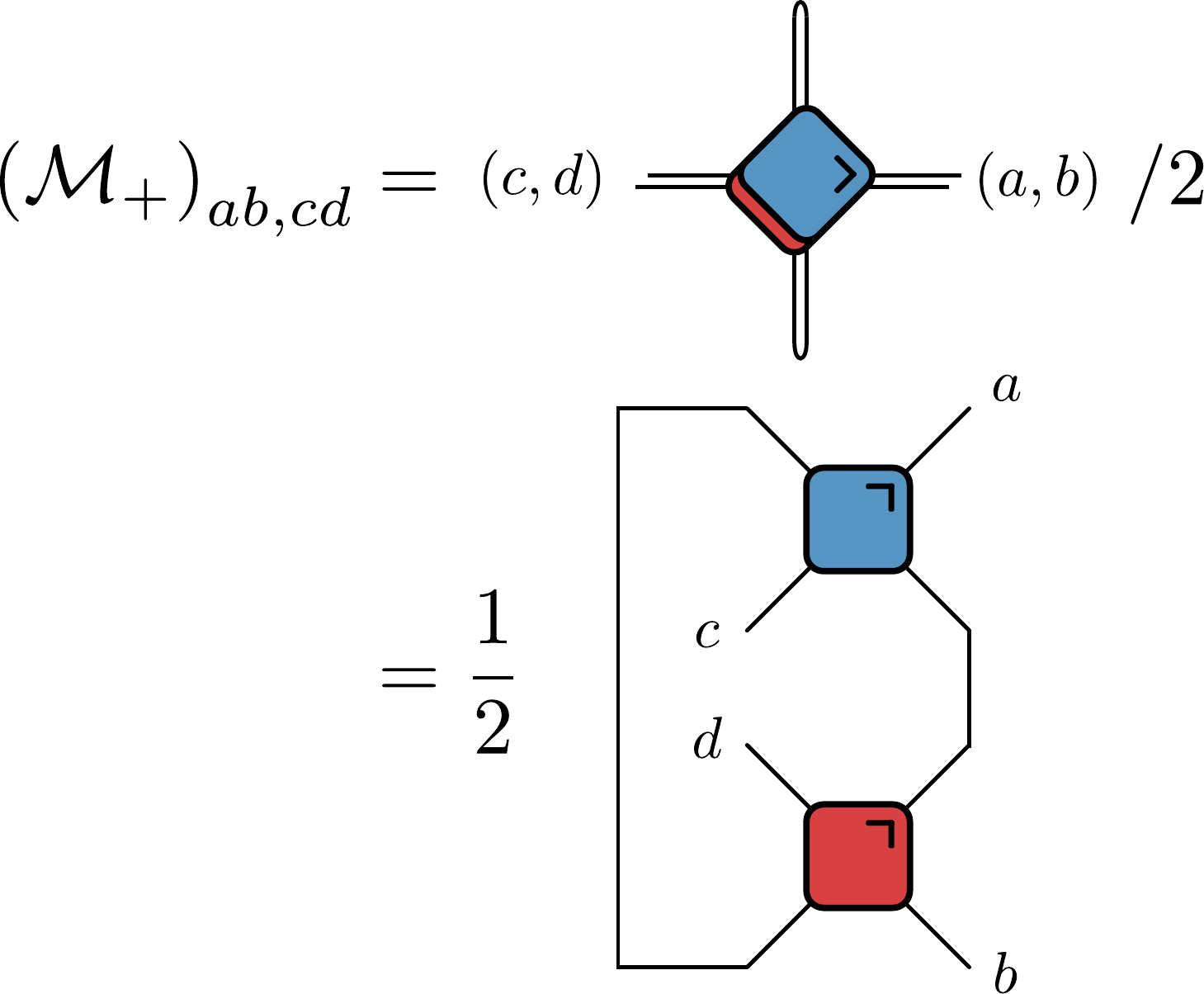}}} \label{eq:Mplus}
\end{align}
as $(\mathcal{M}_{+})_{ab,cd} = \frac{1}{2}\sum_{e,f}U_{ea,cf}(U^{\dagger})_{df,eb}$. Explicitly writing out these matrix elements returns
\begin{align}
&\left(\mathcal{M}_{+}\right)_{ab,cd} = \frac{1}{2} \exp\left[i \frac{h_1}{2}(b-a)+i\frac{h_2}{2}(d-c)\right] \nonumber\\
&\qquad\qquad \times \cos\left(\frac{\pi}{4}\left[(b-a)-(d-c)\right]\right)^2.
\end{align}
From the factor $\cos\left(\frac{\pi}{4}\left[(b-a)-(d-c)\right]\right)^2$ it follows that $\mathcal{M}_{+}$ maps diagonal $2 \times 2$ matrices to diagonal matrices, since a non-zero matrix element for $c=d$ requires $a=b$ (otherwise $a-b = \pm 2$ and the cosine vanishes), and maps off-diagonal matrices to off-diagonal matrices ($c \neq d$ similarly implies $a \neq b$ for a non-zero matrix element). In this way, the matrix can be block-diagonalized by expressing it in the basis $\{(1,1),(-1,-1), (1,-1), (-1,1)\}$,
\begin{align}
\left(\mathcal{M}_{+}\right)_{ab,cd} = \frac{1}{2} 
\begin{bmatrix}
1 & 1 & 0 & 0 \\
1 & 1 & 0 & 0 \\
0 & 0 & e^{-i(h_1+h_2)} & e^{i(h_1-h_2)} \\
0 & 0 & e^{-i(h_1-h_2)} & e^{i(h_1+h_2)}
\end{bmatrix},
\end{align}
where both blocks can be diagonalized to obtain the following eigenvalues $\lambda$ and right eigenoperators (re-expressed as operators rather than states)  
\begin{align}
&\lambda = 1 \rightarrow \begin{bmatrix}
1 & 0 \\
0 & 1
\end{bmatrix}, \qquad \lambda = 0 \rightarrow \begin{bmatrix}
1 & 0 \\
0 & -1
\end{bmatrix},  \\
&\lambda = \cos(h_1+h_2)\rightarrow 
\begin{bmatrix}
0 & e^{-i h_2} \\
e^{ih_2} & 0
\end{bmatrix}, \\
&\lambda=0 \rightarrow
\begin{bmatrix}
0 & -e^{i h_1}\\
e^{-i h_1} & 0
\end{bmatrix},
\end{align}
and an accompanying set of left eigenoperators with the same eigenvalues,
\begin{align}
&\lambda = 1 \rightarrow \begin{bmatrix}
1 & 0 \\
0 & 1
\end{bmatrix}, \qquad \lambda = 0 \rightarrow \begin{bmatrix}
1 & 0 \\
0 & -1
\end{bmatrix},  \\
&\lambda = \cos(h_1+h_2)\rightarrow 
\begin{bmatrix}
0 & e^{-i h_1} \\
e^{i h_1} & 0
\end{bmatrix}, \\
&\lambda=0 \rightarrow
\begin{bmatrix}
0 & -e^{ih_2}\\
e^{-ih_2} & 0
\end{bmatrix}.
\end{align}
Using the eigenvalue decomposition and denoting the eigenoperators with eigenvalue $\cos(h_1+h_2)$ as $|h_R)$ (right) and $(h_L|$ (left), normalized by a factor $\left[2\cos(h_1+h_2)\right]^{1/2}$ such that $(h_L|h_R)=\tr(h_L^T h_R) = 1$, and the eigenoperator from the identity matrix as $|1)$ and $(1|$, normalized by a factor $\sqrt{2}$  such that $(1|1)=1$, we obtain
\begin{align}
&\mathcal{M}_{+} = \cos(h_1+h_2) |h_R)(h_L| + |1)(1|, \nonumber\\
&\mathcal{M}_{-} = \cos(h_1+h_2) |\overline{h}_L)(\overline{h}_R| + |1)(1|,
\end{align}
where $|\overline{h}_l)$ is the Hermitian conjugate of $(h_L|$, with the complex conjugation made explicit. Assuming $n \neq 0$, the action of $\mathcal{M}_n$ follows as
\begin{equation}
\mathcal{M}_n(\sigma_{\beta}) = \frac{1}{2} \cos(h_1+h_2)^{2n} (\sigma_{\beta}|\overline{h}_L)(\overline{h}_R|h_R)(h_L|\sigma_{\beta}),
\end{equation}
for a traceless $\sigma_{\beta}$ with $(1|\sigma_{\beta})=0$, which can be evaluated as
\begin{equation}
\mathcal{M}_n(\sigma_{\beta}) = \left(\beta_x \cos(h_1)-\beta_y\sin(h_1)\right)^2\left(\cos(h_1+h_2)\right)^{2(n-1)}.
\end{equation}
Alternatively, for $n=0$, this leads to $\mathcal{M}_0(\sigma_{\beta}) = 1$.

\section{Identities for the integrable KIM}
\label{app:IdentitiesKIM}
At the integrable point, the KIM satisfies a set of identities of the form $U (\mathbbm{1} \otimes \sigma_{\alpha})U^{\dagger} = \sigma_{\alpha}\otimes \sigma_{\beta}$ and $U (\sigma_x \otimes \sigma_{\alpha}) U^{\dagger} = \sigma_{\alpha}\otimes \sigma_{\beta}$, with $\sigma_{\beta}$ either $\sigma_x$ or $\mathbbm{1}$, and these can be graphically represented as
\begin{align}
\vcenter{\hbox{\includegraphics[width=0.8\linewidth]{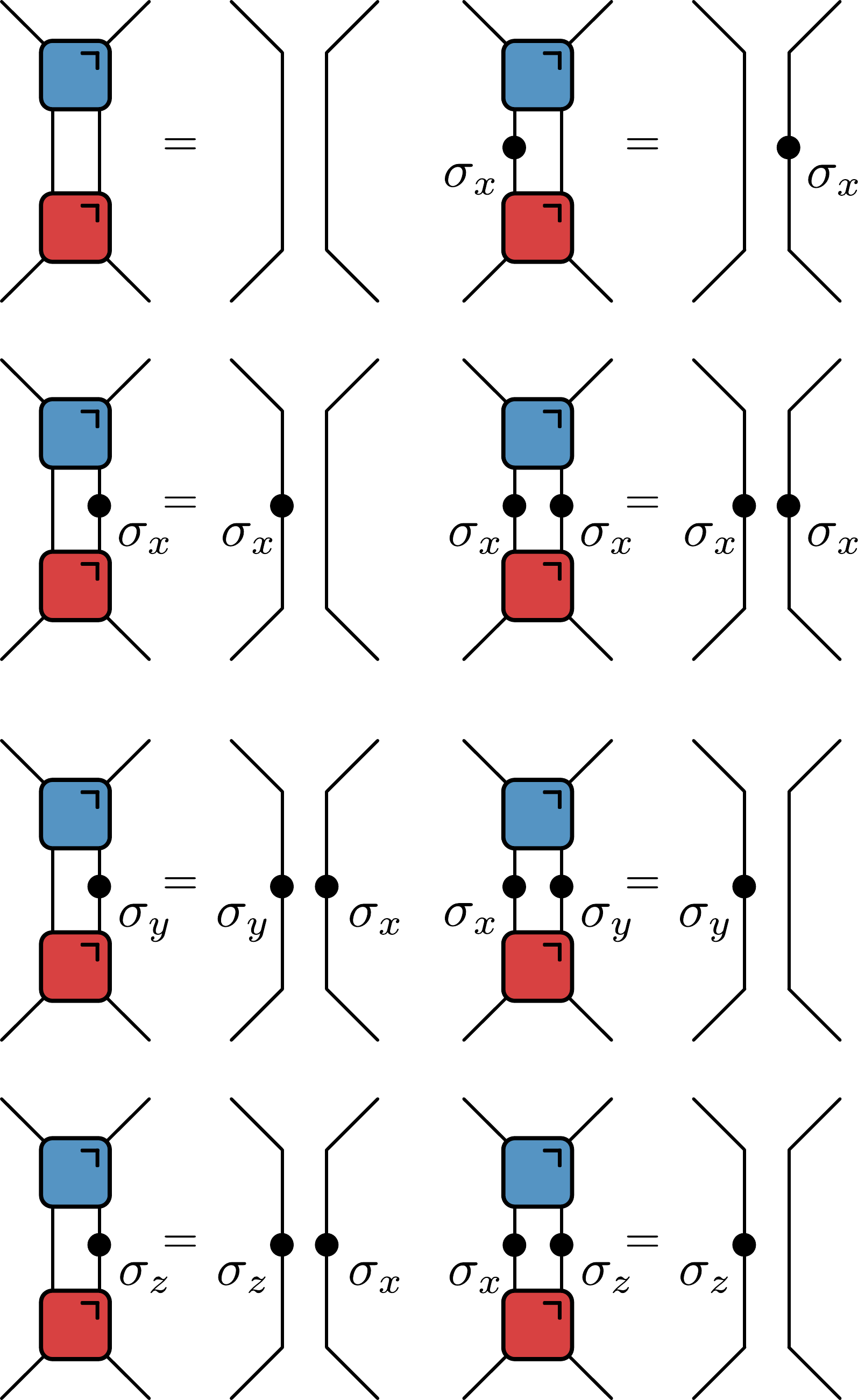}}}\ , \label{eq:diag_U_KIM_int}
\end{align}
or in the folded picture for $U$ as
\begin{align}
\vcenter{\hbox{
\includegraphics[width=0.9\linewidth]{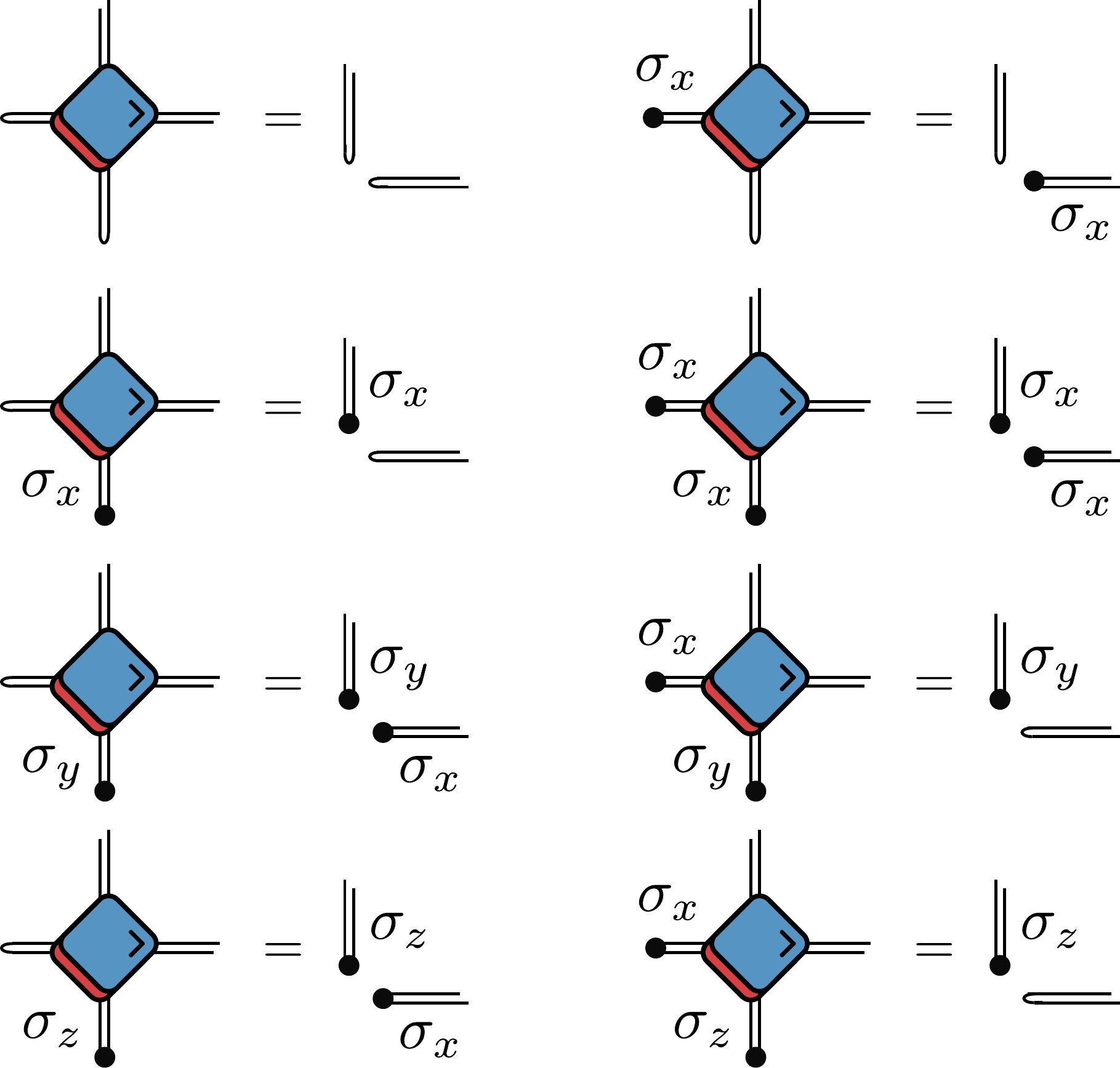}}}.
\end{align}
A similar set of folded identities holds for the folded version of $U^{\dagger}$
\begin{align}
\vcenter{\hbox{
\includegraphics[width=0.9\linewidth]{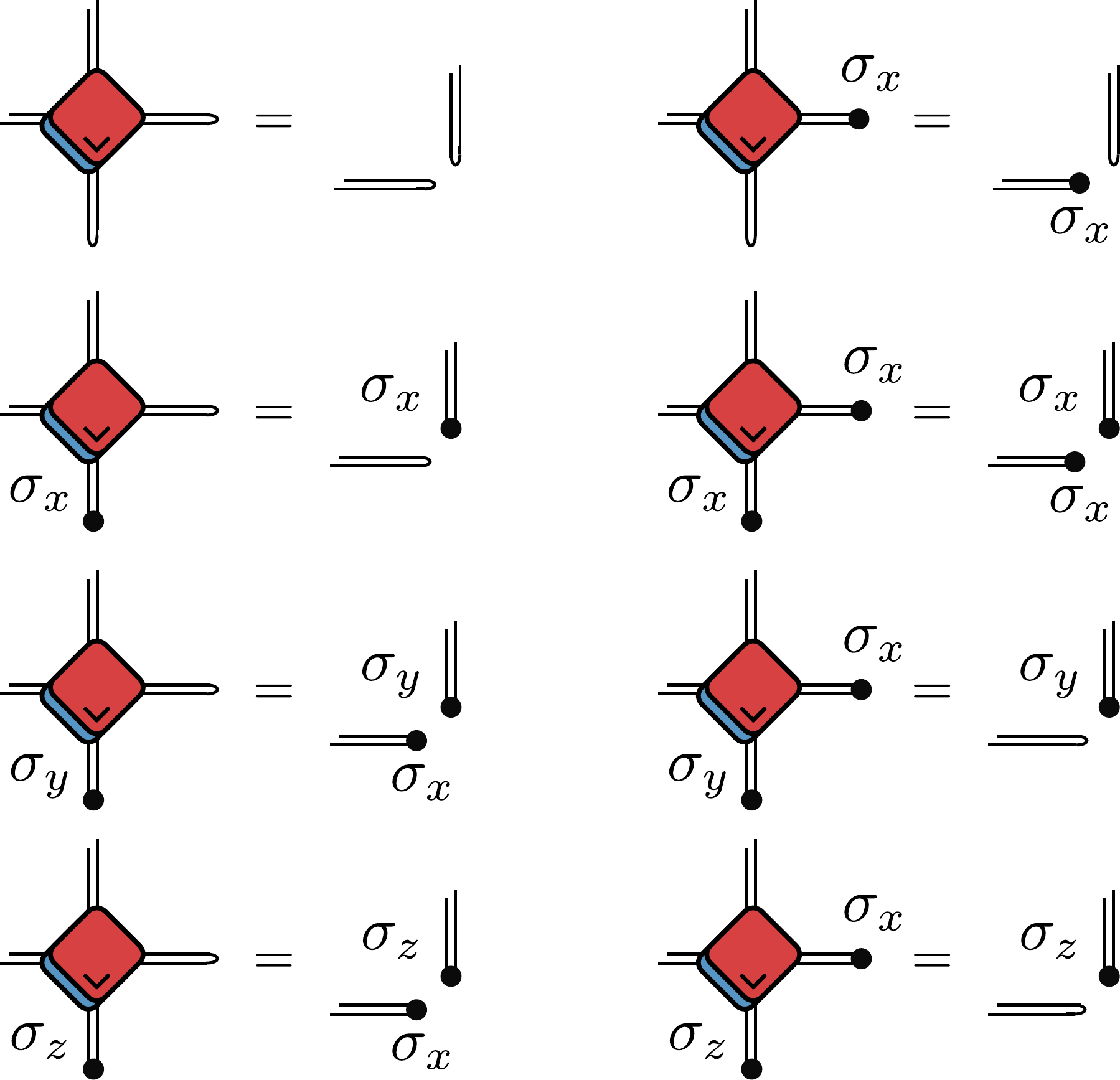}}}.
\end{align}
This way, any action of the transfer matrix on a right `product state' is porportional to the original product state with a prefactor proportional to the contraction of the horizonal loop, which is either $\tr(\sigma_x)/2=0$ if the combined total number of $\sigma_y$ and $\sigma_z$ is odd and $\tr(\mathbbm{1})/2=1$ if this is even.

\subsection*{Eigenoperators of the quantum channels}
The eigenvalues and eigenoperators of the quantum channels immediately follow by either setting $h_1=h_2=0$ in Appendix \ref{app:MapKIM} or from similar identities as presented in \eqref{eq:diag_U_KIM_int}. If $U(\sigma_{\alpha}\otimes \mathbbm{1})U^{\dagger} = \sigma_{\beta}\otimes \sigma_{\alpha}$, then $\sigma_{\alpha}$ is guaranteed to be an eigenoperator of $\mathcal{M}_{+}$, since
\begin{align}
\mathcal{M}_{+}(\sigma_{\alpha}) &=\frac{1}{2}\tr_{1}\left[U(\sigma_{\alpha}\otimes \mathbbm{1})U^{\dagger}\right] \nonumber\\
&=\frac{1}{2}\tr_{1}\left[\sigma_{\beta}\otimes \sigma_{\alpha}\right] = \frac{1}{2} \tr[\sigma_{\beta}] \sigma_{\alpha},
\end{align}
leading to an eigenvalue $\tr(\sigma_{\beta})/2$, which is either $0$ if $\sigma_{\beta}=\sigma_x$ or $1$ if $\sigma_{\beta}=\mathbbm{1}$. The channel $\mathcal{M}_{+}$ acts as a projector, with two eigenvalues $1$ with eigenoperators $\mathbbm{1}$ and $\sigma_x$ and two eigenvalues $0$ with eigenoperators $\sigma_y$ and $\sigma_z$.  Given a traceless $\sigma_{\beta}$ and $t>0$, this results in
\begin{equation}
\mathcal{M}^t_{+}(\sigma_{\beta}) = \beta_x \sigma_x,
\end{equation}
which can be used to evaluate the correlation functions on the light cone (\ref{eq:CorrChannels}) as
\begin{align}
\langle \sigma_{\alpha}(t,t)\sigma_{\beta}(0,0) \rangle &= \frac{1}{2} \beta_x \tr\left(\sigma_{\alpha}\sigma_x\right) = \alpha_x \beta_x,
\end{align}
returning the presented correlation functions from the main text.

\section{Identities for the Kicked XY Model}
\label{app:IdentitiesXY}
A similar, but more restricted, set of identities can be used to construct right operators of the transfer matrix for the kicked XY model. For the right eigenstates, these are given by Eqs.~\eqref{eq:XY_rightIdentities} and can be graphically represented as
\begin{align}
\vcenter{\hbox{
\includegraphics[width=0.9\linewidth]{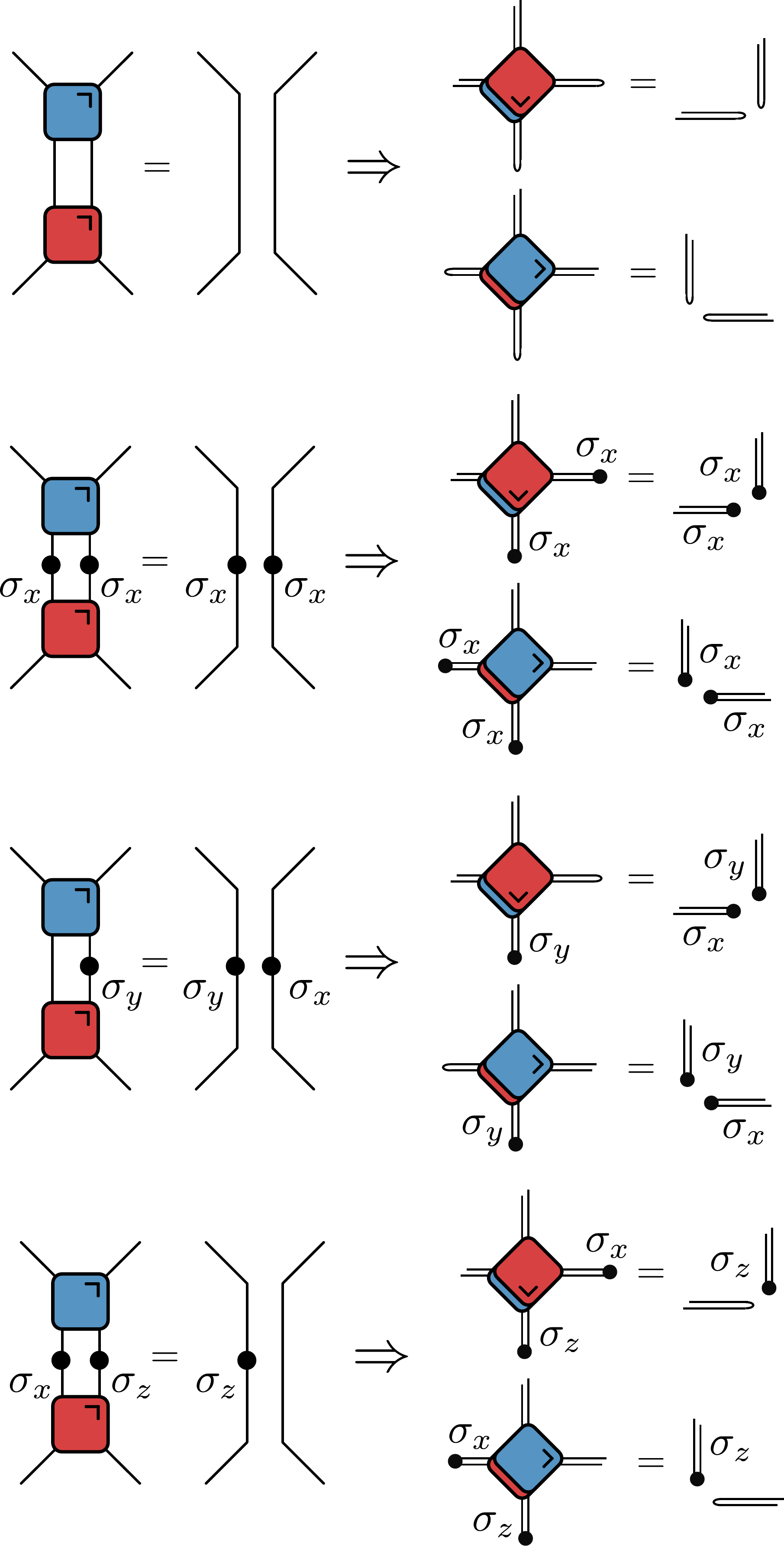}}} \label{eq:XY_identities_b}
\end{align}
Unlike the integrable self-dual KIM, not every product state is an eigenoperator. Considering e.g. a transfer matrix with $n=2$, the total number of unit-eigenvalue eigenoperators is given by $|\circ \circ \circ \circ), |\circ \sigma_{y} \sigma_{y} \circ), |\sigma_y \sigma_z \sigma_z \sigma_y), |\sigma_y \sigma_x \sigma_x \sigma_y), |\circ \circ \ \sigma_z \sigma_y), |\sigma_y \sigma_z \circ \circ), |\circ \sigma_y \sigma_x \sigma_y)$  and $|\sigma_y \sigma_x \sigma_y \circ)$. Of these, only the first four have a non-zero overlap with the left boundary. A systematic construction of these eigenoperators is possible by performing the contraction starting from the outer edges, e.g. the left leg: acting with the transfer matrix on either $\mathbbm{1}$ or $\sigma_{y}$ returns the same state on the (vertical) leg, where the horizontal leg contains the identity respectively $\sigma_x$. Given a horizonal contraction with $\sigma_x$ acting on the left leg, acting on either $\sigma_y$ or $\sigma_x$ returns the same operator on the vertical leg, with either the identity or $\sigma_x$ on the horizontal leg. To illustrate how these identities can be used to construct eigenoperators, consider $n=3$ and the action of the transfer matrix on a right eigenoperator $|\sigma_y \sigma_x \sigma_z \sigma_z \sigma_x \sigma_y)$, contracting from the outer edges in,
\begin{align}
\vcenter{\hbox{\includegraphics[width=0.8\linewidth]{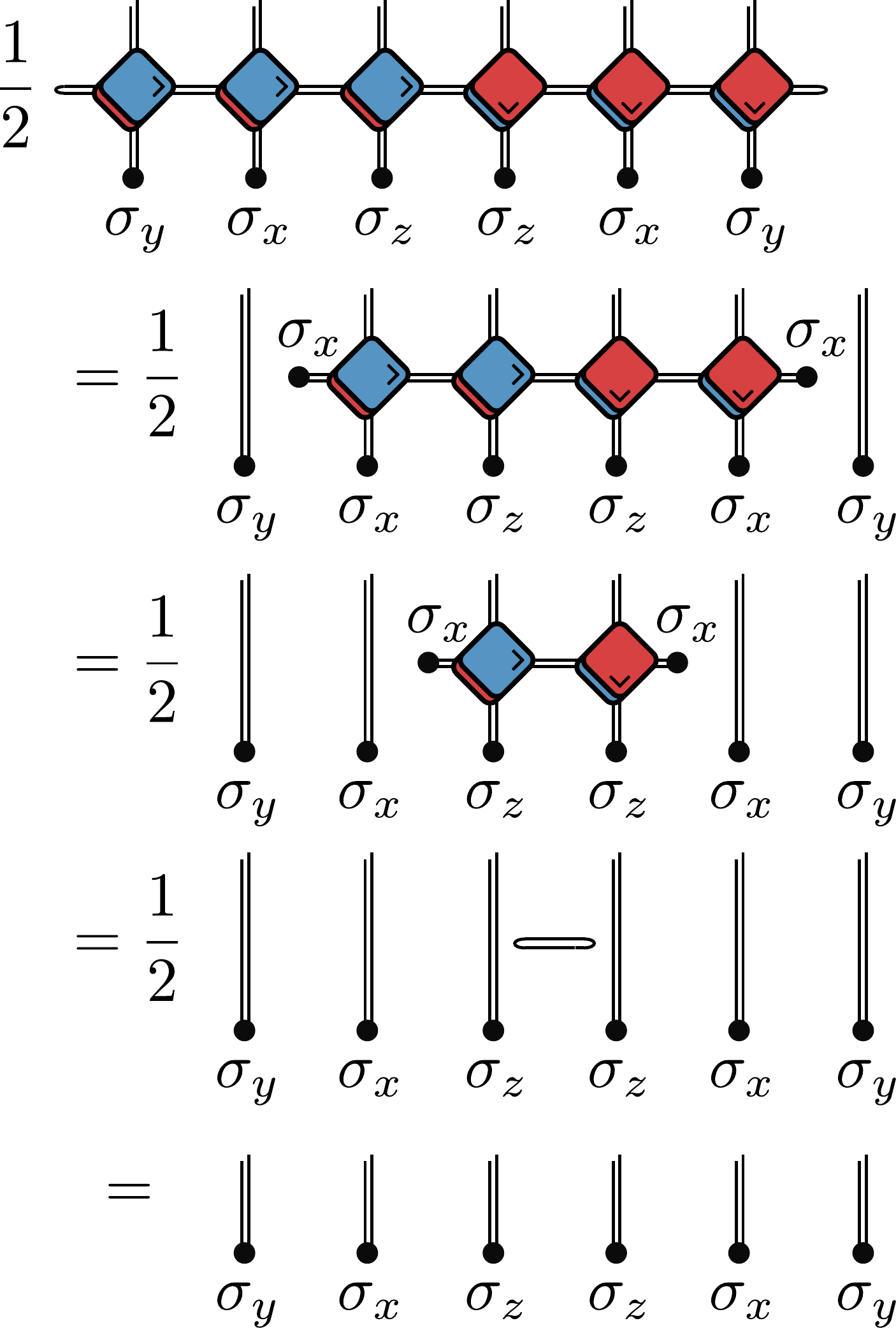}}}\ .
\end{align}

Introducing a notation to capture these constrainst, we can parametrize any symmetric eigenoperator with eigenvalue one by $n$ values $r_i \in \{0,1\}$, where the operator on leg $i$ equals the one on leg $2n+1-i$ and follows from $(r_{i-1}, r_{i})$: $(0,0)$ leads to $\mathbbm{1}$, $(1,0)$ to $\sigma_z$, $(0,1)$ to $\sigma_y$, $(1,1)$ to $\sigma_x$. Contracting the action of the transfer matrix on the product state from the outer edges in, $r_i=0$ denotes the presence of $\mathbbm{1}$ on the horizontal at the $i$-th step of the contraction. If the horizonal contraction is with the identity, $r_{i}=0$ and acting with the identity does not result in a $\sigma_x$, while acting with $\sigma_y$ introduces a $\sigma_x$ on the horizontal, whereas if the horizontal contraction is with $\sigma_x$, then $\sigma_x$ keeps the $\sigma_x$ on the horizontal intact while $\sigma_z$ converts this $\sigma_x$ into the identity, leading to the eigenoperators presented in the main text.

The necessary identities for the left eigenoperators are given by Eqs.~\eqref{eq:XY_leftIdentities}, which can be graphically represented as
\begin{align}
\vcenter{\hbox{\includegraphics[width=0.8\linewidth]{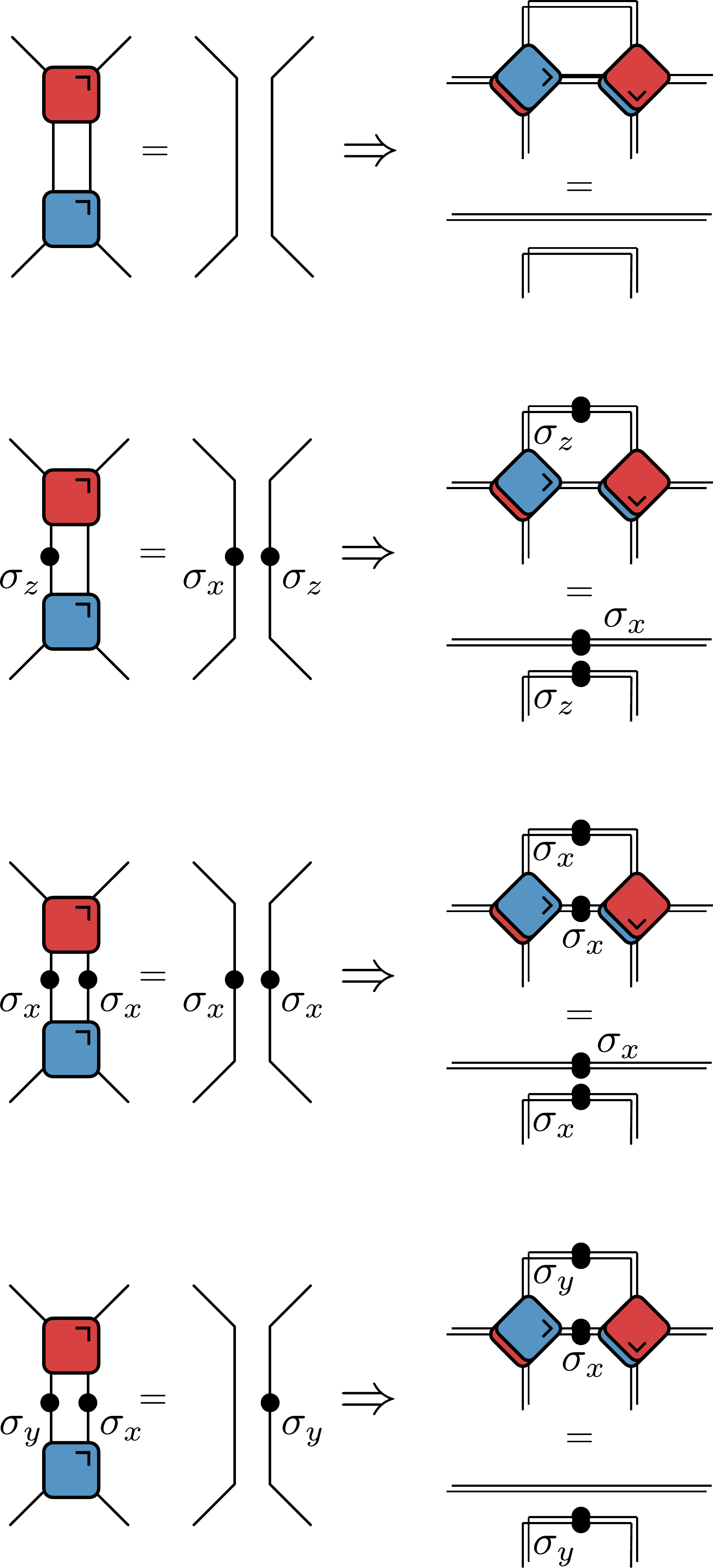}}}\label{eq:XY_identities_t}\ .
\end{align}
These identities now lead to the eigenoperators as presented in the main text, where the contraction is now most easily evaluated starting from the center, e.g. for $(I^x_3 I^z_2 I_1 I_1 I_2^z I_3^x)$,
\begin{align}
\vcenter{\hbox{\includegraphics[width=0.8\linewidth]{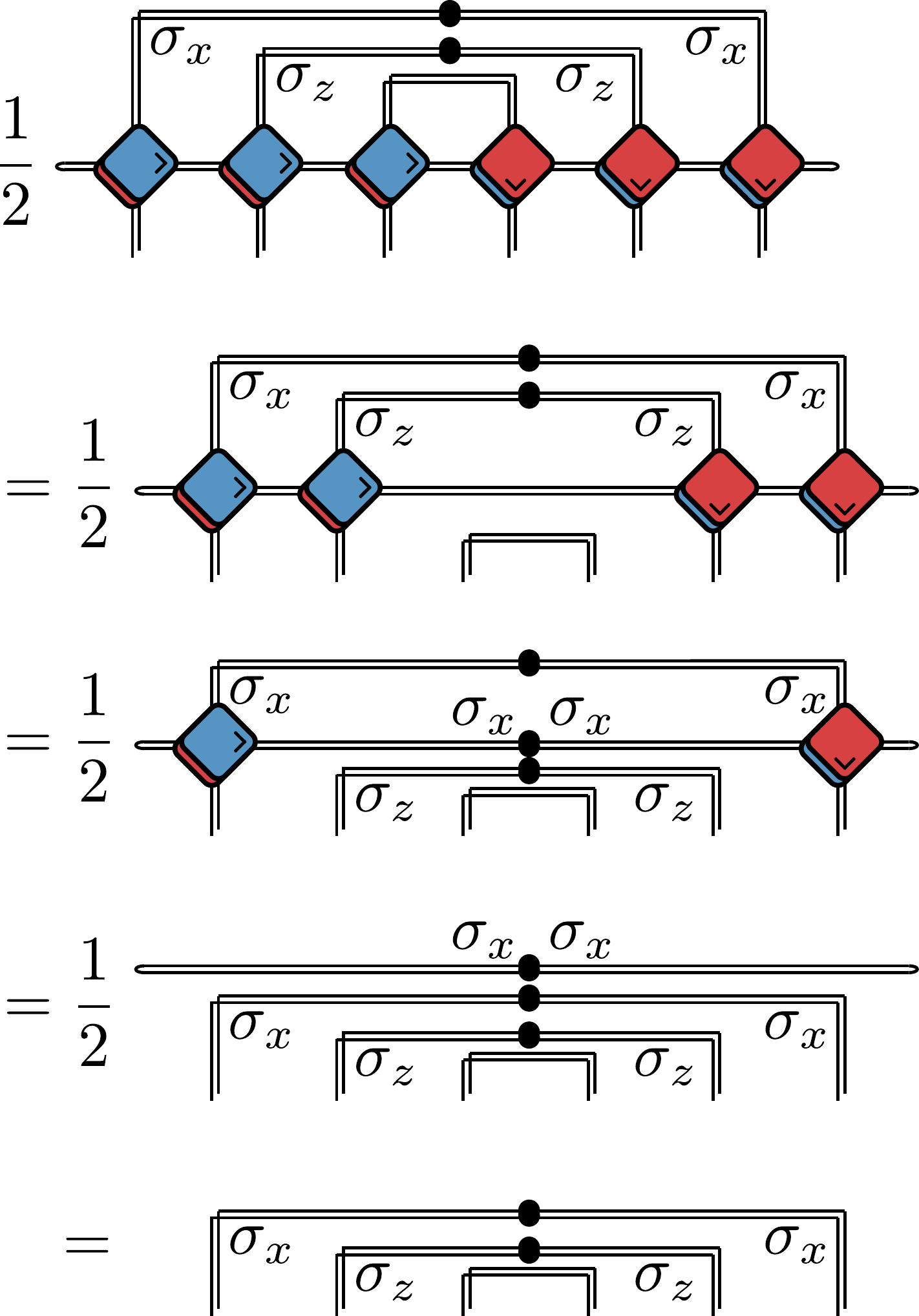}}}\ .
\end{align}

Considering the construction from the main text, these have the same interpretation of either introducing or cancelling a $\sigma_x$ on the horizontal contraction, where the roles of $\sigma_{y}$ and $\sigma_{z}$ have been exchanged, or $(0,1) \leftrightarrow (1,0)$. 

\subsection*{Eigenoperators of the quantum channels}
Using Eqs.~\eqref{eq:XY_identities_b} and \eqref{eq:XY_identities_t}, it follows that 
\begin{align}
\mathcal{M}_{+}(\sigma_y) = \mathcal{M}_{+}(\sigma_{z}) = 0,
\end{align}
which can be combined with $\mathcal{M}_{+}(\mathbbm{1})$ to return three of the four eigenvalues and eigenoperators of $\mathcal{M}_{+}$. The fourth will explicitly depend on $J$ and can be obtained from
\begin{align}
U(\sigma_x \otimes \mathbbm{1}) U^{\dagger} = \sin(2J) \mathbbm{1}\otimes \sigma_x + \cos(2J) \sigma_z \otimes \sigma_x,
\end{align}
such that
\begin{equation}
\mathcal{M}_{+}(\sigma_x) = \frac{1}{2}\tr_{1}\left[U(\sigma_x \otimes \mathbbm{1}) U^{\dagger}\right] = \sin(2J) \sigma_x,
\end{equation}
returning the fourth eigenvalue and eigenoperator as $\sigma_x$. For any traceless $\sigma_{\beta}$ and $t>0$, we then have
\begin{equation}
\mathcal{M}^t_{+}(\sigma_{\beta}) = \sin(2J)^t \beta_x \sigma_x,
\end{equation}
which can be used to evaluate the correlation functions on the light cone (\ref{eq:CorrChannels}) as
\begin{align}
\langle \sigma_{\alpha}(t,t)\sigma_{\beta}(0,0) \rangle &= \frac{1}{2} \sin(2J)^t \beta_x \tr\left(\sigma_{\alpha}\sigma_x\right) \nonumber\\
&= \sin(2J)^t \alpha_x \beta_x,
\end{align}
returning the presented correlation functions from the main text.

\subsection*{Orthogonality of the overlap matrix}
Using the notation of the main text, it can be checked that the overlap matrix between these left and right eigenstates is an orthonormal matrix, since
\begin{align}
&\sum_{r_1, \dots,r_n} (\{l_1\dots l_n\}| \{r_1  \dots r_n\})(\{l_1'  \dots l_n'\}| \{r_1  \dots r_n\}) \nonumber\\
&=2^{2n}\sum_{r_1=0,1} (-1)^{r_1((l_{n-2}-l_n)+(l_{n-2}'-l_n'))} \nonumber\\
&\qquad \times \sum_{r_2=0,1}(-1)^{r_2 ((l_{n-3}-l_{n-1})+(l_{n-3}'-l_{n-1}'))}  \nonumber\\
&\qquad \times \dots  \nonumber\\
&\qquad \times \sum_{r_{n-2}=0,1}(-1)^{r_{n-2}((l_{1}-l_{3})+(l_{1}'-l_{3}'))}  \nonumber\\
&\qquad \times \sum_{r_{n-1}=0,1}(-1)^{r_{n-1}(l_2+l_2')} \nonumber\\
&\qquad \times  \sum_{r_{n}=0,1}(-1)^{r_{n}(l_1+l_1')}.
\end{align}
Each summation results in a term
\begin{align*}
\sum_{r=0,1}(-1)^{r(l+l')} = 1 + (-1)^{l+l'} = 2 \delta_{l,l'},
\end{align*}
since $l,l' \in \{0,1\}$. The first $n-2$ summations vanish unless $(l_i-l_{i-2})=(l_{i}'-l_{i-2}')$ and the final two summations fix $\delta_{l_1, l_1'}$ and $\delta_{l_2, l_2'}$, such that the total summation can be evaluated as
\begin{align}
&\sum_{r_1, \dots,r_n} (\{l_1\dots l_n\}| \{r_1  \dots r_n\})(\{l_1'  \dots l_n'\}| \{r_1  \dots r_n\}) \nonumber\\
&\qquad\qquad \qquad =2^{3n} \delta_{l_{1},l_{1}'}\delta_{l_{2},l_{2}'}\dots \delta_{l_{n},l_{n}'},
\end{align}

\bibliography{LibraryOTOC.bib}

\end{document}